\documentclass[aps,a4paper,prd,amsmath,amssymb]{revtex4}
\usepackage{exscale,latexsym}
\usepackage[dvips]{graphicx}
\begin{document}
\preprint{QG@largeD}
\title{Quantum gravity at a large number of dimensions}
\author{N.~E.~J.~Bjerrum-Bohr}
\email{bjbohr@nbi.dk}
\affiliation{The Niels Bohr Institute,\\
Blegdamsvej 17, DK-2100 Copenhagen, Denmark}
\date{\today}
\begin{abstract}
We consider the large-$D$ limit of Einstein gravity. It is
observed that a consistent leading large-$D$ graph limit exists,
and that it is built up by a subclass of planar diagrams. The
graphs in the effective field theory extension of Einstein gravity
are investigated in the same context, and it is seen that an
effective field theory extension of the basic Einstein-Hilbert
theory will not upset the latter leading large-$D$ graph limit,
{\it i.e.}, the same subclass of planar diagrams will dominate at
large-$D$ in the effective field theory. The effective field
theory description of large-$D$ quantum gravity limit will be
renormalizable, and the resulting theory will thus be completely
well defined up to the Planck scale at $\sim 10^{19}$ GeV. The
$\left(\frac1D\right)$ expansion in gravity is compared to the
successful $\left(\frac1N\right)$ expansion in gauge theory (the
planar diagram limit), and dissimilarities and parallels of the
two expansions are discussed. We consider the expansion of the
effective field theory terms and we make some remarks on explicit
calculations of $n$-point functions.
\end{abstract}\maketitle
\section{Introduction}
A traditional belief is that an adequate description of quantum
gravity is necessarily complicated, and that we are very far from
a satisfying explanation of its true physical content. This might
be correct, however progress on the problems in quantum gravity
are being made, and gravitational interactions are today better
understood than previously. The fate of gravity in the limit of
Planckian scale energies is still unknown. At the physical
transition to such energies $\sim 10^{19}$ GeV, fundamental issues
are believed to arise. They concern the validity of the conceptual
picture of space and time geometry. A completely new understanding
of physics might be the consequence. String theory or the so
called $M$-theory are candidates for such theories in high energy
physics.

Einstein's theory of general relativity on the other hand presents
us with an excellent classical model for the interactions of
matter and space-time in the low energy regime of physics.
Astronomical observations support the theoretical predictions, and
there are no experimental reasons for a disbelief in the physical
foundations of general relativity. The fact that general
relativity is not a quantum theory is however intriguing. A
fundamental description of nature should either be completely
classical or quantum mechanical from a theoretical point of view,
and so far experiments support quantum mechanics as the correct
basis for interpreting physical measurements. It would be nice to
unify Einstein's theory in the classical low energy limit with a
full quantum field theoretic description in the high energy limit.
This seems not to be immediately possible, attempts of making a
quantum field theoretic description of general relativity have
normally proven to be incomplete in regard to the question of
renormalizability of the fundamental action for the theory. The
basic four-dimensional Einstein-Hilbert action:
\begin{equation}
{\cal L}_{\rm EH} = \int d^4x{\sqrt {-g}}\Big(\frac{2R}{\kappa^2}
+ {\cal L}_{\text{matter}}\Big),
\end{equation}
is not renormalizable in the sense that the counter-terms which
appear in the renormalized action, {\it e.g.}, ($\sim R^2$),
($\sim R_{\mu\nu}R^{\mu\nu}$) and ($\sim R^3),\ \ldots$ cannot be
absorbed in the original action~\cite{Veltman,NON,matter}. In the
above equation ($\kappa^2$) is defined as ($32\pi G_D$), where
($G_D$) is the $D$-dimensional Newton constant. The disappointing
conclusion is usually that a field theoretical treatment of
gravity is not possible, and that one should avoid attempting such
a description.

Effective field theory presents a solution to the renormalization
problem of gravity. It initially originates from an idea put
forward by Weinberg~\cite{Weinberg}. Including all possible
invariants into the action, {\it i.e.}, replacing the basic
$D$-dimensional Einstein-Hilbert action with:
\begin{equation}
{\cal L}_{\rm effective\ EH} = \int d^Dx{\sqrt
{-g}}\bigg(\Big(\frac{2R}{\kappa^2}+c_1R^2+c_2R^{\mu\nu}R_{\mu\nu}+\ldots\Big)+{\cal
L}_{\text{eff. matter}}\bigg),
\end{equation}
solves the issue of renormalizability. No counter terms can be
generated in a covariant renormalization scheme, which are not
already present in the action. The effective field theory approach
will always lead to consistent renormalizable theories in any
dimension, {\it e.g.}, for ($D=4$), ($D=10$) or even ($D=127$). It
is important to note that in regard to its renormalizability, an
effective field theory in the large-$D$ limit of gravity is just
as well defined as the planar diagram large-$N$ limit, in e.g., a
$(D=4)$ Yang-Mills theory. A Yang-Mills theory, at ($D > 4$), will
also need to be treated by means of effective field theory in
order to be renormalizable. What we give up in the effective field
treatment of gravity or in any other theory, is the finite number
of terms in the action. The Einstein action is replaced with an
action which has an infinite number of coupling constants. These
have to be renormalized at each loop order. The coupling constants
of the effective expansion of the action, {\it i.e.}, ($c_1, c_2,
\ldots$), will have to be determined by experiment and are
running, {\it i.e.}, they will depend on the energy scale.
However, the advantage of the effective approach is quite clear.
Quantum gravity is an effectively renormalizable theory up to the
energies at the Planck scale!

The effective field theory description of general relativity has
directly shown its worth by explicit calculations of quantum
corrections to the metric and to the Newtonian potential of large
masses. This work was pioneered by Donoghue,
refs.~\cite{Donoghue:dn,B1,B2}, and as an effective field theory
combined with QED, refs.~\cite{DB,QED}. The conceptual picture of
Einstein gravity is not affected by the effective field theory
treatment of the gravitational interactions. The equations of
motion will of course be slightly changed by the presence of
higher derivative terms in the action, however the contributions
coming from such terms can at normal energies be shown to be
negligible~\cite{Stelle,Simon:ic}. Effective field theory thus
provides us with a very solid framework in the study of
gravitational interactions and quantum effects at normal energies,
and solves some of the essential problems in building field
theoretical quantum gravity models.

An intriguing paper by Strominger~\cite{Strominger:1981jg} deals
with the perspectives of basic Einstein gravity at an infinite
number of dimensions. The basic idea is to let the spatial
dimension be a parameter in which one is allowed to expand the
theory. Each graph of the theory is then associated with a
dimensional factor. Formally one can expand every Greens function
of the theory as a series in $\left(\frac1D\right)$ and the
gravitational coupling constant ($\kappa$):
\begin{equation}
G = \sum_{i,j}(\kappa)^i\left(\frac1D\right)^j {\cal G}_{i,j}.
\end{equation}
The contributions with highest dimensional dependence will be the
leading ones in the large-$D$ limit. Concentrating on these graphs
only simplifies the theory and makes explicit calculations easier.
In the effective field theory we expect an expansion of the theory
of the type:
\begin{equation}
G = \sum_{i,j,k}(\kappa)^i\left(\frac1D\right)^j ({\cal
E})^{2k}{\cal G}_{i,j,k},
\end{equation}
where (${\cal E}$) represents a parameter for the effective
expansion of the theory in terms of the energy, {\it i.e.}, each
derivative acting on a massless field will correspond to a factor
of (${\cal E}$). The basic Einstein-Hilbert scale will correspond
to the (${\cal E}^2\sim
\partial^2 g$) contribution, while higher order effective
contributions will be of order (${\cal E}^4\sim\partial^4 g$),
(${\cal E}^6\sim \partial^6 g$), $\ldots$ etc.

The idea of making a large-$D$ expansion in gravity is somewhat
similar to the large-$N$ Yang-Mills planar diagram limit
considered by 't Hooft~\cite{tHooft}. In large-$N$ gauge theories,
one expands in the internal symmetry index ($N$) of the gauge
group, {\it e.g.}, $SU(N)$ or $SO(N)$. The physical
interpretations of the two expansions are of course completely
dissimilar, {\it e.g.}, the number of dimensions in a theory
cannot really be compared to the internal symmetry index of a
gauge theory. A comparison of the Yang Mills large-$N$ limit (the
planar diagram limit), and the large-$D$ expansion in gravity is
however still interesting to perform, and as an example of a
similarity between the two expansions the leading graphs in the
large-$D$ limit in gravity consists of a subset of planar
diagrams.

Higher dimensional models for gravity are well known from string
and supergravity theories. The mysterious $M$-theory has to exist
in an 11-dimensional space-time. So there are many good reasons to
believe that on fundamental scales, we might experience additional
space-time dimensions. Additional dimensions could be treated as
free or as compactified below the Planck scale, as in the case of
a Kaluza-Klein mechanism. One application of the large-$D$
expansion in gravity could be to approximate Greens functions at
finite dimension, {\it e.g.}, ($D=4$). The successful and various
uses of the planar diagram limit in Yang-Mills theory might
suggest other possible scenarios for the applicability of the
large-$D$ expansion in gravity. For example, is effective quantum
gravity renormalizable in its leading large-$D$ limit, {\it i.e.},
is it renormalizable in the same way as some non-renormalizable
theories are renormalizable in their planar diagram limit? It
could also be, that quantum gravity at large-$D$ is a completely
different theory, than Einstein gravity. A planar diagram limit is
essentially a string theory at large distances, {\it i.e.}, could
gravity at large-$D$ be interpreted as a large distance truncated
string limit?

We will here combine the successful effective field theory
approach which holds in any dimension, with the expansion of
gravity in the large-$D$ limit. The treatment will be mostly
conceptual, but we will also address some of the phenomenological
issues of this theory. The structure of the paper will be as
follows. First we will discuss the basic quantization of the
Einstein-Hilbert action, and then we will go on with the large-$D$
behavior of gravity, {\it i.e.}, we will show how to derive a
consistent limit for the leading graphs. The effective extension
of the theory will be taken up in the large-$D$ framework, and we
will especially focus on the implications of the effective
extension of the theory in the large-$D$ limit. The
$D$-dimensional space-time integrals will then be briefly looked
upon, and we will make a conceptual comparison of the large-$D$ in
gravity and the large-$N$ limit in gauge theory. Here we will
point out some similarities and some dissimilarities of the two
theories. We will also discuss some issues in the original paper
by Strominger, ref.~\cite{Strominger:1981jg}.

We will work in units: $(\hbar=c=1)$, and metric: ${\rm
diag}(\eta_{\mu\nu})=(1,-1,-1,-1,\dots,-1)$, {\it i.e.}, with
$(D-1)$ minus signs.

\section{Review of the large-$D$ limit of Einstein gravity}
In this section we will review the main idea. To begin we will
formally describe how to quantize a gravitational theory.

As is well known the $D$-dimensional Einstein-Hilbert Lagrangian
has the form:
\begin{equation}
{\cal L} = \int d^Dx \sqrt{-g}\Big(\frac{2R}{\kappa^2}\Big).
\end{equation}
If we neglect the renormalization problems of this action, it is
possible to make a formal quantization using the path integral
approach. Introducing a gauge breaking term in the action will fix
the gauge, and because gravity is a non-abelian theory we have to
introduce a proper Faddeev-Popov ghost action as well.

The action for the quantized theory will consequently be:
\begin{equation}
{\cal L} = \int d^Dx \sqrt{-g}\Big(\frac{2R}{\kappa^2}+{\cal
L}_{\rm gauge\ fixing}+ {\cal L}_{\rm ghosts}\Big),
\end{equation}
where (${\cal L}_{\rm gauge\ fixing}$) is the gauge fixing term,
and (${\cal L}_{\rm ghosts}$) is the ghost contribution. In order
to generate vertex rules for this theory an expansion of the
action has to be carried out, and vertex rules have to be
extracted from the gauge fixed action. The vertices will depend on
the gauge choice and on how we define the gravitational field.

It is conventional to define:
\begin{equation}
g_{\mu\nu} \equiv \eta_{\mu\nu}+\kappa h_{\mu\nu},
\end{equation}
and work in harmonic gauge: ($\partial^\lambda h_{\mu\lambda} =
\frac12\partial_\mu h^\lambda_\lambda$). For this gauge choice the
vertex rules for the 3- and 4-point Einstein vertices can be found
in~\cite{Dewitt,Sannan}.

Yet another possibility is to use the background field method,
here we define:
\begin{equation}
g_{\mu\nu} \equiv \tilde g_{\mu\nu}+\kappa h_{\mu\nu},
\end{equation}
and expand the quantum fluctuation: ($h_{\mu\nu}$) around an
external source, the background field: ($\tilde g_{\mu\nu} \equiv
\eta_{\mu\nu} + \kappa H_{\mu\nu}$). In the background field
approach we have to differ between vertices with internal lines
(quantum lines: $\sim h_{\mu\nu}$) and external lines (external
sources: $\sim H_{\mu\nu}$). There are no real problems in using
either method; it is mostly a matter of notation and conventions.
Here we will concentrate our efforts on the conventional approach.
In many loop calculations the background field method is the
easiest to employ. See ref.~\cite{Donoghue:dn,B1,B2,NON} for
additional details about, {\it e.g.}, the vertex rules in the
background field method.

Another possibility is to define:
\begin{equation}
\tilde g^{\mu\nu} \equiv \sqrt{-g}g^{\mu\nu} = \eta^{\mu\nu} +
\kappa h^{\mu\nu},
\end{equation}
this field definition makes a transformation of the
Einstein-Hilbert action into the following form:
\begin{equation}\begin{split}
{\cal L} &= \int d^D x \frac{1}{2\kappa^2}(\tilde
g^{\rho\sigma}\tilde g_{\lambda\alpha}\tilde g_{\kappa\tau}\tilde
g^{\alpha\kappa}_{\ \ \ , \rho}\tilde g^{\lambda\tau}_{\ \ \ ,
\sigma} -\frac1{(D-2)}\tilde g^{\rho\sigma}\tilde
g_{\alpha\kappa}\tilde g^{\lambda\tau}\tilde g^{\lambda\tau}_{\ \
\ , \sigma} -2\tilde g_{\alpha\tau}\tilde g^{\alpha\kappa}_{\ \ \
, \rho}\tilde g^{\rho\tau}_{\ \ \ , \kappa}),
\end{split}\end{equation}
in, ($D=4$), this form of the Einstein-Hilbert action is known as
the Goldberg action~\cite{Capper:pv}. In this description of the
Einstein-Hilbert action its dimensional dependence is more
obvious. It is however more cumbersome to work with in the course
of practical large-$D$ considerations.

In the standard expansion of the action the propagator for
gravitons take the form:
\begin{equation}
D_{\alpha\beta,\gamma\delta}(k) =
-\frac{i}{2}\frac{\Big[\eta_{\alpha\gamma}\eta_{\beta\delta}
+\eta_{\alpha\delta}\eta_{\beta\gamma}-\frac{2}{D-2}
\eta_{\alpha\beta}\eta_{\gamma\delta}\Big]}{k^2-i\epsilon}.
\end{equation}
The 3- and 4-point vertices for the standard expansion of the
Einstein-Hilbert action can be found in~\cite{Dewitt,Sannan} and
has the following form:
\begin{equation}\begin{split}
V^{(3)}_{\mu\alpha,\nu\beta,\sigma\gamma}(k_1,k_2,k_3) &= \kappa\,{\rm sym} \Big[
-\frac12P_3(k_1\cdot k_2\,\eta_{\mu\alpha}\eta_{\nu\beta}\eta_{\sigma\gamma})
 -\frac12P_6(k_{1\nu}k_{1\beta}\eta_{\mu\alpha}\eta_{\sigma\gamma})
+ \frac12P_3(k_1\cdot
k_2\,\eta_{\mu\nu}\eta_{\alpha\beta}\eta_{\sigma\gamma})\\&
+P_6(k_1\cdot
k_2\,\eta_{\mu\alpha}\eta_{\nu\sigma}\eta_{\beta\gamma})
+2P_3(k_{1\nu}k_{1\gamma}\eta_{\mu\alpha}\eta_{\beta\sigma})
-P_3(k_{1\beta}k_{2\mu}\eta_{\alpha\nu}\eta_{\sigma\gamma})
+P_3(k_{1\sigma}k_{2\gamma}\eta_{\mu\nu}\eta_{\alpha\beta})\\&
+P_6(k_{1\sigma}k_{1\gamma}\eta_{\mu\nu}\eta_{\alpha\beta})
+2P_6(k_{1\nu}k_{2\gamma}\eta_{\beta\mu}\eta_{\alpha\sigma})
+2P_3(k_{1\nu}k_{2\mu}\eta_{\beta\sigma}\eta_{\gamma\alpha})
-2P_3(k_1\cdot
k_2\,\eta_{\alpha\nu}\eta_{\beta\sigma}\eta_{\gamma\mu})\Big],
\end{split}\end{equation}
and
\begin{equation}\begin{split}
V^{(4)}_{\mu\alpha,\nu\beta,\sigma\gamma,\rho\lambda}(k_1,k_2,k_3,k_4)&=\kappa^2\,{\rm sym}\Big[
-\frac14P_6(k_1\cdot k_2\,\eta_{\mu\alpha}\eta_{\nu\beta}\eta_{\sigma\gamma}\eta_{\rho\lambda})
-\frac14P_{12}(k_{1\nu}k_{1\beta}\eta_{\mu\alpha}\eta_{\sigma\gamma}\eta_{\rho\lambda})
-\frac12P_6(k_{1\nu}k_{2\mu}\eta_{\alpha\beta}\eta_{\sigma\gamma}\eta_{\rho\lambda})\\&
+\frac14P_6(k_1\cdot k_2\, \eta_{\mu\nu}\eta_{\alpha\beta}\eta_{\sigma\gamma}\eta_{\rho\lambda})
+\frac12P_6(k_1\cdot k_2\, \eta_{\mu\alpha}\eta_{\nu\beta}\eta_{\sigma\rho}\eta_{\gamma\lambda})
+\frac12P_{12}(k_{1\nu}k_{1\beta}\eta_{\mu\alpha}\eta_{\sigma\rho}\eta_{\gamma\lambda})\\&
+P_6(k_{1\nu}k_{2\mu}\eta_{\alpha\beta}\eta_{\sigma\rho}\eta_{\gamma\lambda})
-\frac12P_6(k_1\cdot k_2\,\eta_{\mu\nu}\eta_{\alpha\beta}\eta_{\sigma\rho}\eta_{\gamma\lambda})
+\frac12P_{24}(k_1\cdot k_2\,\eta_{\mu\alpha}\eta_{\nu\sigma}\eta_{\beta\gamma}\eta_{\rho\lambda})
\\&+\frac12P_{24}(k_{1\nu}k_{1\beta}\eta_{\mu\sigma}\eta_{\alpha\gamma}\eta_{\rho\lambda}
+\frac12P_{12}(k_{1\sigma}k_{2\gamma}\eta_{\mu\nu}\eta_{\alpha\beta}\eta_{\rho\lambda})
+P_{24}(k_{1\nu}k_{2\sigma}\eta_{\beta\mu}\eta_{\alpha\gamma}\eta_{\rho\lambda})\\&
-P_{12}(k_1\cdot k_2\,\eta_{\alpha\nu}\eta_{\beta\sigma}\eta_{\gamma\mu}\eta_{\rho\lambda})
+P_{12}(k_{1\nu}k_{2\mu}\eta_{\beta\sigma}\eta_{\gamma\alpha}\eta_{\rho\lambda})
+P_{12}(k_{1\nu}k_{1\sigma}\eta_{\beta\gamma}\eta_{\mu\alpha}\eta_{\rho\lambda})
\\&-P_{24}(k_1\cdot k_2\,\eta_{\mu\alpha}\eta_{\beta\sigma}\eta_{\gamma\rho}\eta_{\lambda\nu})
-2P_{12}(k_{1\nu}k_{1\beta}\eta_{\alpha\sigma}\eta_{\gamma\rho}\eta_{\lambda\mu})
-2P_{12}(k_{1\sigma}k_{2\gamma}\eta_{\alpha\rho}\eta_{\lambda\nu}\eta_{\beta\mu})\\&
-2P_{24}(k_{1\nu}k_{2\sigma}\eta_{\beta\rho}\eta_{\lambda\mu}\eta_{\alpha\gamma})
-2P_{12}(k_{1\sigma}k_{2\rho}\eta_{\gamma\nu}\eta_{\beta\mu}\eta_{\alpha\lambda})
+2P_{6}(k_1\cdot
k_2\,\eta_{\alpha\sigma}\eta_{\gamma\nu}\eta_{\beta\rho}\eta_{\lambda\mu})\\&
-2P_{12}(k_{1\nu}k_{1\sigma}\eta_{\mu\alpha}\eta_{\beta\rho}\eta_{\lambda\gamma})
-P_{12}(k_1\cdot
k_2\,\eta_{\mu\sigma}\eta_{\alpha\gamma}\eta_{\nu\rho}\eta_{\beta\lambda})
-2P_{12}(k_{1\nu}k_{1\sigma}\eta_{\beta\gamma}\eta_{\mu\rho}\eta_{\alpha\lambda}\\&
-P_{12}(k_{1\sigma}k_{2\rho}\eta_{\gamma\lambda}\eta_{\mu\nu}\eta_{\alpha\beta})
-2P_{24}(k_{1\nu}k_{2\sigma}\eta_{\beta\mu}\eta_{\alpha\rho}\eta_{\lambda\gamma})
-2P_{12}(k_{1\nu}k_{2\mu}\eta_{\beta\sigma}\eta_{\gamma\rho}\eta_{\lambda\alpha})\\&
+4P_{6}(k_1\cdot
k_2\,\eta_{\alpha\nu}\eta_{\beta\sigma}\eta_{\gamma\rho}\eta_{\lambda\mu})
\Big],
\end{split}\end{equation}
in the above two expressions, '${\rm sym}$', means that each pair
of indices: ($\mu\alpha$), ($\nu\beta$), $\ldots$ will have to be
symmetrized. The momenta factors: ($k_1$, $k_2$, $\ldots$) are
associated with the index pairs: ($\mu\alpha$, $\nu\beta$,
$\ldots$) correspondingly. The symbol: ($P_{\#}$) means that a
$\#$-permutation of indices and corresponding momenta has to
carried out for this particular term. As seen the algebraic
structures of the 3- and 4-point vertices are already rather
involved and complicated. 5-point vertices will not be considered
explicitly in this paper.

The explicit prefactors of the terms in the vertices will not be
essential in this treatment. The various algebraic structures
which constitutes the vertices will be more important. Different
algebraic terms in the vertex factors will generate dissimilar
traces in the final diagrams. It is useful to adapt an index line
notation for the vertex structure similar to that used in
large-$N$ gauge theory.

\begin{figure}[h]
\begin{tabular}{llll}\vspace{0.05cm}
$\left(\parbox{1cm}{\includegraphics[height=0.8cm]{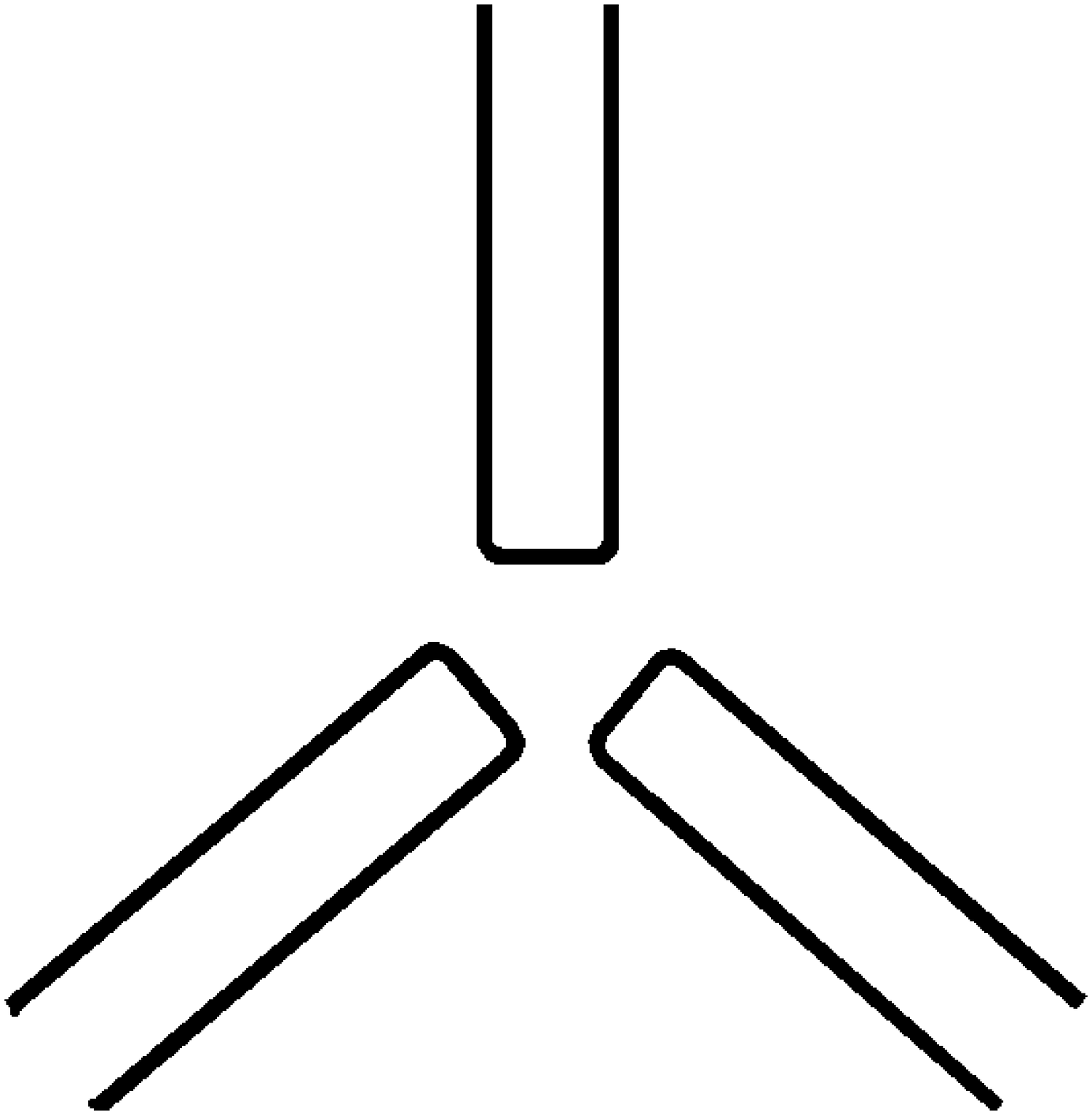}}\right)_{\rm
3A}$&{$\sim {\rm sym}[-\frac12P_3(k_1\cdot
k_2\eta_{\mu\alpha}\eta_{\nu\beta}\eta_{\sigma\gamma})],$}
&$\left(\parbox{1cm}{\includegraphics[height=0.8cm]{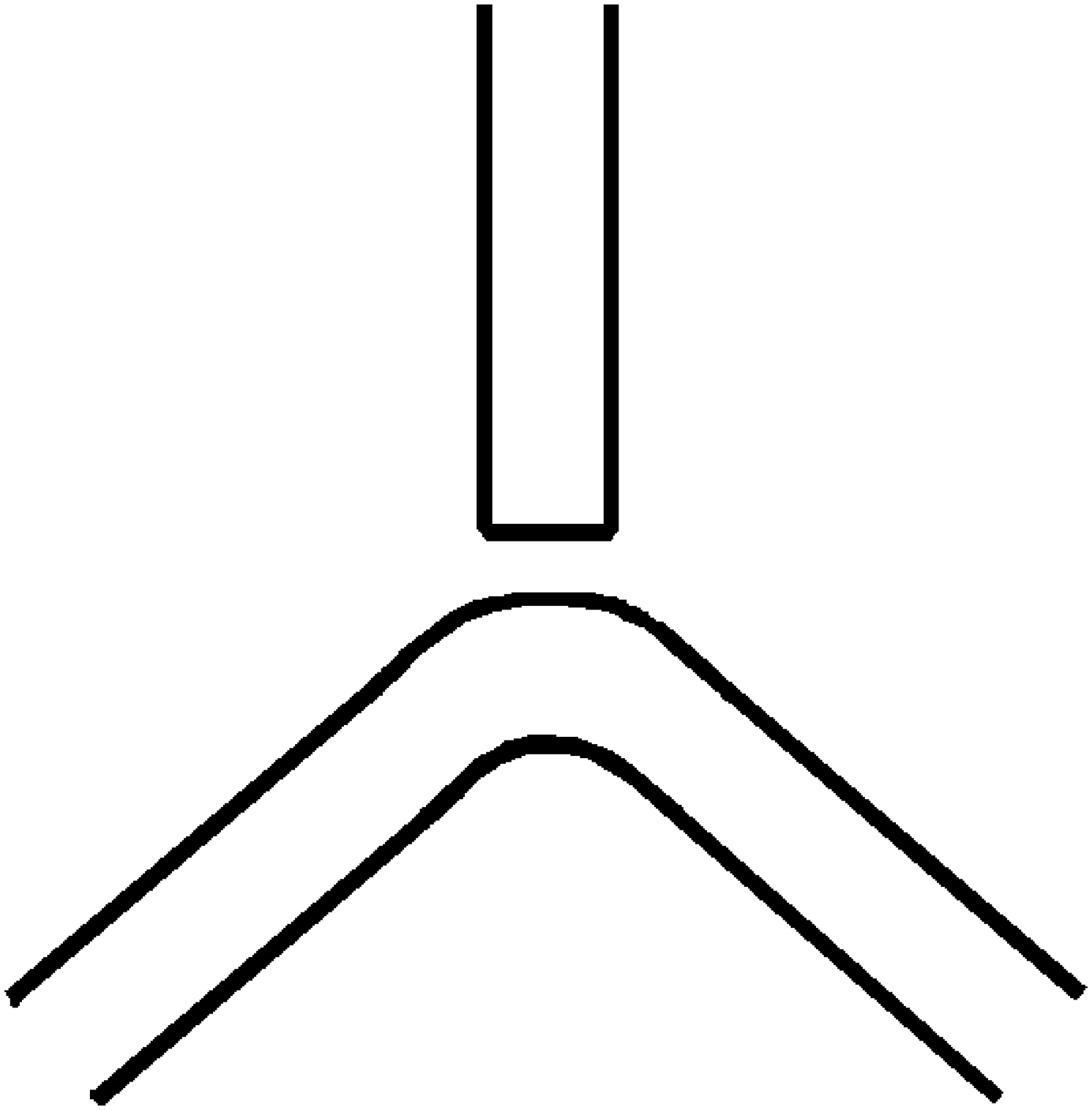}}\right)_{\rm
3B}$&$\sim {\rm sym}[ \frac12P_3(k_1\cdot k_2
\eta_{\mu\nu}\eta_{\alpha\beta}\eta_{\sigma\gamma})+P_6(k_1\cdot
k_2 \eta_{\mu\alpha}\eta_{\nu\sigma}\eta_{\beta\gamma})],$
\\ \vspace{0.05cm}
$\left(\parbox{1cm}{\includegraphics[height=0.8cm]{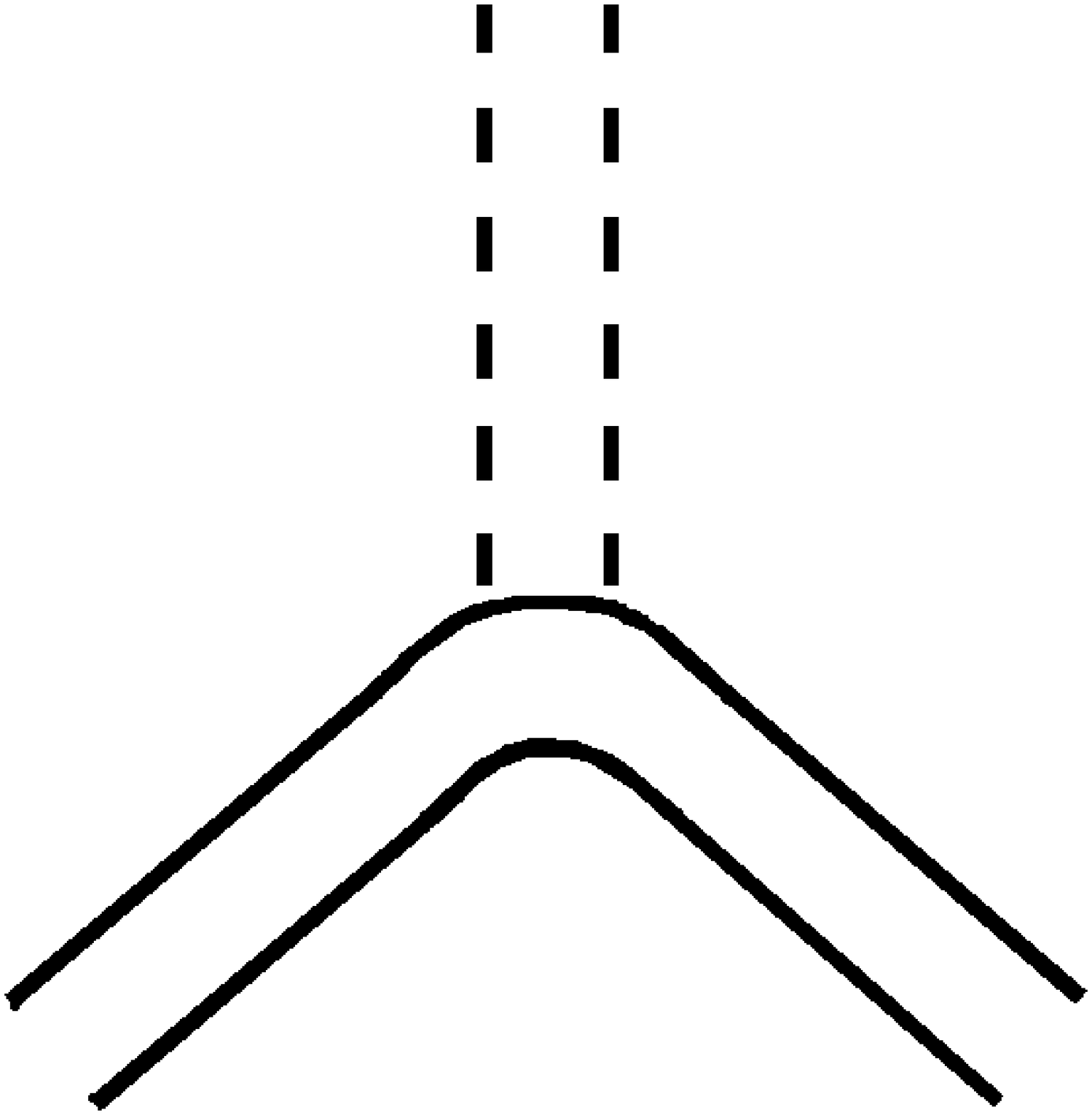}}\right)_{\rm
3C}$&$\sim {\rm sym}[
P_3(k_{1\sigma}k_{2\gamma}\eta_{\mu\nu}\eta_{\alpha\beta})],$
&$\left(\parbox{1cm}{\includegraphics[height=0.8cm]{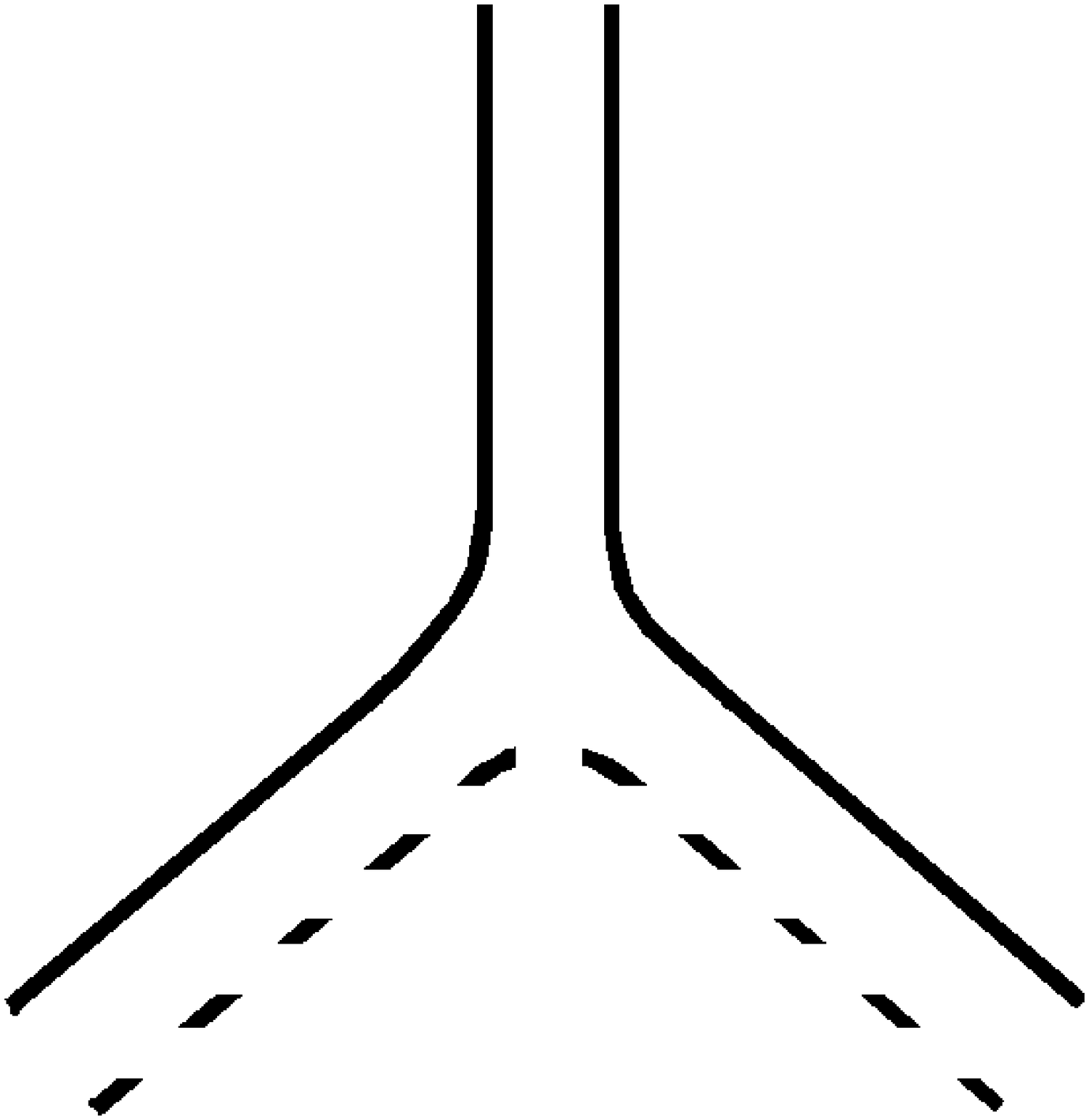}}\right)_{\rm
3D}$&$\sim {\rm sym}[
2P_6(k_{1\nu}k_{2\gamma}\eta_{\beta\mu}\eta_{\alpha\sigma})+2P_3(k_{1\nu}k_{2\mu}\eta_{\beta\sigma}\eta_{\gamma\alpha})],$\\
\vspace{0.1cm}
$\left(\parbox{1cm}{\includegraphics[height=0.8cm]{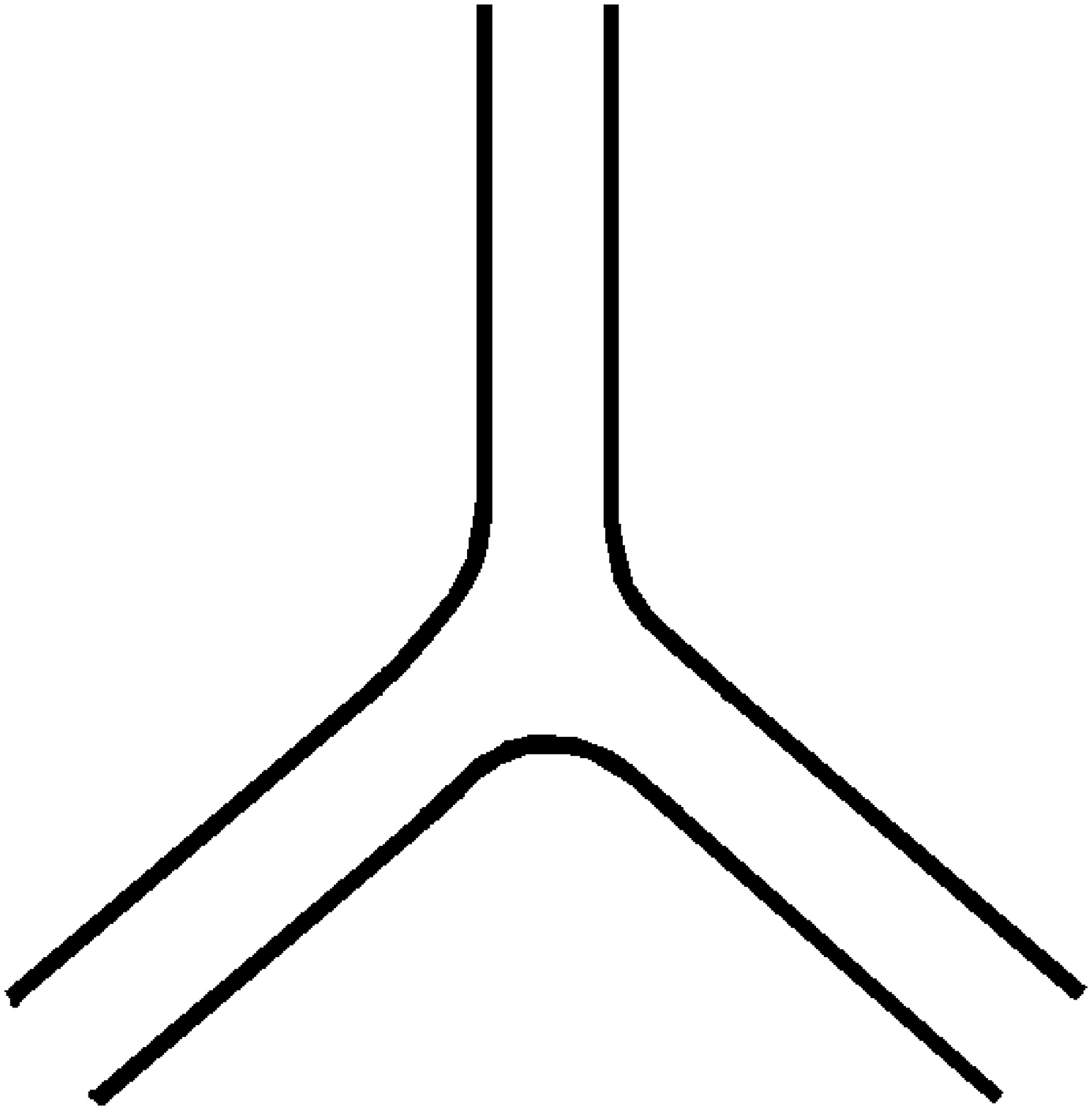}}\right)_{\rm
3E}$&$\sim{\rm sym}[ -2P_3(k_1\cdot
k_2\eta_{\alpha\nu}\eta_{\beta\sigma}\eta_{\gamma\mu})],$
&$\left(\parbox{1cm}{\includegraphics[height=0.8cm]{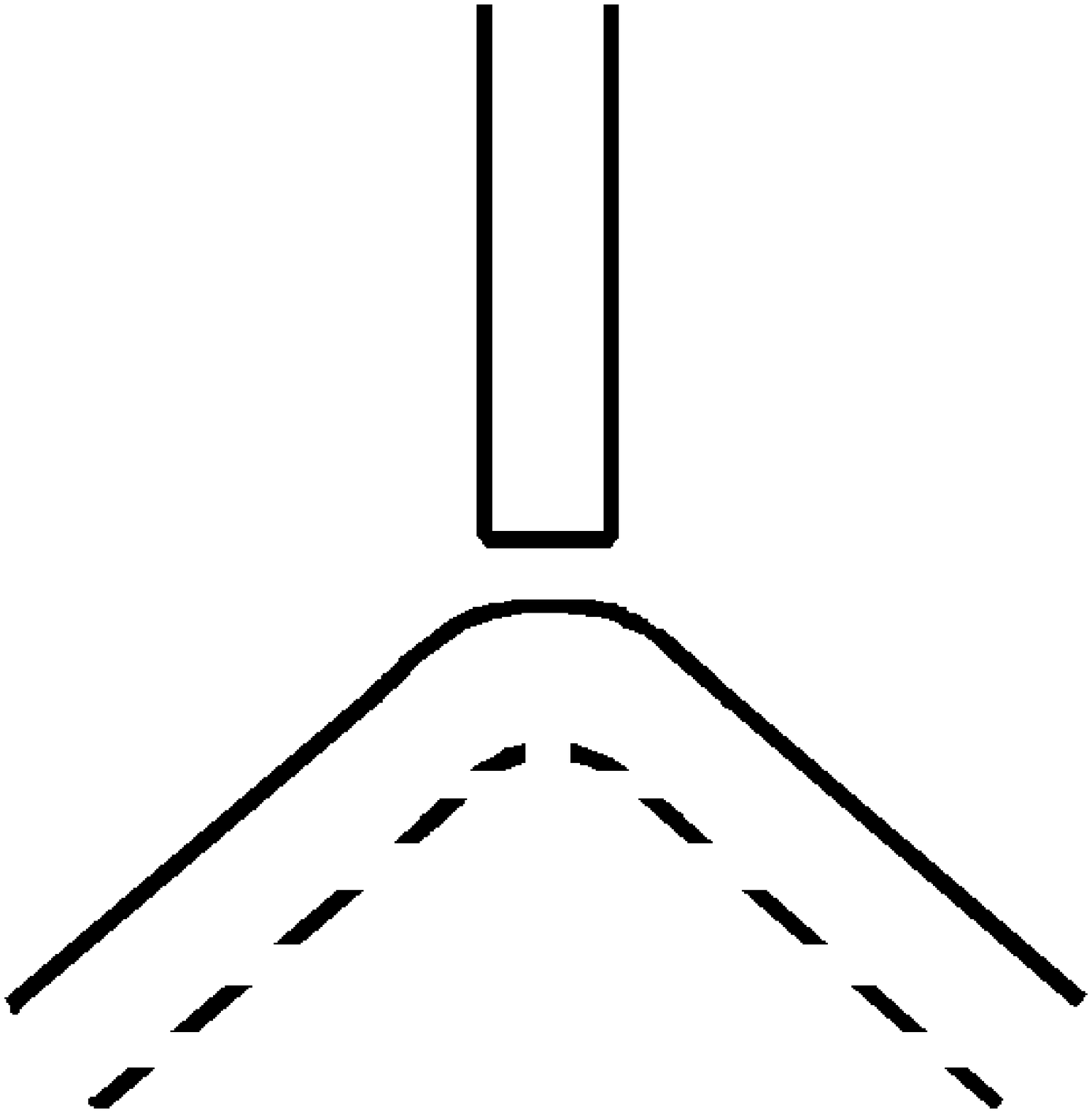}}\right)_{\rm
3F}$&$\sim {\rm
sym}[2P_3(k_{1\nu}k_{1\gamma}\eta_{\mu\alpha}\eta_{\beta\sigma})
-P_3(k_{1\beta}k_{2\mu}\eta_{\alpha\nu}\eta_{\sigma\gamma}) ],$\\
\vspace{0.05cm}
$\left(\parbox{1cm}{\includegraphics[height=0.8cm]{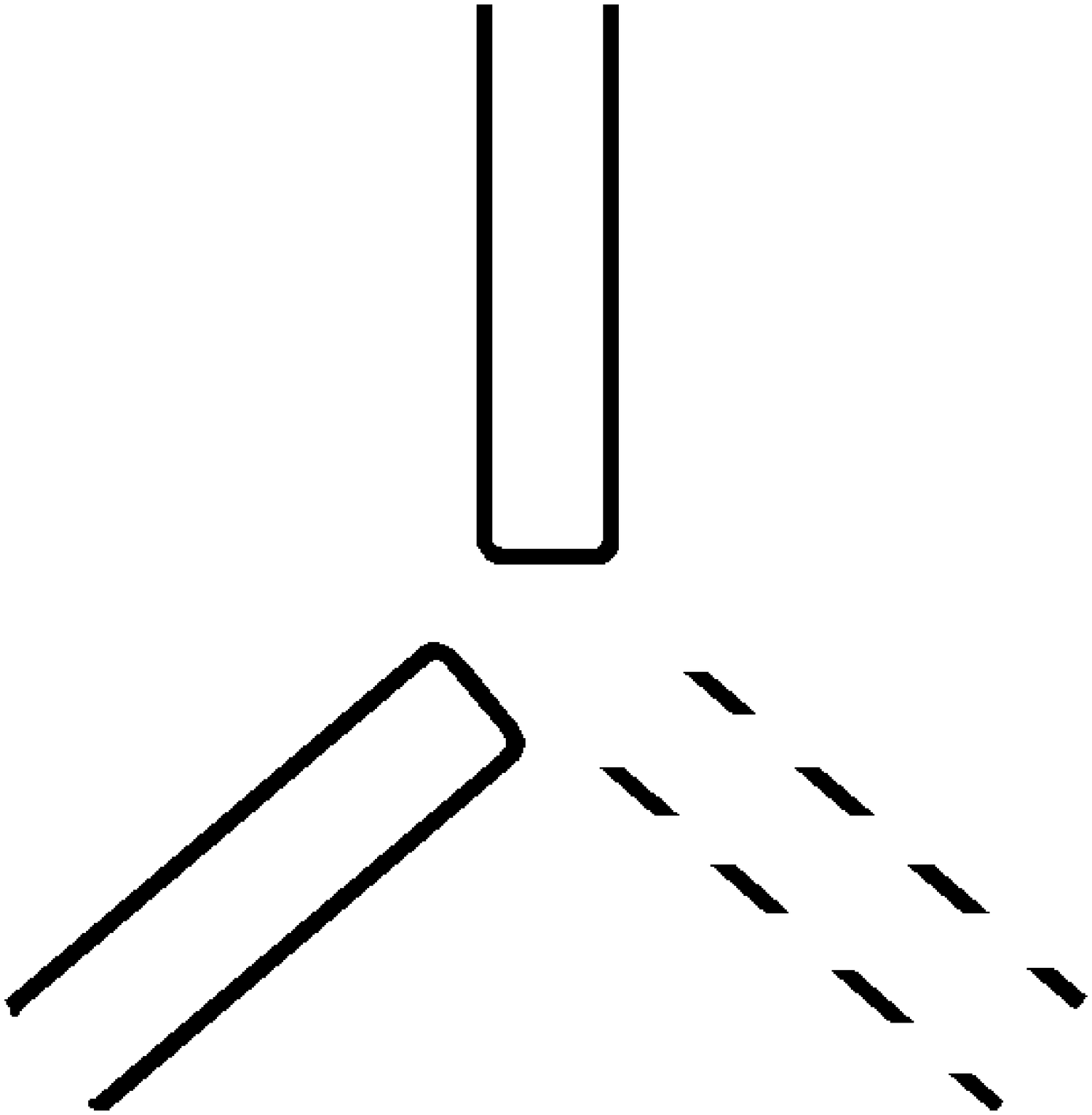}}\right)_{\rm
3G}$&$\sim {\rm
sym}[-\frac12P_6(k_{1\nu}k_{1\beta}\eta_{\mu\alpha}\eta_{\sigma\gamma})].$&
\end{tabular}
\caption{A graphical representation of the various terms in the
3-point vertex factor. A dashed line represents a contraction of a
index with a momentum line. A full line means a contraction of two
index lines. The above vertex notation for the indices and momenta
also apply here.\label{3vert}}
\end{figure}

In this notation we can represent the different algebraic terms of
the 3-point vertex (see figure \ref{3vert}).

\begin{figure}[h]
\begin{tabular}{ll}\vspace{0.05cm}
$\left(\parbox{1cm}{\includegraphics[height=0.8cm]{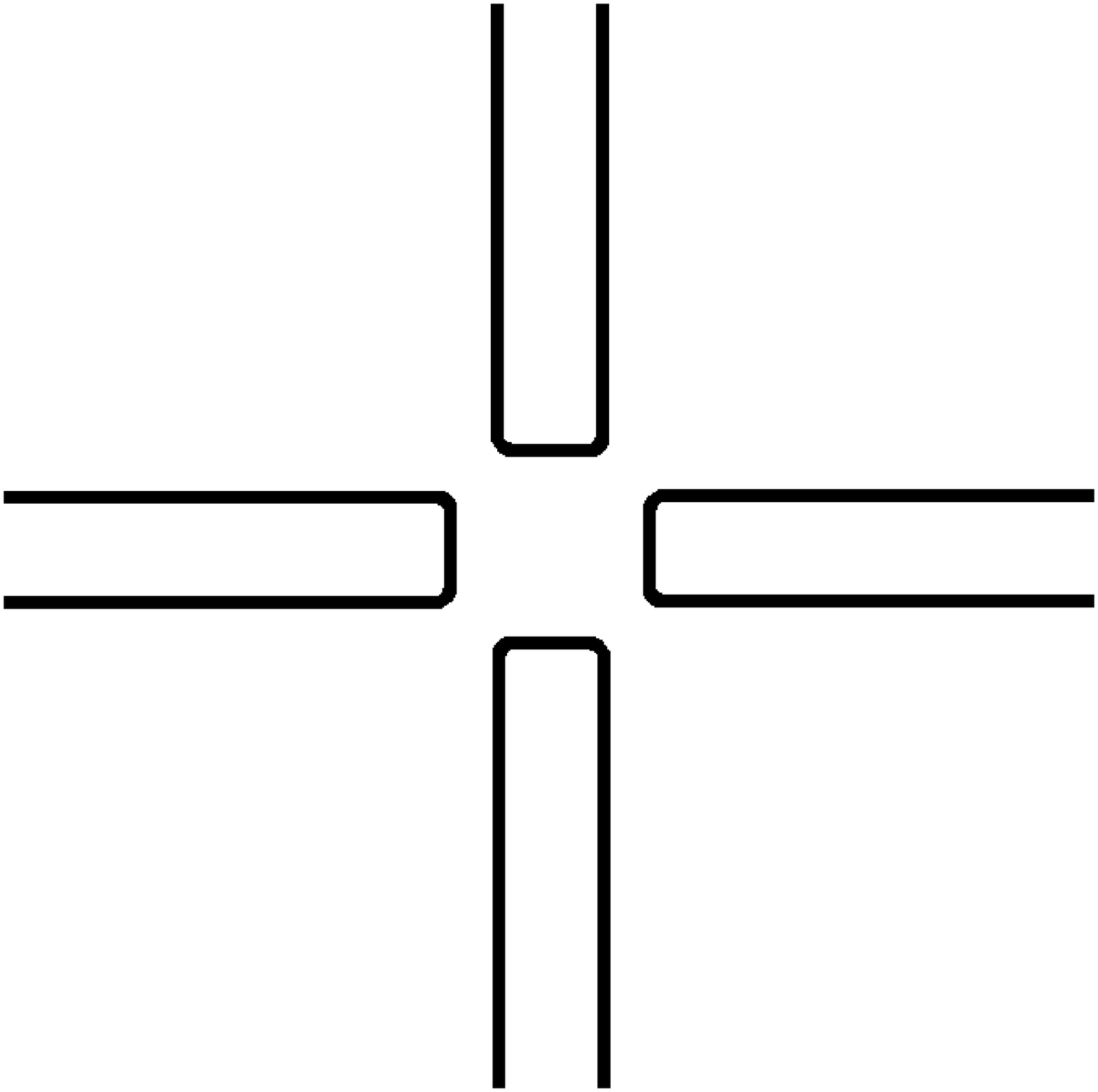}}\right)_{\rm
4A}$&$\sim {\rm sym} [-\frac14P_6(k_1\cdot k_2
\eta_{\mu\alpha}\eta_{\nu\beta}\eta_{\sigma\gamma}\eta_{\rho\lambda})],$\\
\vspace{0.05cm}
$\left(\parbox{1cm}{\includegraphics[height=0.8cm]{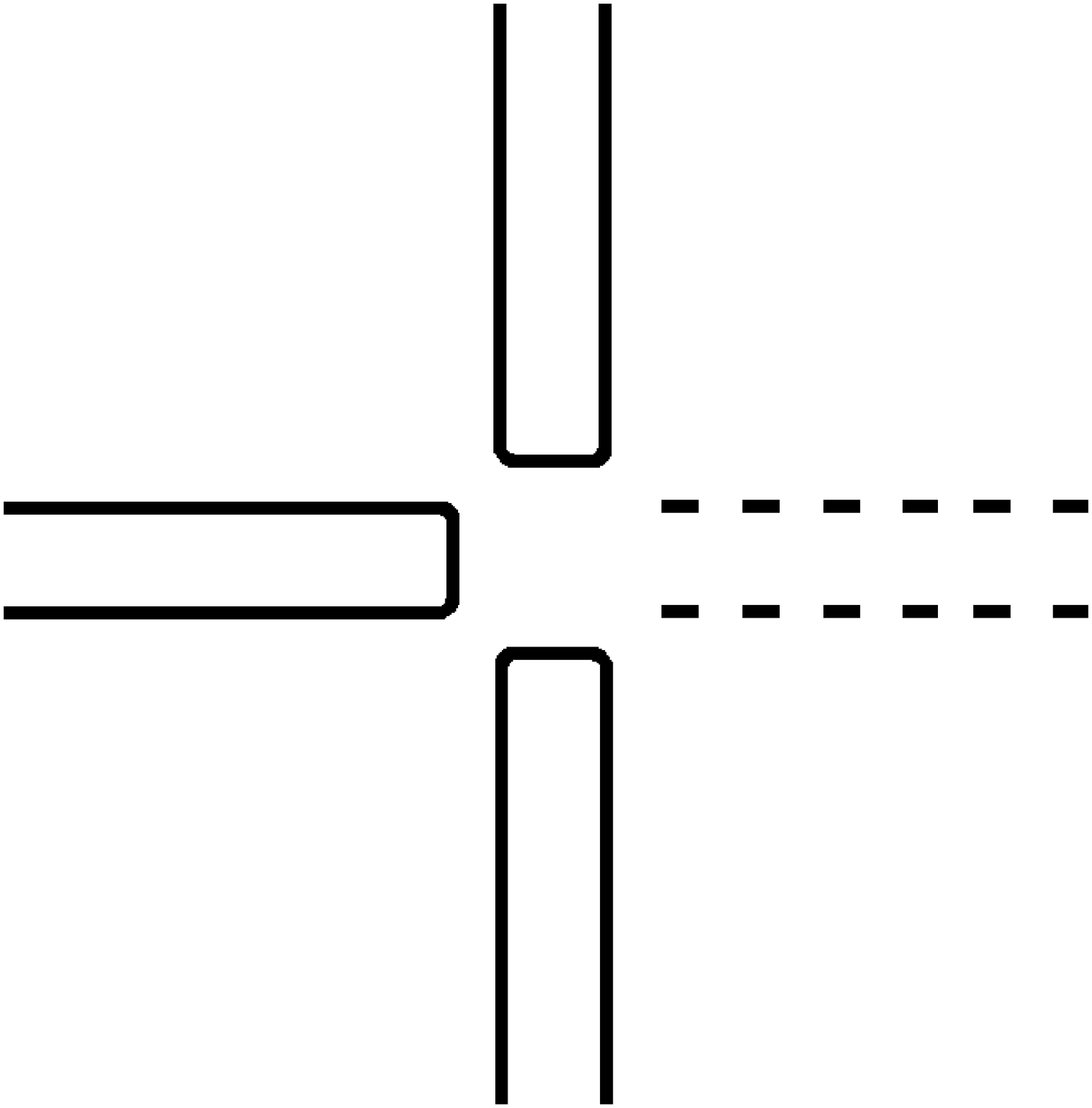}}\right)_{\rm
4B}$&$\sim {\rm sym}
[-\frac14P_{12}(k_{1\nu}k_{1\beta}\eta_{\mu\alpha}\eta_{\sigma\gamma}\eta_{\rho\lambda})],$\\
\vspace{0.05cm}
$\left(\parbox{1cm}{\includegraphics[height=0.8cm]{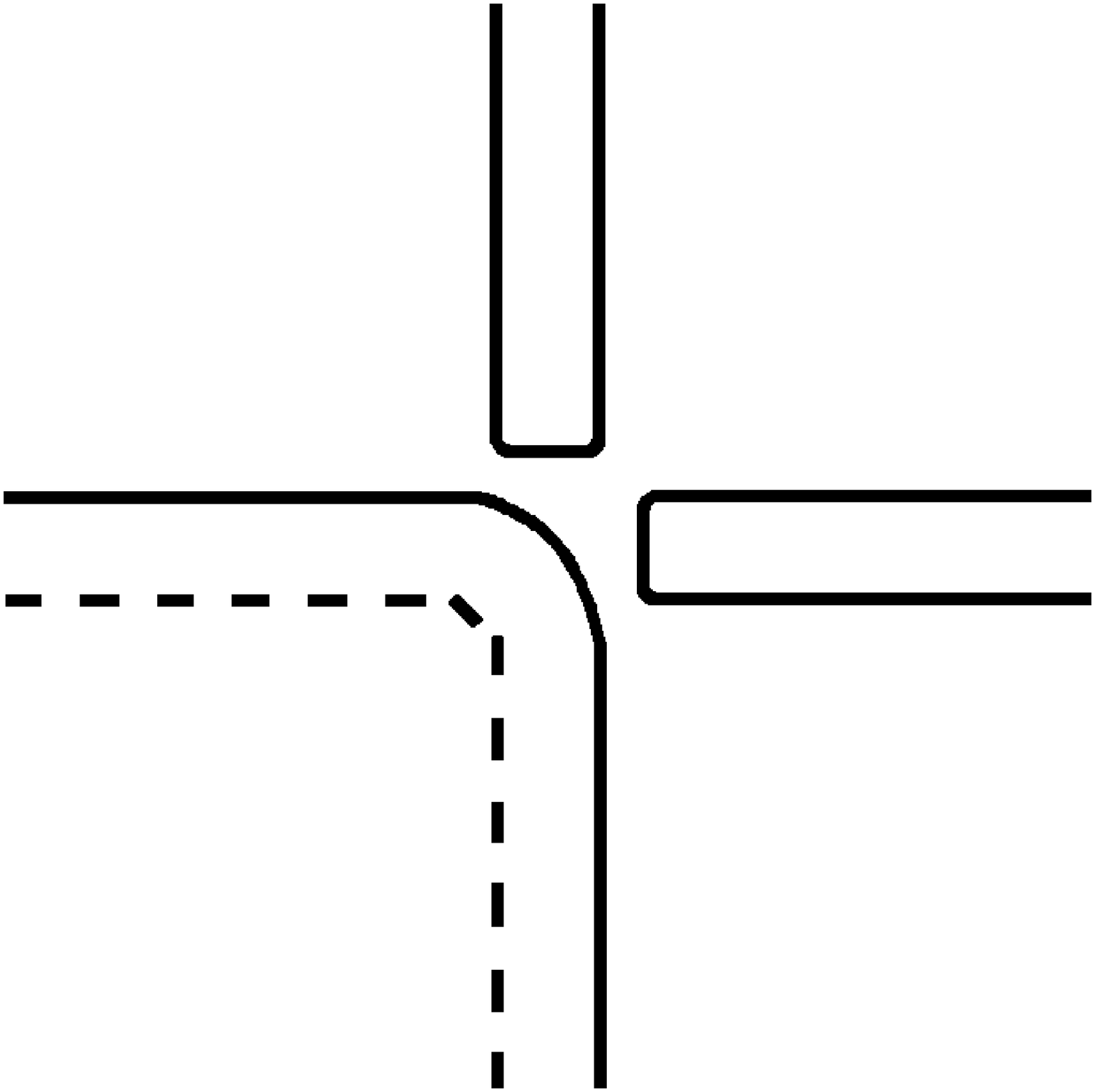}}\right)_{\rm
4C}$&$\sim {\rm sym}
[-\frac12P_6(k_{1\nu}k_{2\mu}\eta_{\alpha\beta}\eta_{\sigma\gamma}\eta_{\rho\lambda})],$
\\ \vspace{0.05cm}
$\left(\parbox{1cm}{\includegraphics[height=0.8cm]{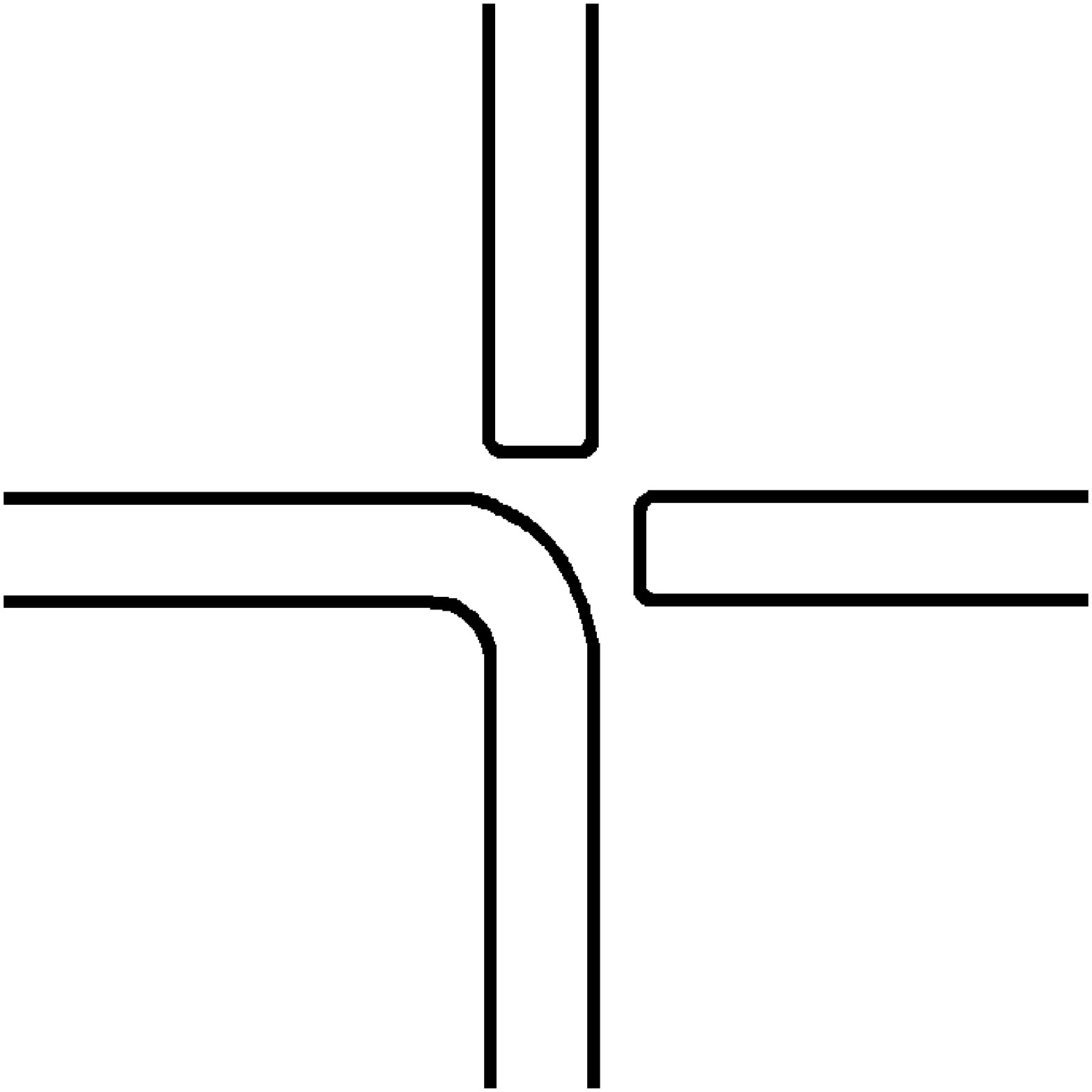}}\right)_{\rm
4D}$&$\sim {\rm sym} [\frac14P_6(k_1\cdot k_2
\eta_{\mu\nu}\eta_{\alpha\beta}\eta_{\sigma\gamma}\eta_{\rho\lambda})
+\frac12P_6(k_1\cdot k_2
\eta_{\mu\alpha}\eta_{\nu\beta}\eta_{\sigma\rho}\eta_{\gamma\lambda})
+\frac12P_{24}(k_1\cdot k_2
\eta_{\mu\alpha}\eta_{\nu\sigma}\eta_{\beta\gamma}\eta_{\rho\lambda})],$\\
\vspace{0.05cm}
$\left(\parbox{1cm}{\includegraphics[height=0.8cm]{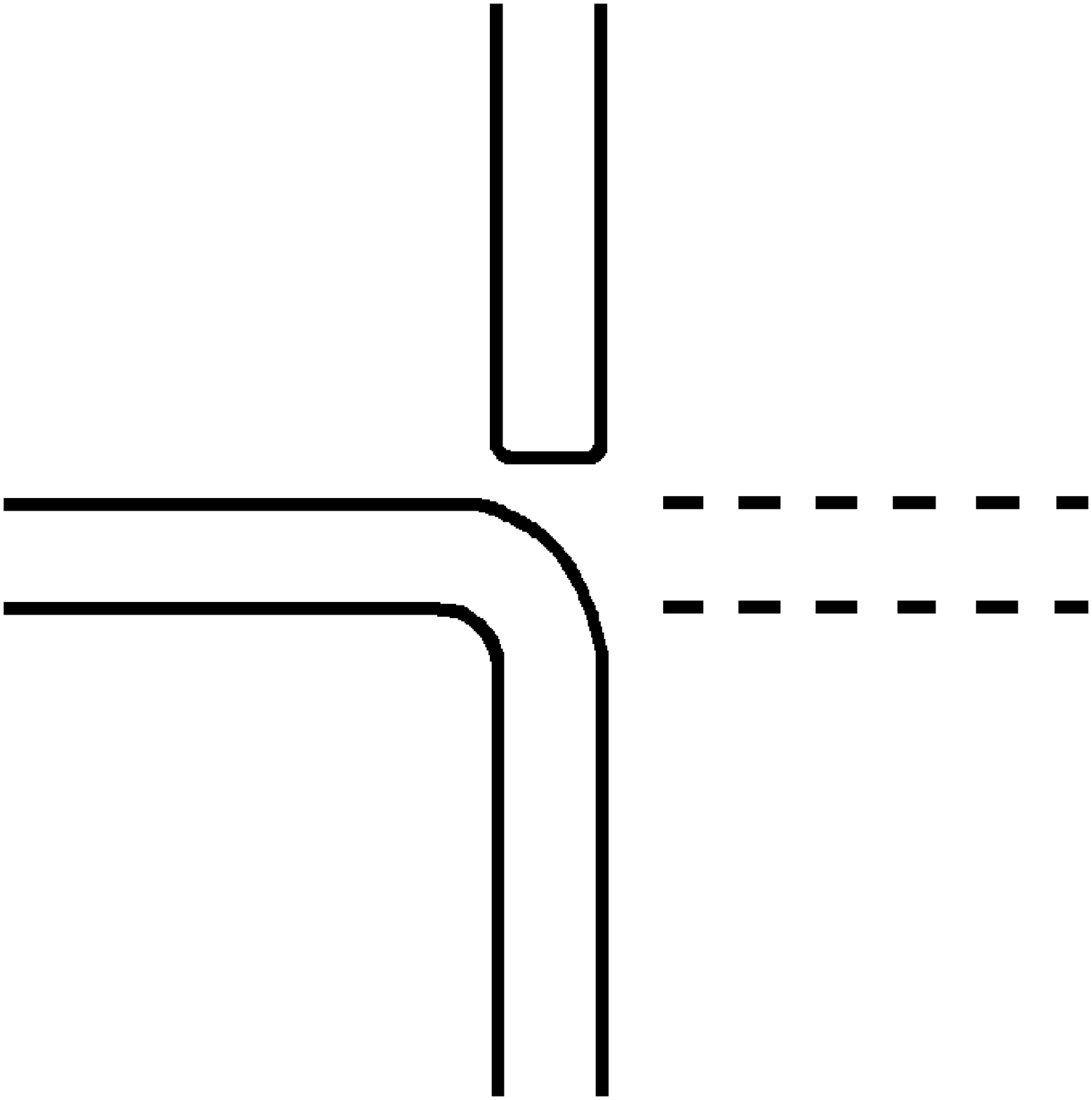}}\right)_{\rm
4E}$&$\sim {\rm sym}
[\frac12P_{12}(k_{1\nu}k_{1\beta}\eta_{\mu\alpha}\eta_{\sigma\rho}\eta_{\gamma\lambda})
+\frac12P_{24}(k_{1\nu}k_{1\beta}\eta_{\mu\sigma}\eta_{\alpha\gamma}\eta_{\rho\lambda})
+\frac12P_{12}(k_{1\sigma}k_{2\gamma}\eta_{\mu\nu}\eta_{\alpha\beta}\eta_{\rho\lambda})],$
\\ \vspace{0.05cm}
$\left(\parbox{1cm}{\includegraphics[height=0.8cm]{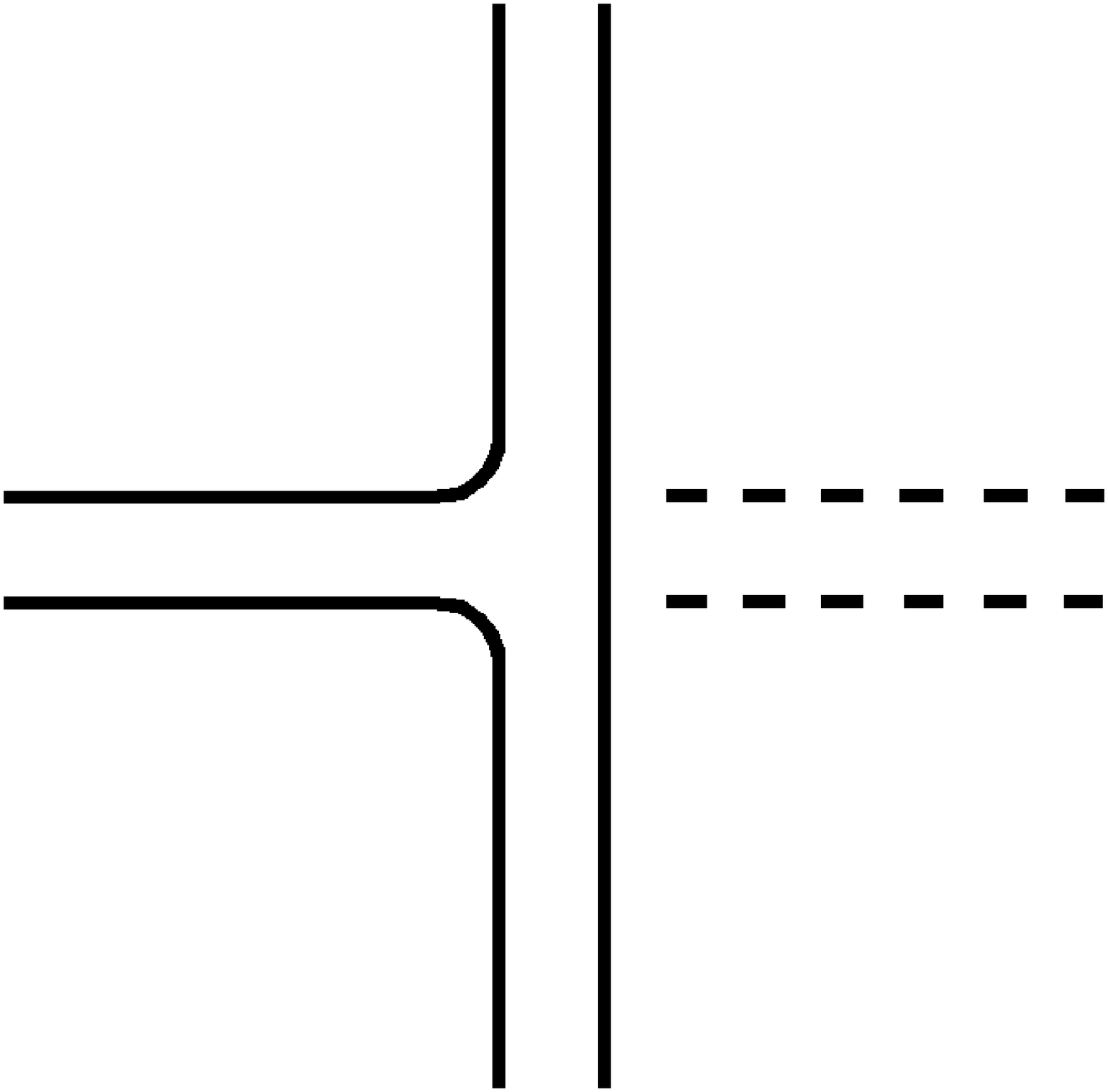}}\right)_{\rm
4F}$&$\sim {\rm sym}
[-2P_{12}(k_{1\nu}k_{1\beta}\eta_{\alpha\sigma}\eta_{\gamma\rho}\eta_{\lambda\mu})
-2P_{12}(k_{1\sigma}k_{2\gamma}\eta_{\alpha\rho}\eta_{\lambda\nu}\eta_{\beta\mu})],$\\
\vspace{0.05cm}
$\left(\parbox{1cm}{\includegraphics[height=0.8cm]{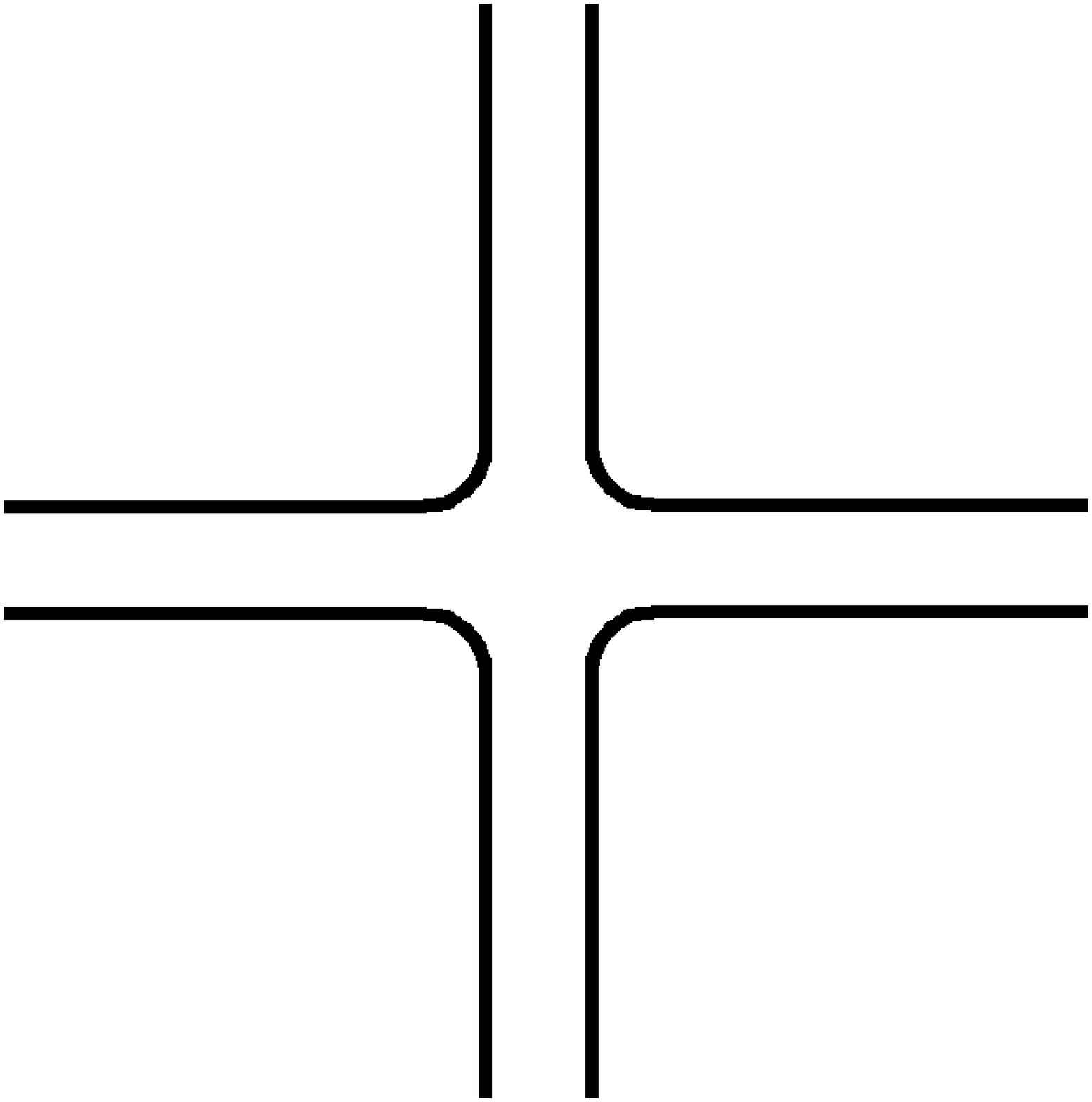}}\right)_{\rm
4G}$&$\sim {\rm sym} [2P_6(k_1\cdot k_2
\eta_{\alpha\sigma}\eta_{\gamma\nu}\eta_{\beta\rho}\eta_{\lambda\mu})
+4P_6(k_1\cdot k_2
\eta_{\alpha\nu}\eta_{\beta\sigma}\eta_{\gamma\rho}\eta_{\lambda\mu})],$
\\ \vspace{0.05cm}
$\left(\parbox{1cm}{\includegraphics[height=0.8cm]{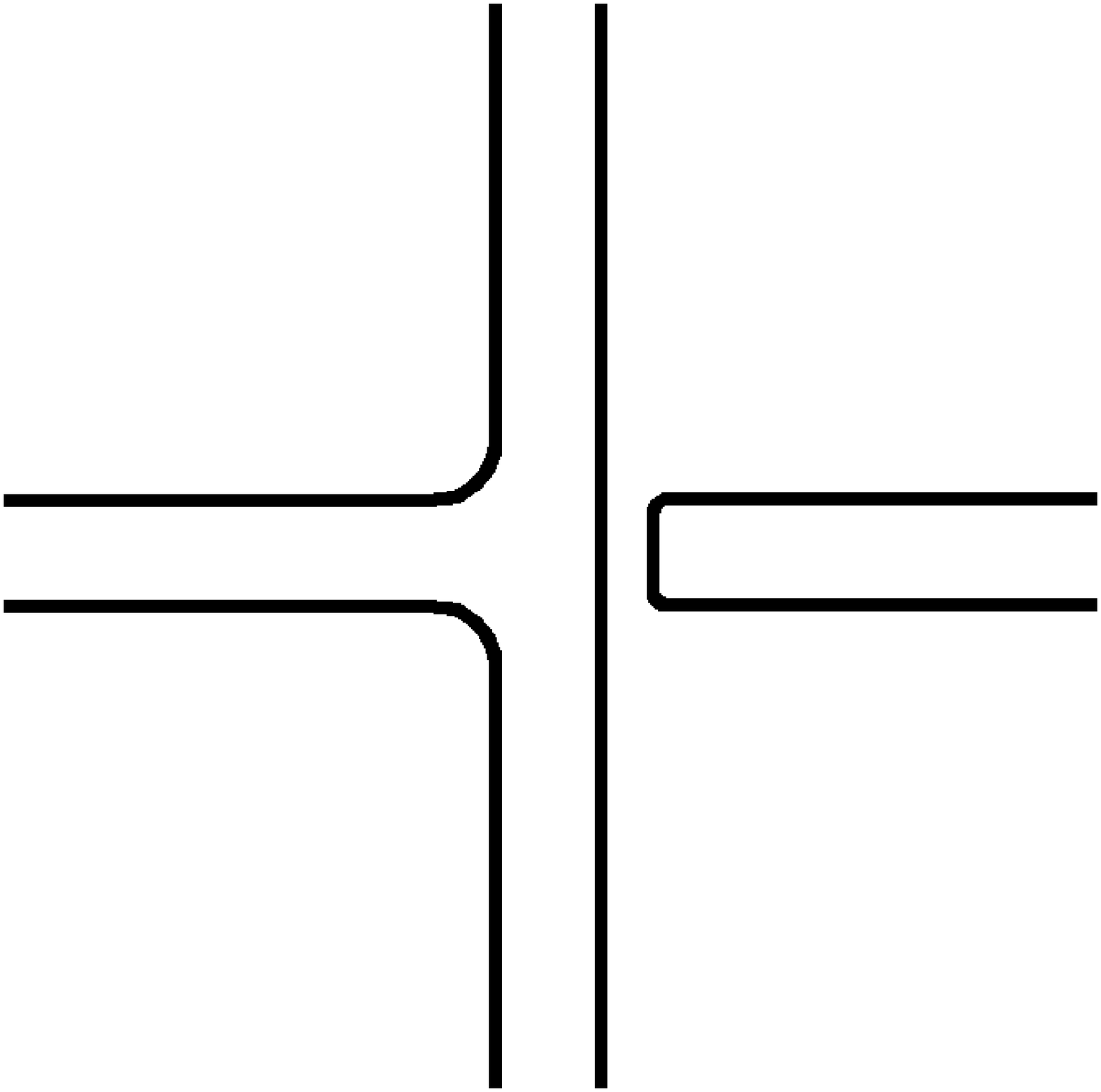}}\right)_{\rm
4H}$&$\sim {\rm sym} [-P_{12}(k_1\cdot k_2
\eta_{\alpha\nu}\eta_{\beta\sigma}\eta_{\gamma\mu}\eta_{\rho\lambda})
-P_{24}(k_1\cdot k_2
\eta_{\mu\alpha}\eta_{\beta\sigma}\eta_{\gamma\rho}\eta_{\lambda\nu})],$\\
\vspace{0.05cm}
$\left(\parbox{1cm}{\includegraphics[height=0.8cm]{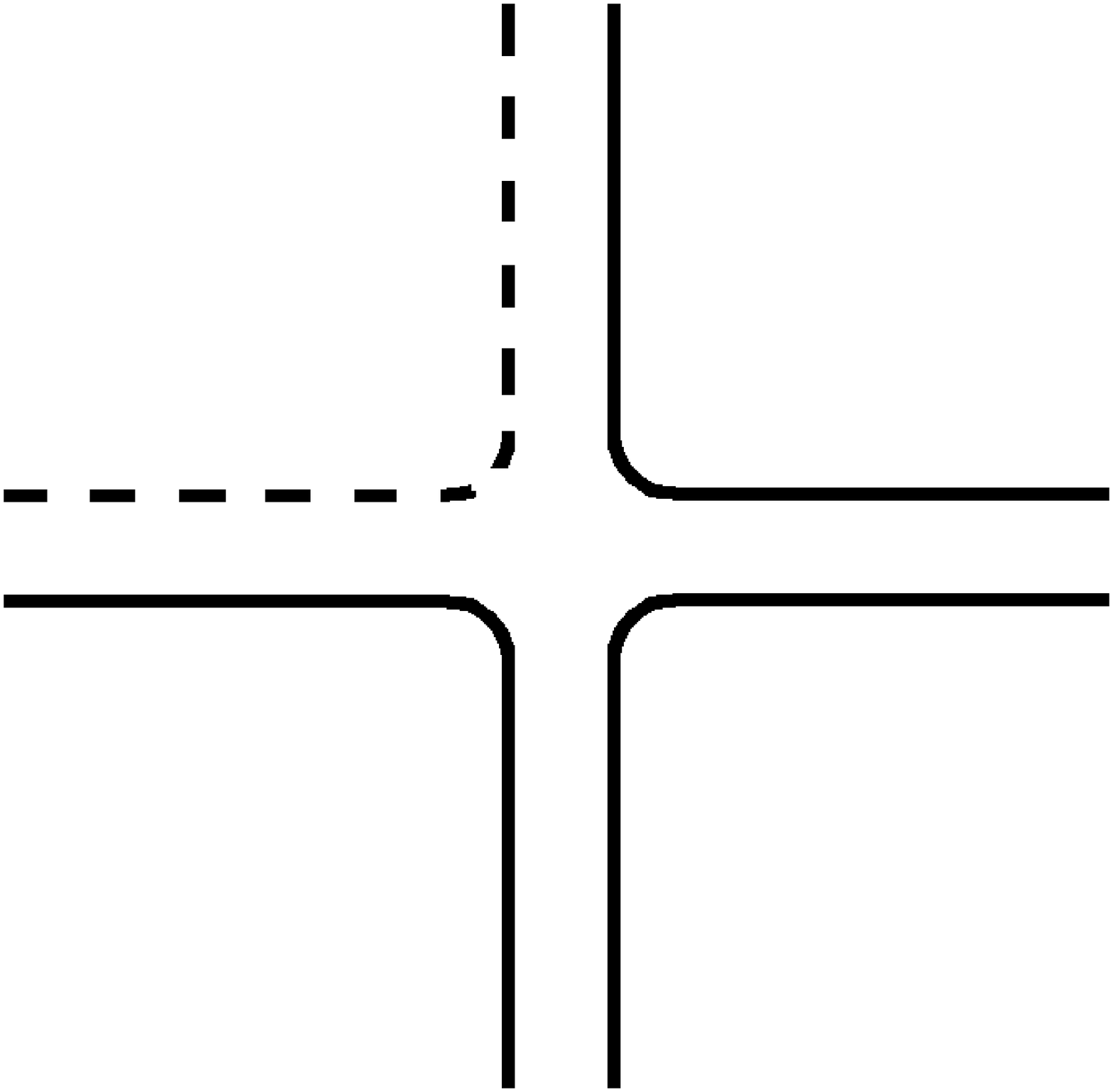}}\right)_{\rm
4I}$&$\sim {\rm sym}
[-2P_{24}(k_{1\nu}k_{2\sigma}\eta_{\beta\rho}\eta_{\lambda\mu}\eta_{\alpha\gamma})
-2P_{12}(k_{1\sigma}k_{2\rho}\eta_{\gamma\nu}\eta_{\beta\mu}\eta_{\alpha\lambda})
-2P_{12}(k_{1\sigma}k_{2\rho}\eta_{\gamma\lambda}\eta_{\mu\nu}\eta_{\alpha\beta})
$\\&$-2P_{24}(k_{1\nu}k_{2\sigma}\eta_{\beta\mu}\eta_{\alpha\rho}\eta_{\lambda\gamma})
-2P_{12}(k_{1\nu}k_{2\mu}\eta_{\beta\sigma}\eta_{\gamma\rho}\eta_{\lambda\alpha})],$\\
\vspace{0.05cm}
$\left(\parbox{1cm}{\includegraphics[height=0.8cm]{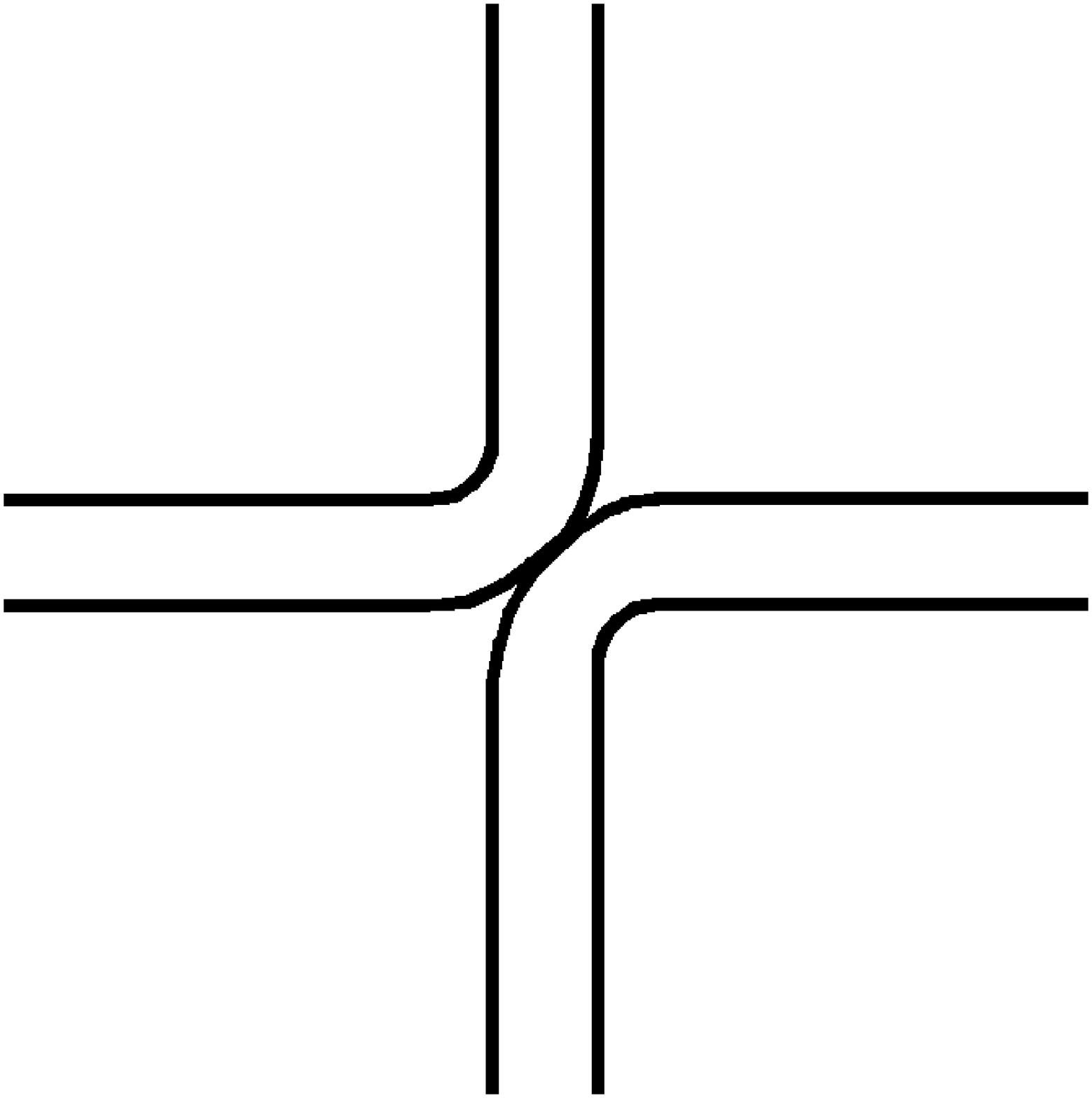}}\right)_{\rm
4J}$&$\sim {\rm sym} [-\frac12P_6(k_1\cdot k_2
\eta_{\mu\nu}\eta_{\alpha\beta}\eta_{\sigma\rho}\eta_{\gamma\lambda})
-P_{12}(k_1\cdot k_2
\eta_{\mu\sigma}\eta_{\alpha\gamma}\eta_{\nu\rho}\eta_{\beta\lambda})],$\\
\vspace{0.05cm}
$\left(\parbox{1cm}{\includegraphics[height=0.8cm]{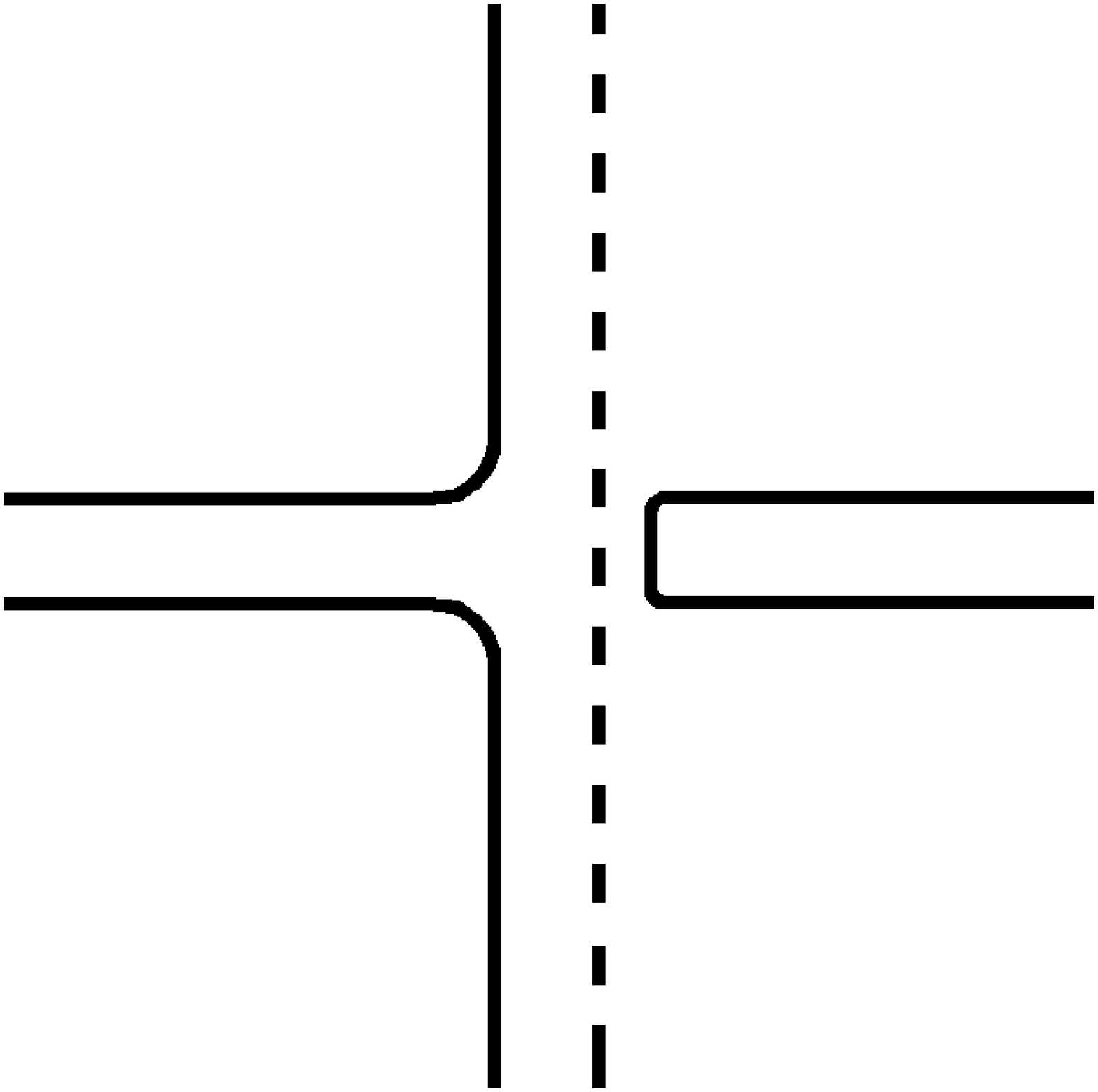}}\right)_{\rm
4K}$&$\sim {\rm sym}
[P_{24}(k_{1\nu}k_{2\sigma}\eta_{\beta\mu}\eta_{\alpha\gamma}\eta_{\rho\lambda})
+P_{12}(k_{1\nu}k_{2\mu}\eta_{\beta\sigma}\eta_{\gamma\alpha}\eta_{\rho\lambda})
-2P_{12}(k_{1\nu}k_{1\sigma}\eta_{\mu\alpha}\eta_{\beta\rho}\eta_{\lambda\gamma})],$\\
\vspace{0.05cm}
$\left(\parbox{1cm}{\includegraphics[height=0.8cm]{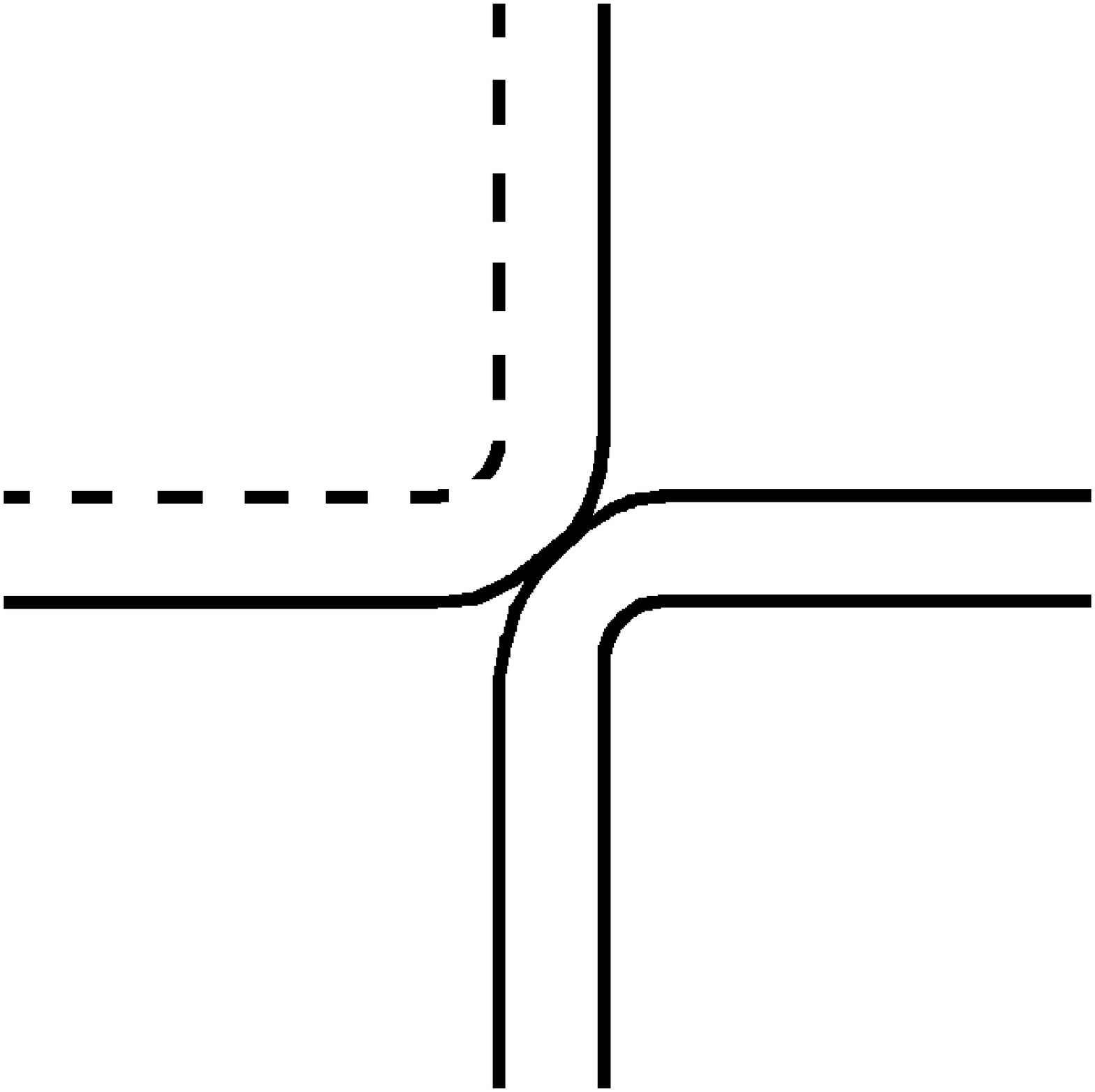}}\right)_{\rm
4L}$&$\sim {\rm sym}
[P_6(k_{1\nu}k_{2\mu}\eta_{\alpha\beta}\eta_{\sigma\rho}\eta_{\gamma\lambda}
-2P_{12}(k_{1\nu}k_{1\sigma}\eta_{\beta\gamma}\eta_{\mu\rho}\eta_{\alpha\lambda})
-P_{12}(k_{1\sigma}k_{2\rho}\eta_{\gamma\lambda}\eta_{\mu\nu}\eta_{\alpha\beta})].$
\end{tabular}
\caption{A graphical representation of the terms in the 4-point
vertex factor. The notation here is the same as the above for the
3-point vertex factor.\label{4-point}}
\end{figure}

For the 4-point vertex we can use a similar diagrammatic notation
(see figure \ref{4-point}).

There are only two sources of factors of dimension in a diagram. A
dimension factor can arise either from the algebraic contractions
in that particular diagram or from the space-time integrals. No
other possibilities exists. In order to arrive at the large-$D$
limit in gravity both sources of dimension factors have to be
considered and the leading contributions will have to be derived
in a systematic way.

Comparing to the case of the large-$N$ planar diagram expansion,
only the algebraic structure of the diagrams has to be considered.
The symmetry index: ($N$) of the gauge group does not go into the
evaluation of the integrals. Concerns with the $D$-dimensional
integrals only arises in gravity. This is a great difference
between the two expansions.

A dimension factor will arise whenever there is a trace over a
tensor index in a diagram, {\it e.g.},
($\eta_{\mu\alpha}\eta^{\alpha\mu},\
\eta_{\mu\alpha}\eta^{\alpha\beta}
\eta_{\beta\gamma}\eta^{\gamma\mu},\ \ldots \sim D$). The diagrams
with the most traces will dominate the large-$D$ limit in gravity.
Traces of momentum lines will never generate a factor of ($D$).

It is easy to be assured that some diagrams naturally will
generate more traces than other diagrams. The key result of
ref.~\cite{Strominger:1981jg} is that only a particular class of
diagrams will constitute the large-$D$ limit, and that only
certain contributions from these diagrams will be important there.
The large-$D$ limit of gravity will correspond to a truncation of
the Einstein theory of gravity, where in fact only a subpart,
namely the leading contributions of the graphs will dominate. We
will use the conventional expansion of the gravitational field,
and we will not separate conformal and traceless parts of the
metric tensor.

\begin{figure}[h]
\includegraphics[height=0.5cm]{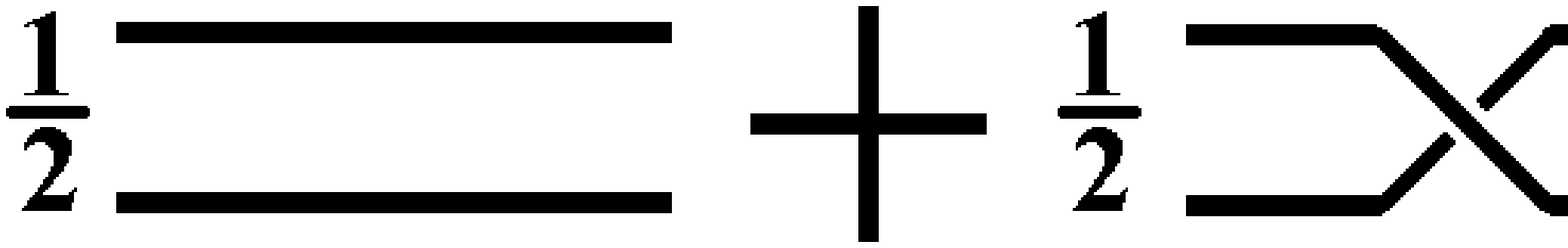}
\caption{A graphical representation of the graviton propagator,
where a full line corresponds to a contraction of two
indices.\label{prop}}
\end{figure}

To justify this we begin by looking on how traces occur from contractions of the propagator.\\
We can graphically write the propagator in the way shown in figure
\ref{prop}.

\begin{figure}[h]
\parbox{8.7cm}{\includegraphics[height=2.2cm]{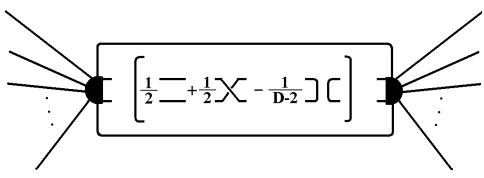}}
\caption{A graphical representation of a propagator contraction in
an amplitude. The black half dots represent an arbitrary internal
structure for the amplitude, the full lines between the two half
dots are internal contractions of indices, the outgoing lines
represents index contractions with external sources.
\label{figprop1}}
\end{figure}

The only way we can generate index loops is through a propagator.
We can graphically represent a propagator contraction of two
indices in a particular amplitude in the following way (see figure
\ref{figprop1}).

\begin{figure}[h]
\parbox{8.7cm}{\includegraphics[height=2.2cm]{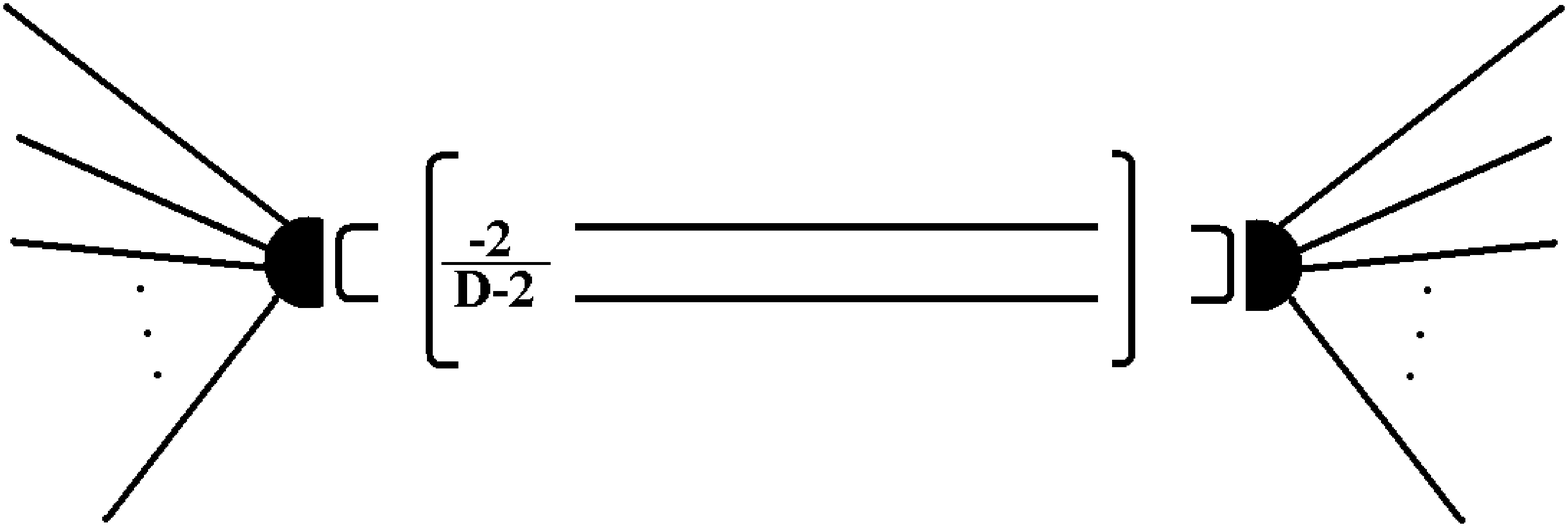}}
\parbox{8.7cm}{\includegraphics[height=2.2cm]{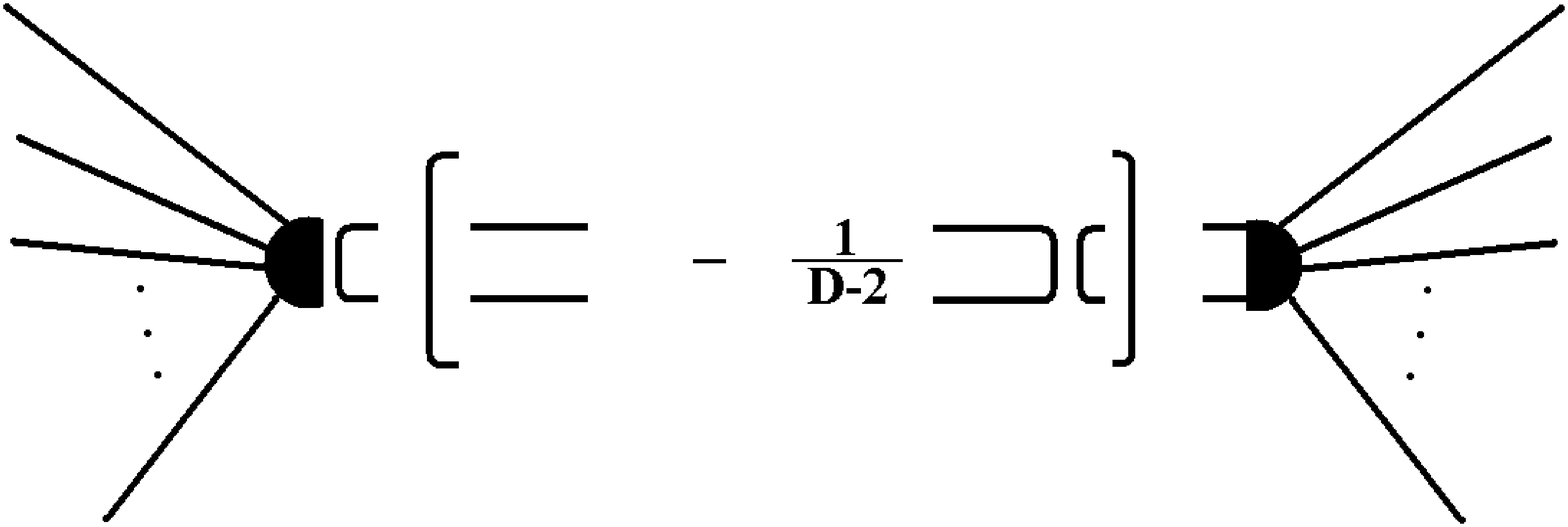}}
\caption{Two possible types of contractions for the propagator.
Whenever we have an index loop we have a contraction of indices.
It is seen that none of the above contractions will generate
something with a positive power of ($D$).\label{figprop2}}
\end{figure}

\begin{figure}[h]
\parbox{8.7cm}{\includegraphics[height=2.2cm]{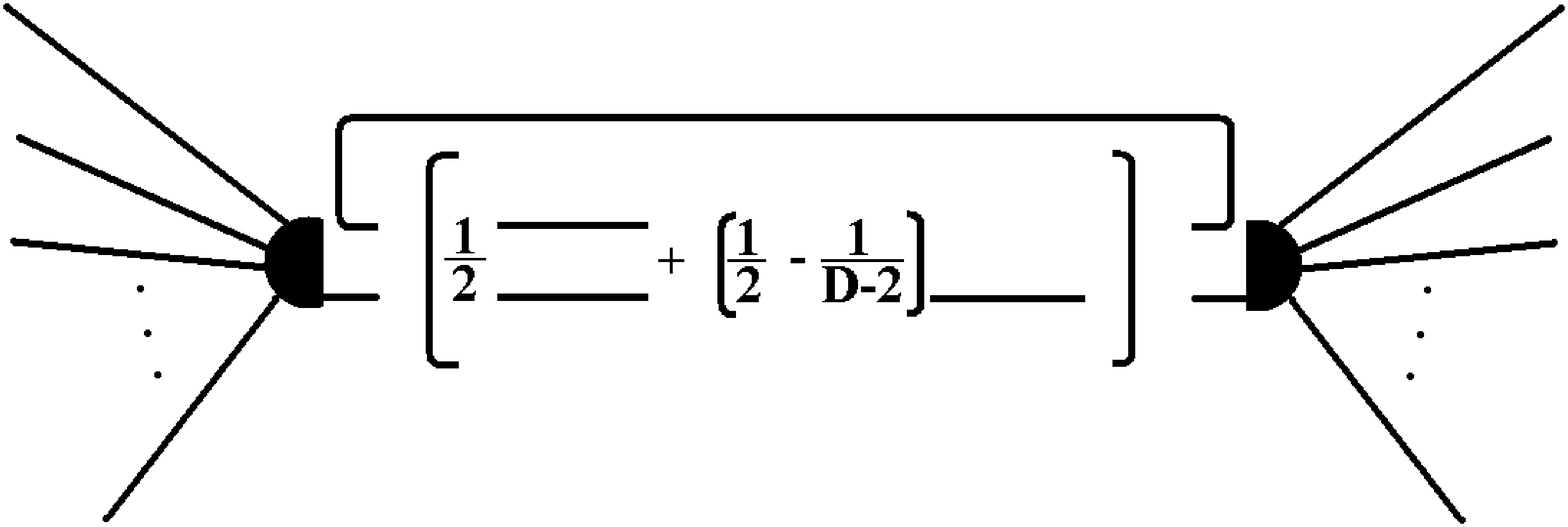}}
\parbox{8.7cm}{\includegraphics[height=2.2cm]{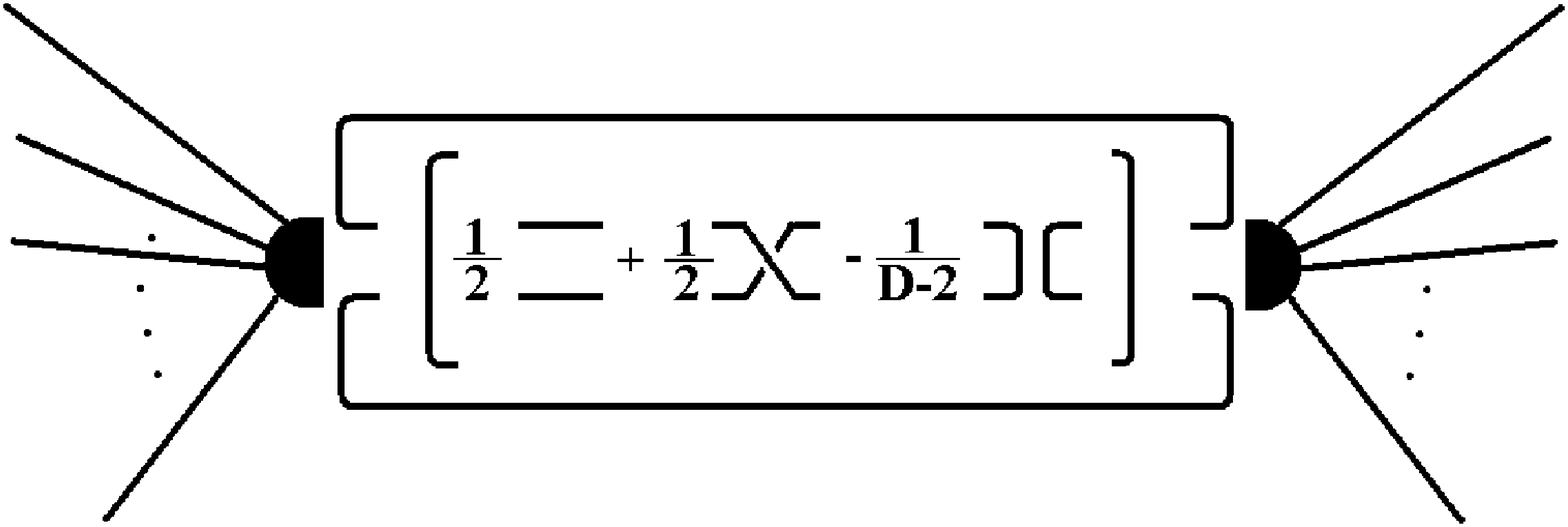}}
\caption{Two other types of index contractions for the propagator.
Again it is seen that only the
$(\eta_{\alpha\gamma}\eta_{\beta\delta}
+\eta_{\alpha\delta}\eta_{\beta\gamma})$ part of the propagator is
important in large-$D$ considerations.\label{figprop3}}
\end{figure}

Essentially (disregarding symmetrizations of indices etc) only the
following contractions of indices in the conventional graviton
propagator can occur in an given amplitude, we can contract, {\it
e.g.}, ($\alpha$ and $\beta$) or both ($\alpha$ and $\beta$) and
($\gamma$ and $\delta$). The results of such contractions are
shown below and also graphically depicted in figure
\ref{figprop2}.
\begin{equation}\begin{split}
\frac12[\eta_{\alpha\gamma}\eta_{\beta\delta}+\eta_{\alpha\delta}\eta_{\beta\gamma}-\frac{2}{D-2}\eta_{\alpha\beta}\eta_{\gamma\delta}\Big]\eta^{\alpha\beta} = \eta_{\gamma\delta}
-\frac{D}{D-2}\eta_{\gamma\delta} \\
\frac12[\eta_{\alpha\gamma}\eta_{\beta\delta}+\eta_{\alpha\delta}\eta_{\beta\gamma}-\frac{2}{D-2}\eta_{\alpha\beta}\eta_{\gamma\delta}\Big]\eta^{\alpha\beta}\eta^{\gamma\delta}
=-\frac{2D}{D-2}
\end{split}\end{equation}
We can can also contract, {\it e.g.}, ($\alpha$ and $\gamma$) or
both ($\alpha$ and $\gamma$) and ($\beta$ and $\delta$) again the
results are shown below and depicted graphically in figure
\ref{figprop3}.
\begin{equation}\begin{split}
\frac12[\eta_{\alpha\gamma}\eta_{\beta\delta}+\eta_{\alpha\delta}\eta_{\beta\gamma}-\frac{2}{D-2}\eta_{\alpha\beta}\eta_{\gamma\delta}\Big]\eta^{\alpha\gamma} =
\frac12\eta_{\beta\delta}D+\eta_{\beta\delta}\left(\frac12+\frac{1}{D-2}\right)\\
\frac12[\eta_{\alpha\gamma}\eta_{\beta\delta}+\eta_{\alpha\delta}\eta_{\beta\gamma}-\frac{2}{D-2}\eta_{\alpha\beta}\eta_{\gamma\delta}\Big]\eta^{\alpha\gamma}\eta^{\beta\delta}
= D^2 - \frac{D}{D-2}\end{split}\end{equation}

As it is seen, only the
$(\eta_{\alpha\gamma}\eta_{\beta\delta}+\eta_{\alpha\delta}\eta_{\beta\gamma})$
part of the propagator will have the possibility to contribute
with a positive power of ($D$). Whenever the
$\left(\frac{\eta_{\mu\nu}\eta_{\alpha\beta}}{D-2}\right)$ term in
the propagator is in play, {\it e.g.}, what remains is something
that goes as $\left(\sim \frac{D}{D-2}\right)$ and which
consequently will have no support to the large-$D$ leading loop
contributions.

\begin{figure}[h]
\parbox{3.3cm}{\includegraphics[height=0.7cm]{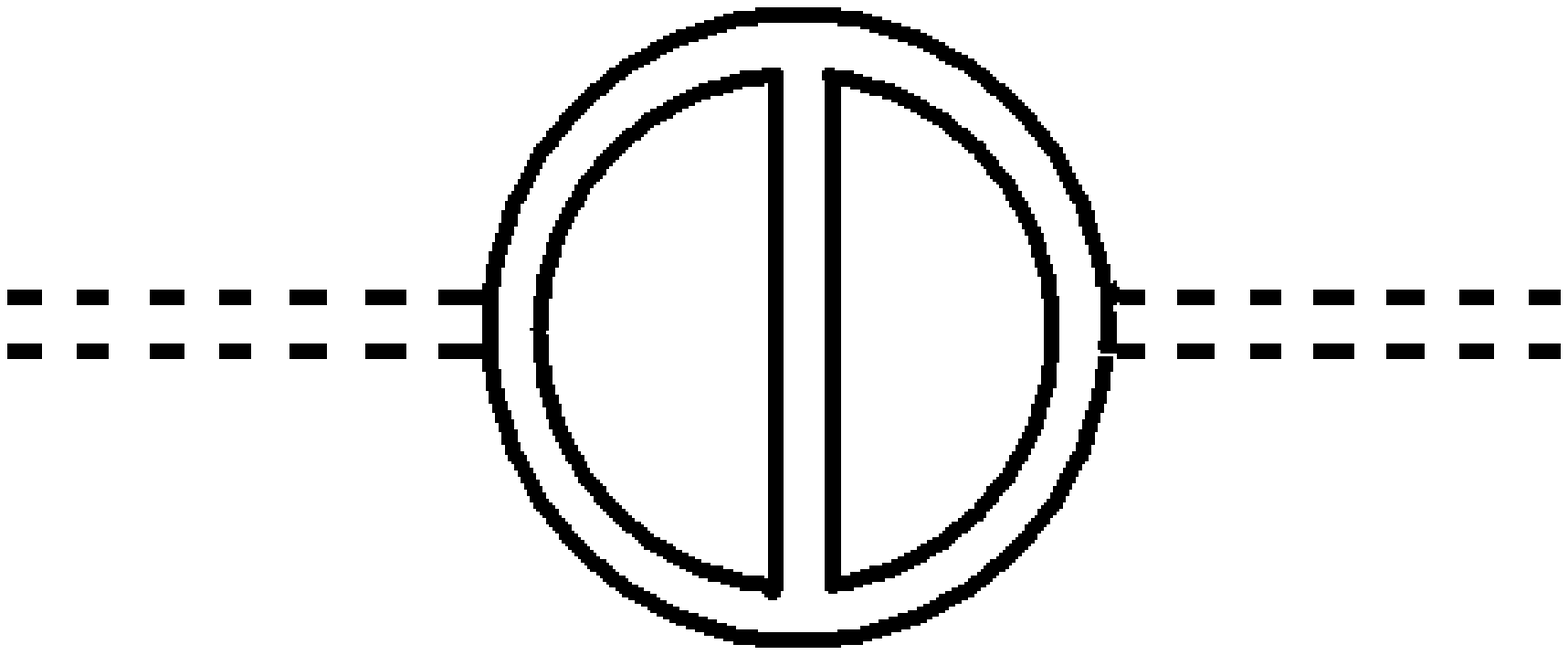}}
\parbox{3.3cm}{\includegraphics[height=0.7cm]{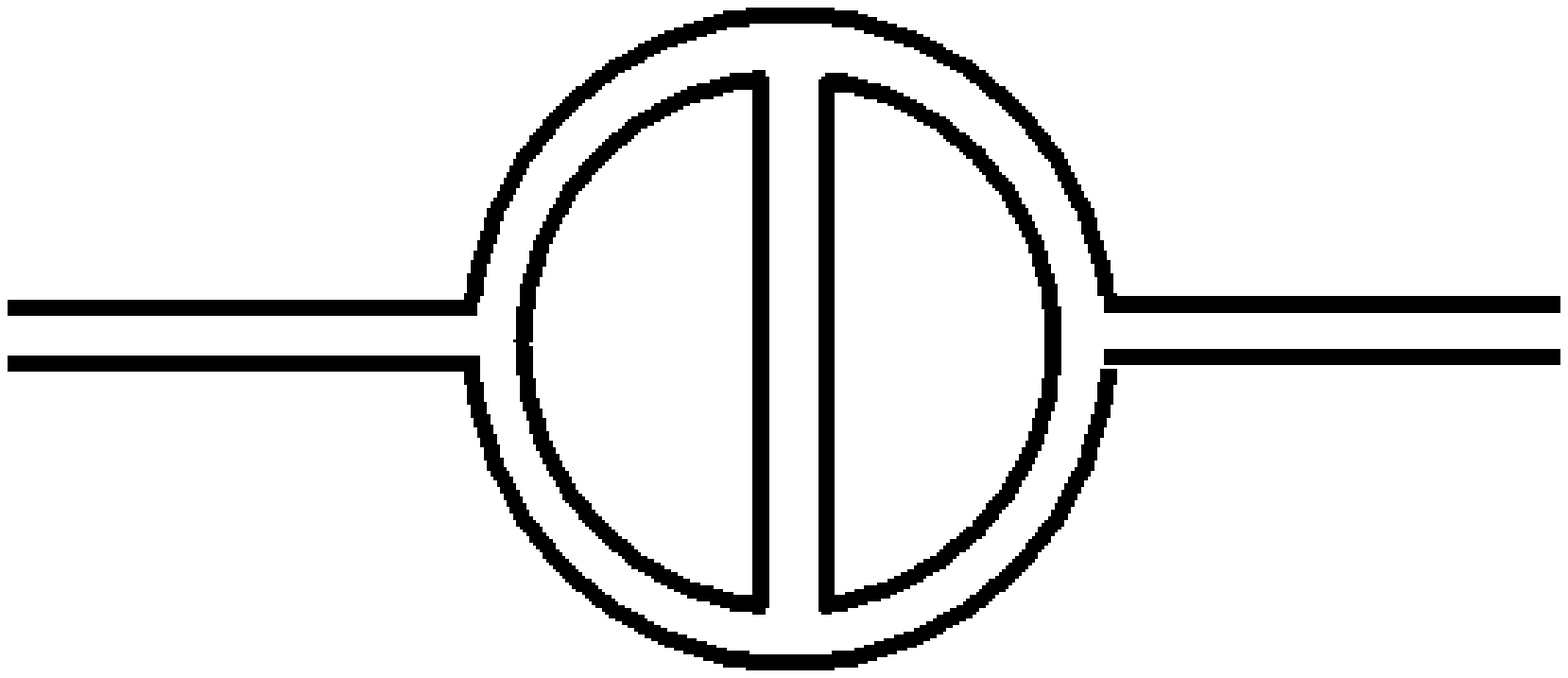}}
\caption{Different trace structures in the same diagram, generated
by different index structures in the vertex (i.e., (3C) and (3E)
respectively), the first diagram have three traces, $\sim D^3$,
the other diagram have only two traces, $\sim D^2$.\label{dif3}}
\end{figure}

We have seen that different index structures go into the same
vertex factor. Below (see figure \ref{dif3}), we depict the same
2-loop diagram, but for different index structures of the external
3-point vertex factor (i.e., (3C) and (3E)) for the external
lines.
Different types of trace structures will hence occur in the same
loop diagram. Every diagram will consist of a sum of contributions
with a varying number of traces. It is seen that the graphical
depiction of the index structure in the vertex factor is a very
useful tool in determining the trace structure of various
diagrams.

\begin{figure}[h]
\begin{minipage}{\linewidth}
\parbox{3.2cm}{\includegraphics[height=0.7cm]{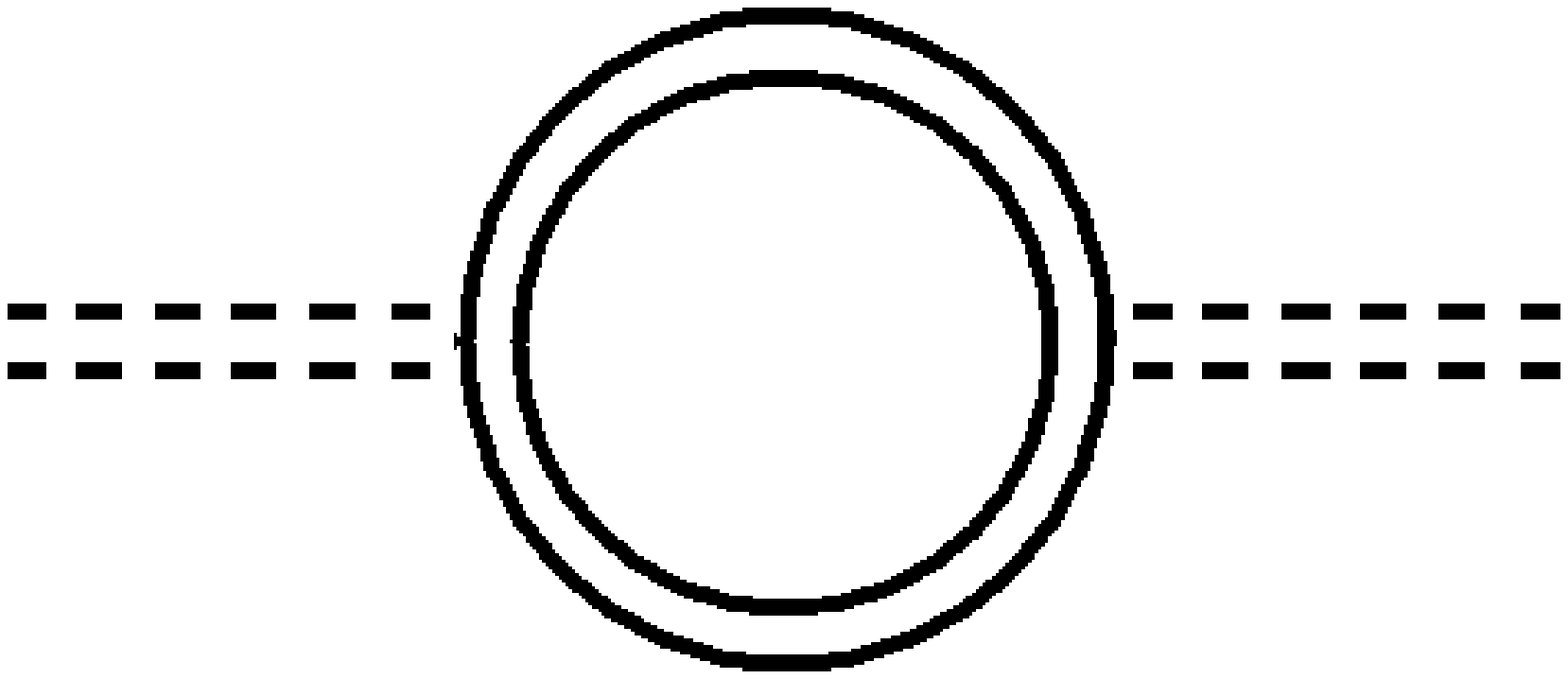}}\parbox{3.2cm}
{\includegraphics[height=0.7cm]{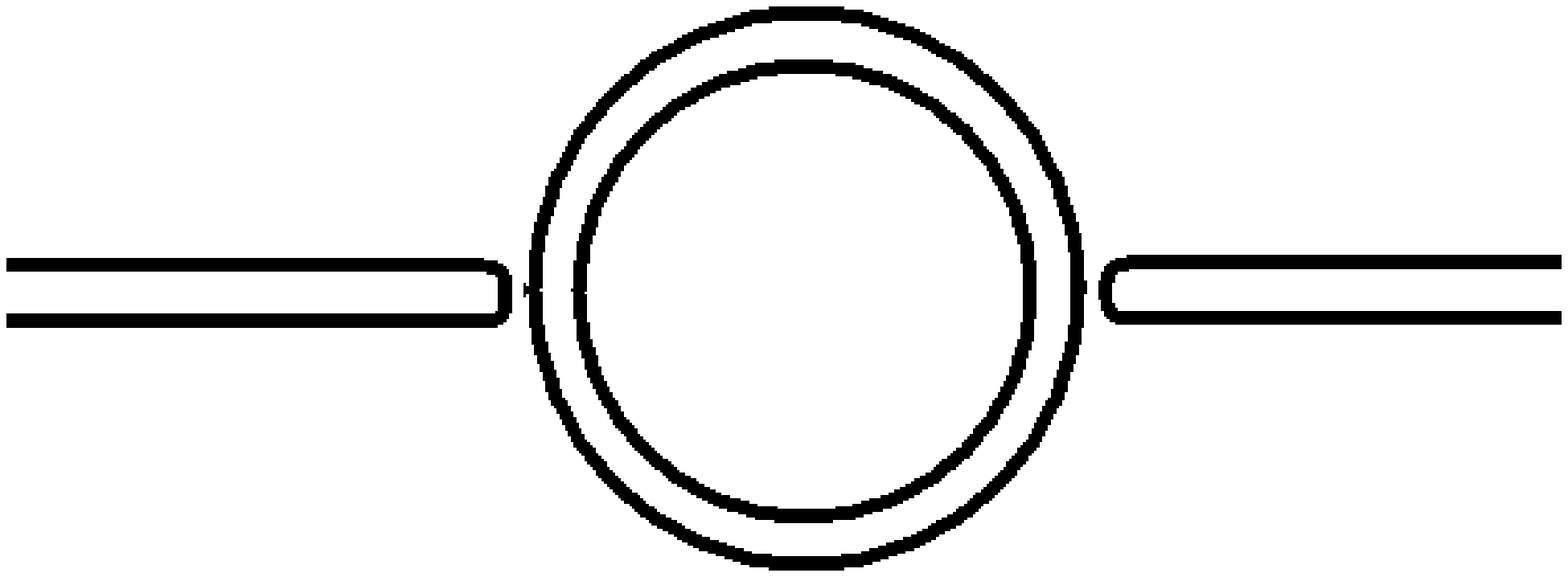}} \caption{A
diagrammatic representation of some leading contributions to the
2-point 1-loop graph in the large-$D$ limit.\label{1-loop}}
\end{minipage}\vspace{0.5cm}
\begin{minipage}{\linewidth}
\parbox{3.2cm}{\includegraphics[height=0.7cm]{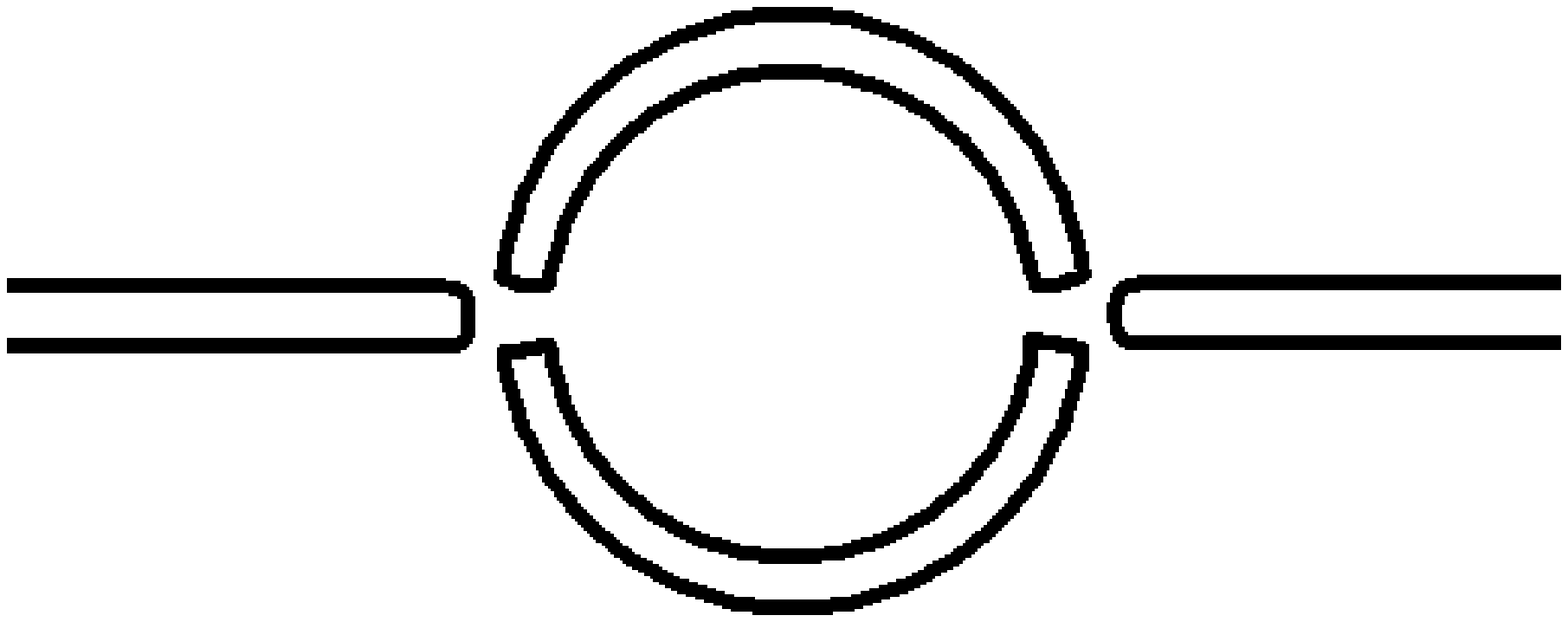}}
\parbox{3.2cm}{\includegraphics[height=0.7cm]{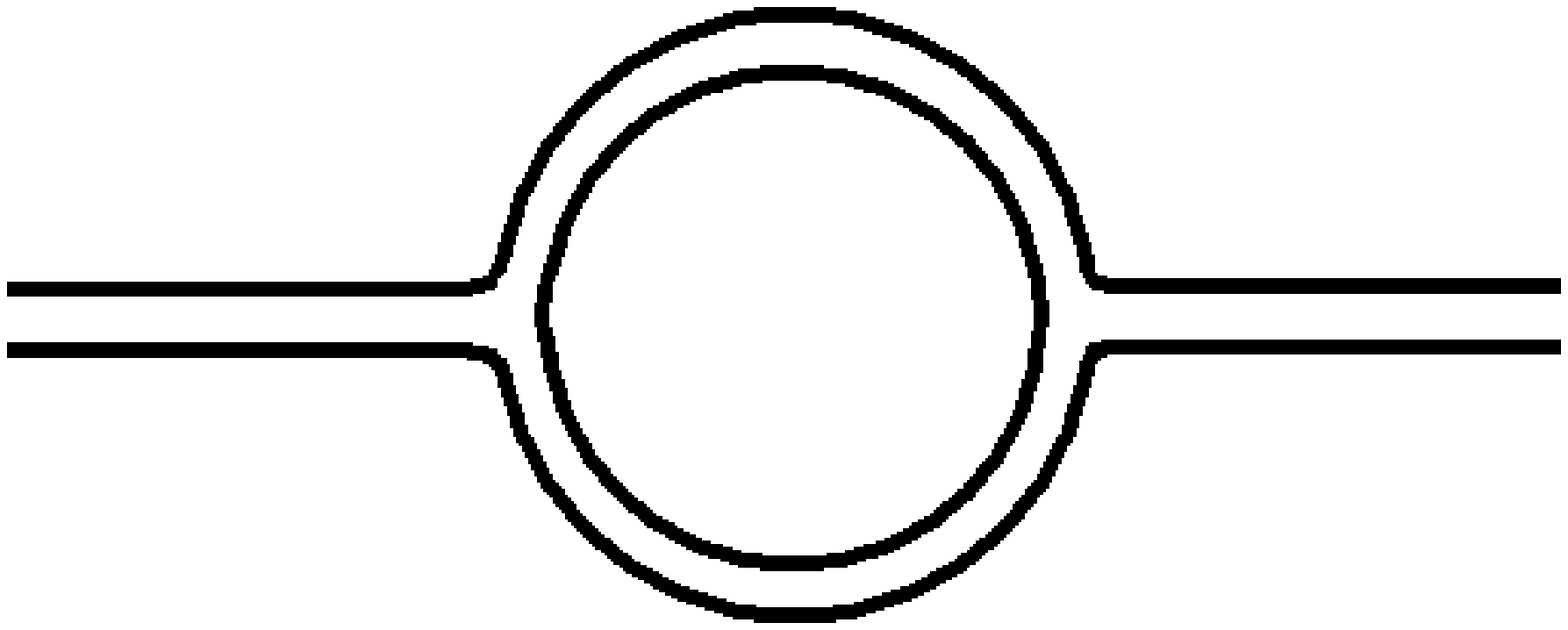}}
\caption{Some non-leading contributions to the 2-point 1-loop
graph in the large-$D$ limit.\label{1-loop2}}
\end{minipage}\end{figure}

The following type of two-point diagram (see figure \ref{1-loop})
build from the (3B) or (3C) parts of the 3-point vertex will
generate a contribution which carries a maximal number of traces
compared to the number of vertices in the diagram.
The other parts of the 3-point vertex factor, {\it e.g.}, the (3A)
or (3E) parts respectively will not generate a leading
contribution in the large-$D$ limit, however such contributions
will go into the non-leading contributions of the theory (see
figure \ref{1-loop2}). The diagram built from the (3A) index
structure will, {\it e.g.}, not carry a leading contribution due
to our previous considerations regarding contractions of the
indices in the propagator (see figure \ref{figprop2}).

It is crucial that we obtain a consistent limit for the graphs,
when we take ($D \rightarrow \infty$). In order to cancel the
factor of ($D^2$) from the above type of diagrams, {\it i.e.}, to
put it on equal footing in powers of ($D$) as the simple
propagator diagram, it is seen that we need to rescale
$\left(\kappa \rightarrow \frac{\kappa}{D}\right)$. In the
large-$D$ limit $\left(\frac{\kappa}{D}\right)$ will define the
'new' redefined gravitational coupling. A finite limit for the
2-point function will be obtained by this choice, i.e, the leading
$D$-contributions will not diverge, when the limit ($D\rightarrow
\infty$) is imposed. By this choice all leading $n$-point diagrams
will be well defined and go as $\left(\sim
\left(\frac{\kappa}{D}\right)^{(n-2)}\right)$ in a
$\left(\frac1D\right)$ expansion. Each external line will carry a
gravitational coupling ~\footnote{\footnotesize{A tadpole diagram
will thus not be well defined by this choice but instead go as
$(\sim D)$ however because we can dimensionally regularize any
tadpole contribution away, any idea to make the tadpoles well
defined in the large-$D$ limit would occur to be strange.}}.
However, it is important to note that our rescaling of $(\kappa)$
is solely determined by the requirement of creating a finite
consistent large-$D$ limit, where all $n$-point tree diagrams and
their loop corrections are on an equal footing. The above
rescaling seems to be the obvious to do from the viewpoint of
creating a physical consistent theory. However, we note that other
limits for the graphs are in fact possible by other redefinitions
of $(\kappa)$ but such rescalings will, {\it e.g.}, scale away the
loop corrections to the tree diagrams, thus such rescalings must
be seen as rather unphysical from the viewpoint of traditional
quantum field theory.

\begin{figure}[h]
\begin{minipage}{\linewidth}
\parbox{3.2cm}{\includegraphics[height=0.7cm]{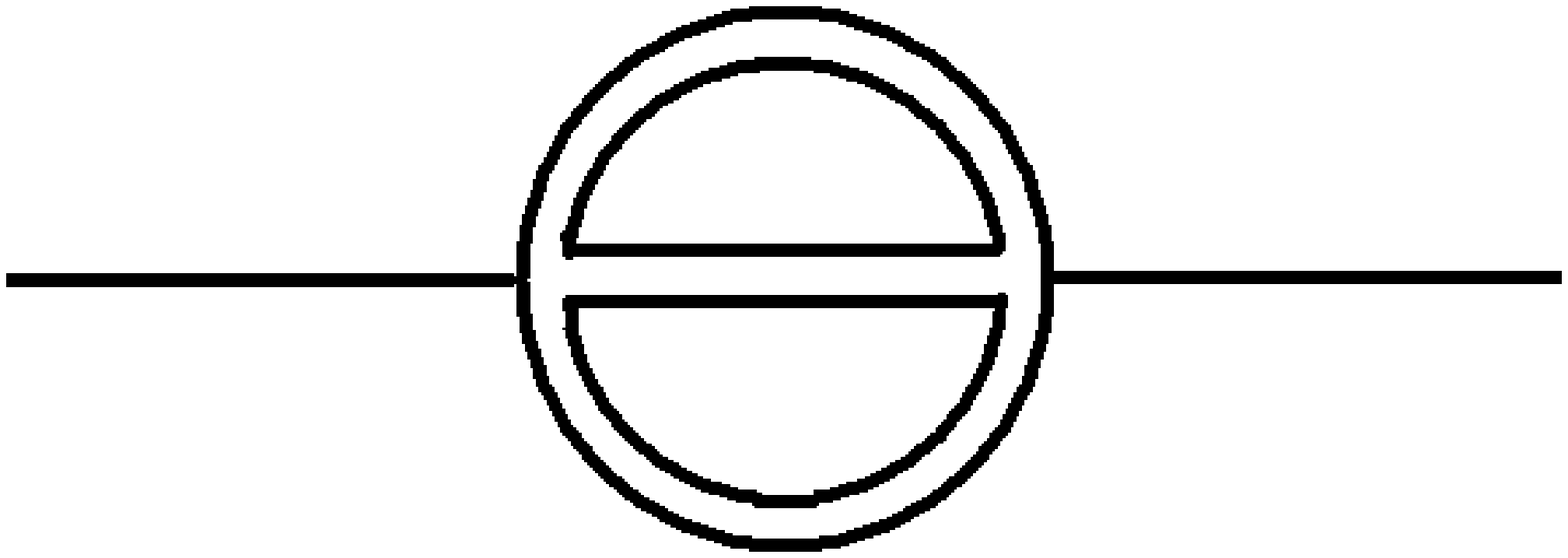}}
\parbox{3.2cm}{\includegraphics[height=0.7cm]{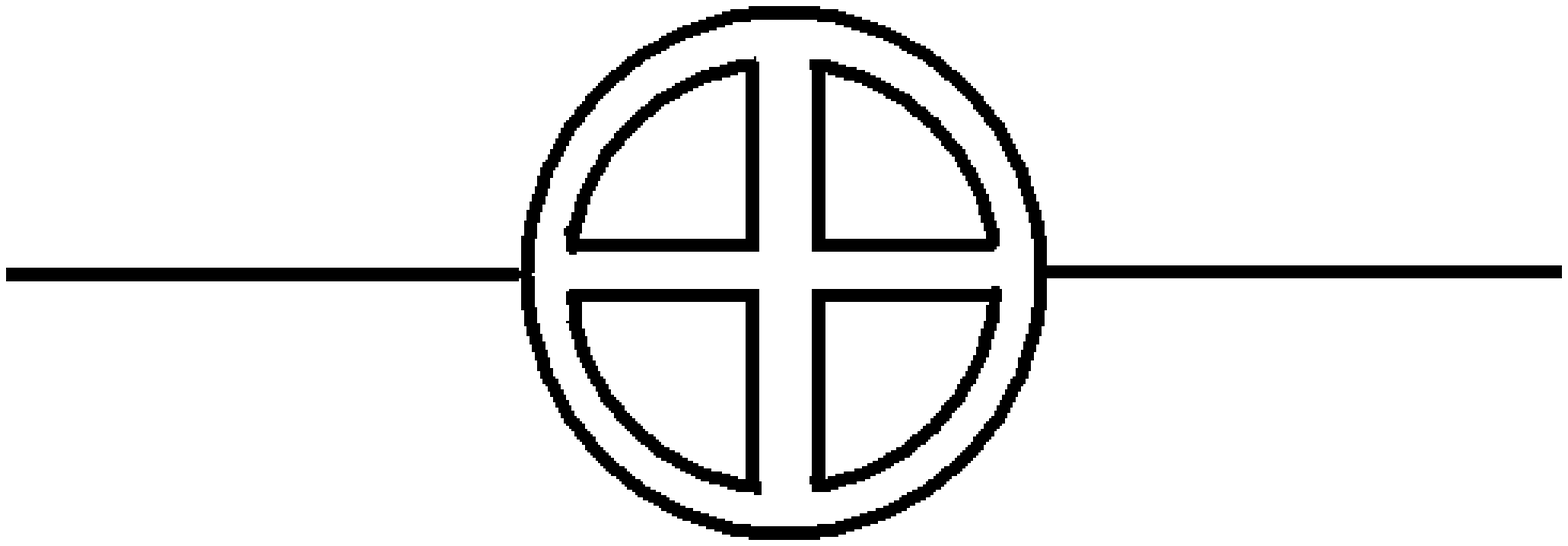}}
\caption{Two multi-loop diagrams with many traces. The rescaling
of $(\kappa)$ will however render both diagrams as non-leading
contributions compared to diagrams with separate loops. Here and
in the next coming figures we will employ the following notation
for the index structure of the lines originating from loop
contributions. Whenever an line originates from a loop
contribution it should be understood that there are only two
possibilities for its index structure, {\it i.e.}, the lines has
to be of the same type as in the contracted index lines in the
index structure (3B) or as the dashed lines in (3C). No
assumptions on the index structures for tree external lines are
made, here any viable index structures are present, but we still
represent such lines with a single full line for simplicity.
\label{other}}
\end{minipage}
\end{figure}

For any loop with two propagators the maximal number of traces are
obviously two. However, we do not know which diagrams will
generate the maximal number of powers of $D$ in total. Some
diagrams will have more traces than others, however because of the
rescaling of $(\kappa)$ they might not carry a leading
contribution after all. For example,  if we look at the two graphs
depicted in figure \ref{other}, they will in fact go as $\left(D^3
\times \frac{\kappa^4}{D^4}\times D^3 \sim \frac{1}{D}\right)$ and
$\left(D^5 \times \frac{\kappa^8}{D^8}\times D^5 \sim
\frac{1}{D^3}\right)$ respectively and thus not be leading
contributions.

In order to arrive at the result of ref.~\cite{Strominger:1981jg},
we have to consider the rescaling of $(\kappa)$ which will
generate negative powers of $(D)$ as well. Let us now discuss the
main result of ref.~\cite{Strominger:1981jg}.

Let ($P_{\rm \max}$) be the maximal power of ($D$) associated with
a graph. Next we can look at the function ($\Pi$):
\begin{equation}
\Pi = 2L - \sum_{m = 3}^\infty(m-2)V_m.
\end{equation}
Here ($L$) is the number of loops in a given diagram, and ($V_m$)
counts the number of $m$-point vertices. It is obviously true that
($P_{\rm max}$) must be less than or equal to ($\Pi$).

The above function counts the maximal number of positive factors
of ($D$) occurring for a given graph. The first part of the
expression simply counts the maximal number of traces ($\sim D^2$)
from a loop, the part which is subtracted counts the powers of
($\kappa$ $\sim \frac1D$) arising from the rescaling of
($\kappa$).

Next, we employ the two well-known formulas for the topology of
diagrams:
\begin{equation}
L = I - V + 1,
\end{equation}
\begin{equation}
\sum_{m=3}^\infty m V_m = 2I + E,
\end{equation}
where ($I$) is the number of internal lines, $\left(V =
\sum_{m=3}^\infty V_m\right)$ is total number of vertices and
($E$) is the number of external lines.

Putting these three formulas together one arrives at:
\begin{equation}
P_{\rm max} \leq 2 - E - 2\sum_{m=3}^\infty V_m.
\end{equation}
Hence, the maximal power of ($D$) is obtained in the case of
graphs with a minimal number of external lines, and two traces per
diagram loop. In order to generate the maximum of two traces per
loop, separated loops are necessary, because whenever a propagator
index line is shared in a diagram we will not generate a maximal
number of traces. Considering only separated bubble graphs, thus
the following type of $n$-point (see figure \ref{bubble}) diagrams
will be preferred in the large-$D$ limit:
\begin{figure}[h]
\vspace{0cm}\parbox{5cm}{\includegraphics[height=3.4cm]{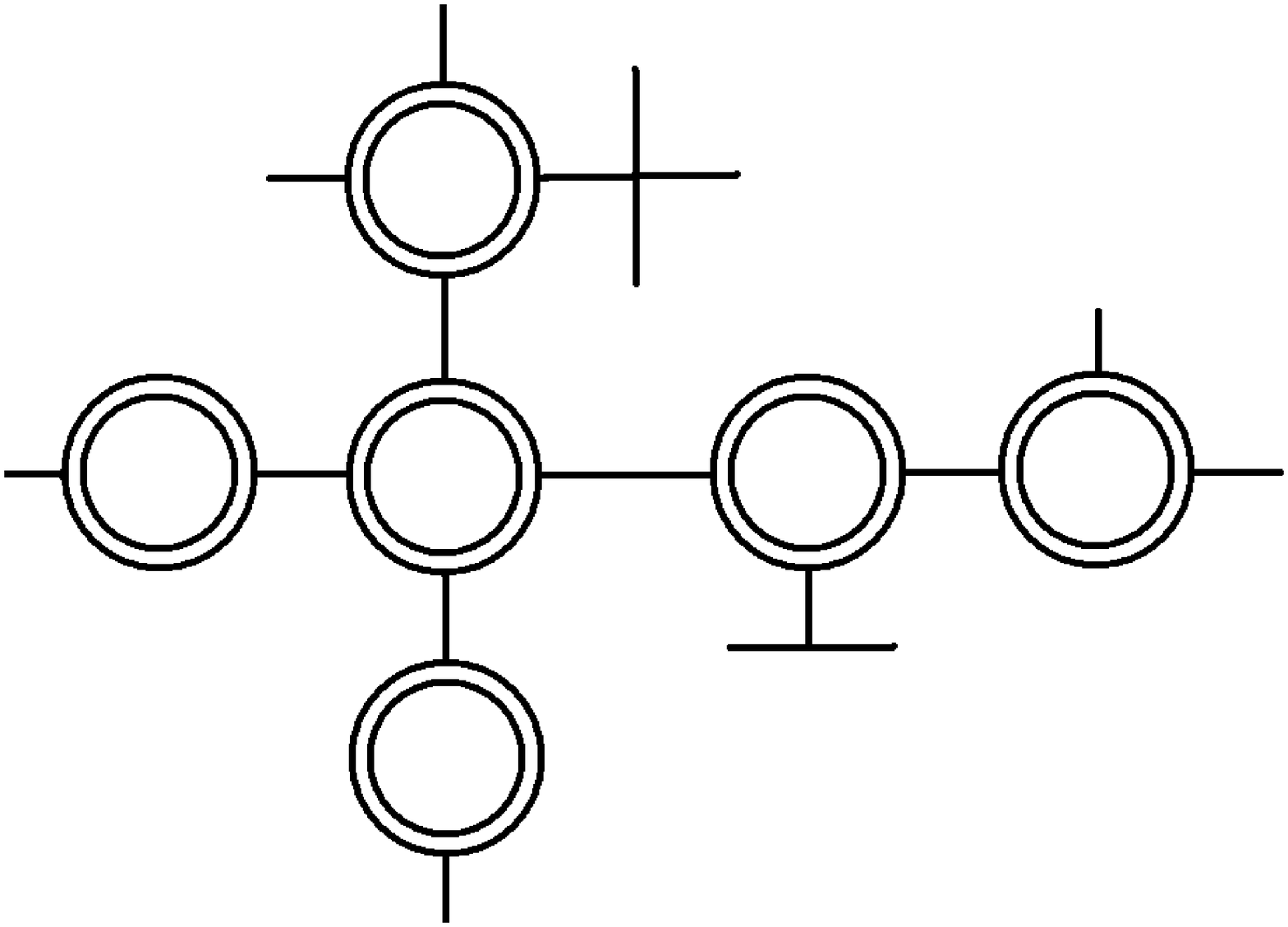}}
\vspace{1.2cm}\parbox{5cm}{\includegraphics[height=3.4cm]{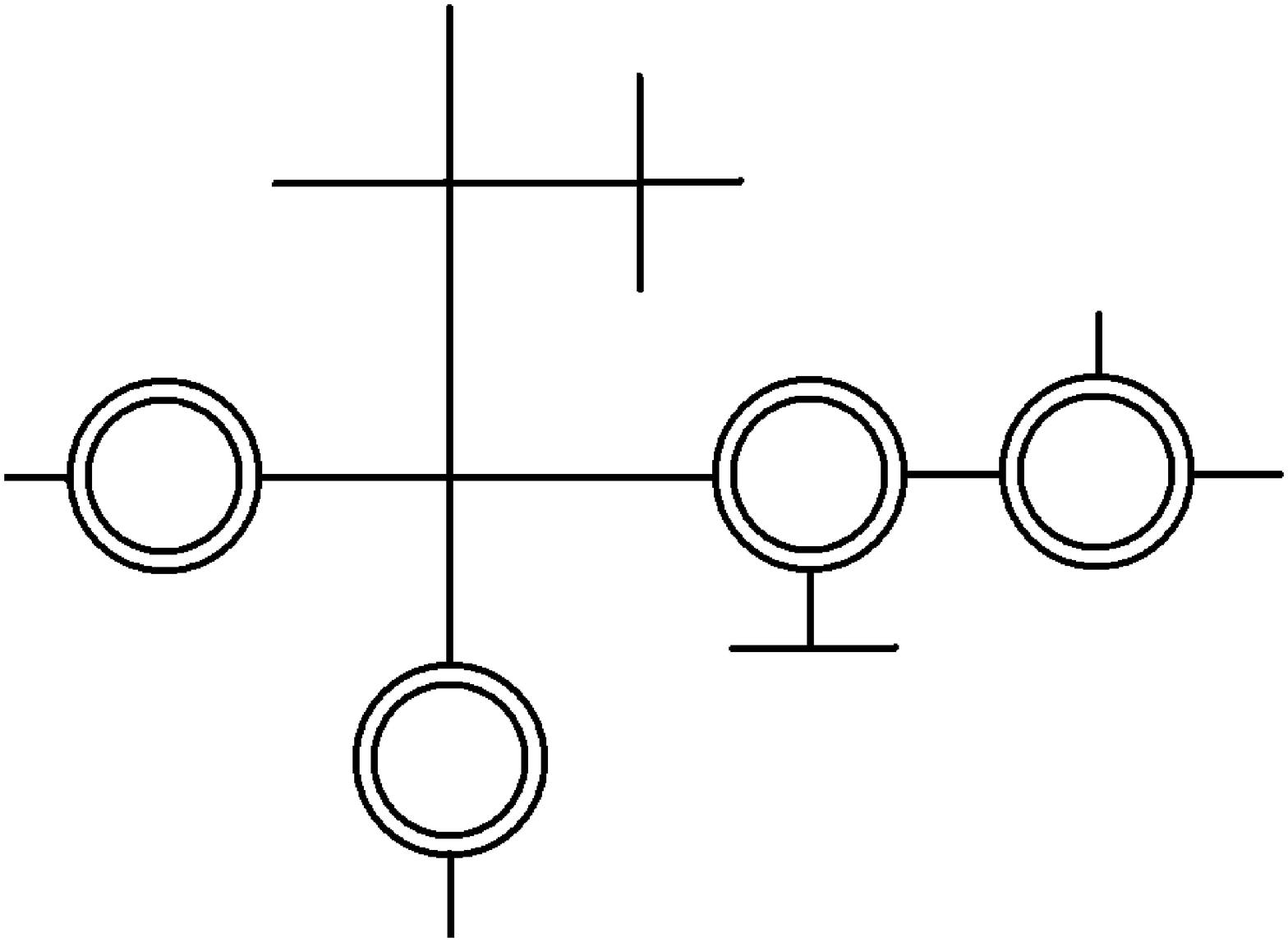}}
\parbox{5cm}{\includegraphics[height=3.4cm]{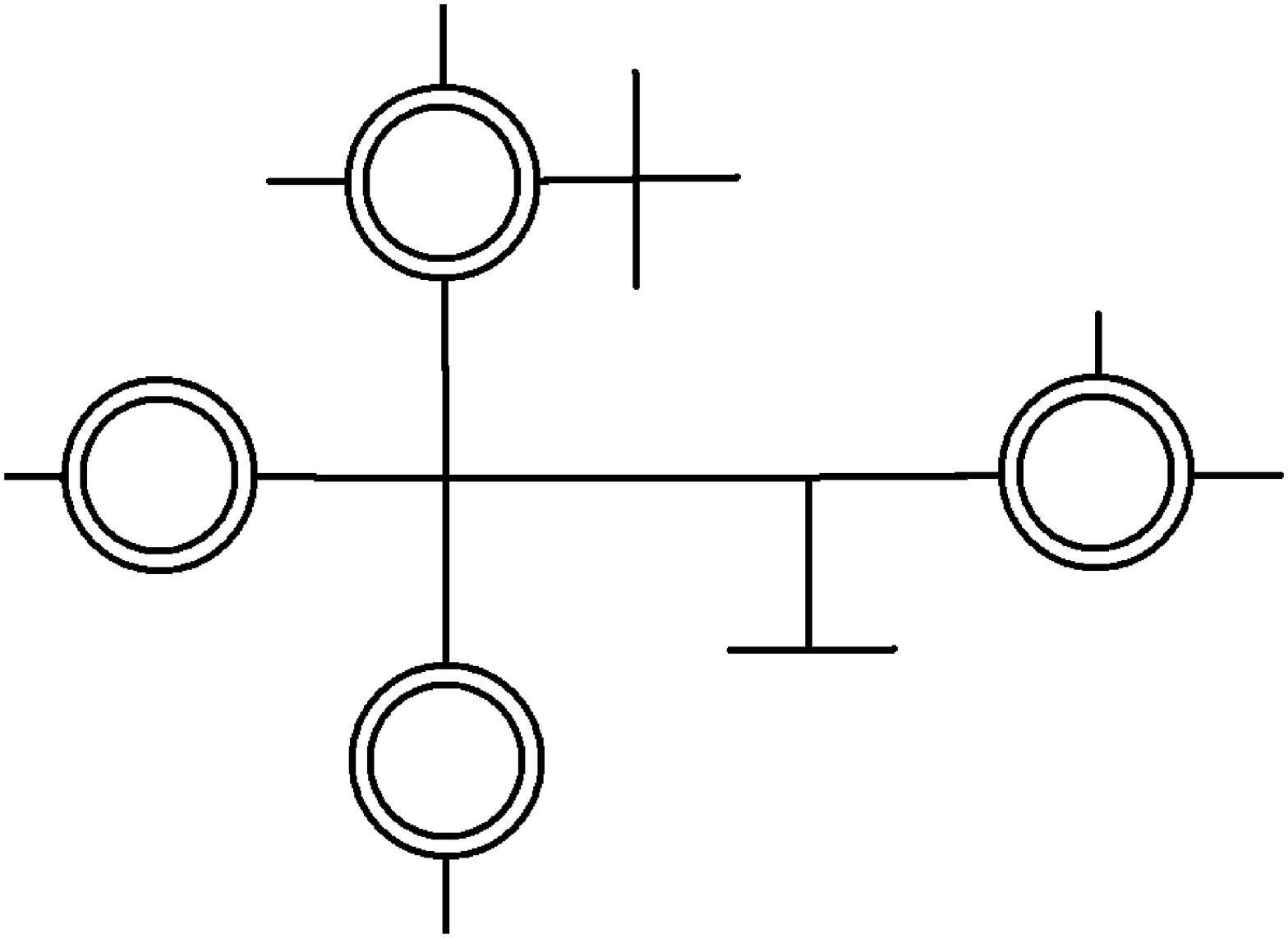}}\vspace{-1.2cm}
\caption{Examples of multi-loop diagrams with leading
contributions in the large-$D$ limit.\label{bubble}}
\end{figure}

\begin{figure}[h]
\parbox{8cm}{\includegraphics[height=0.7cm]{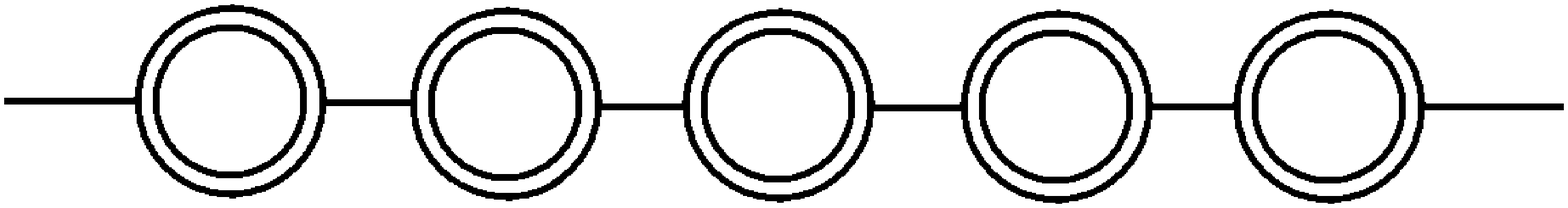}}
\parbox{8cm}{\includegraphics[height=1.6cm]{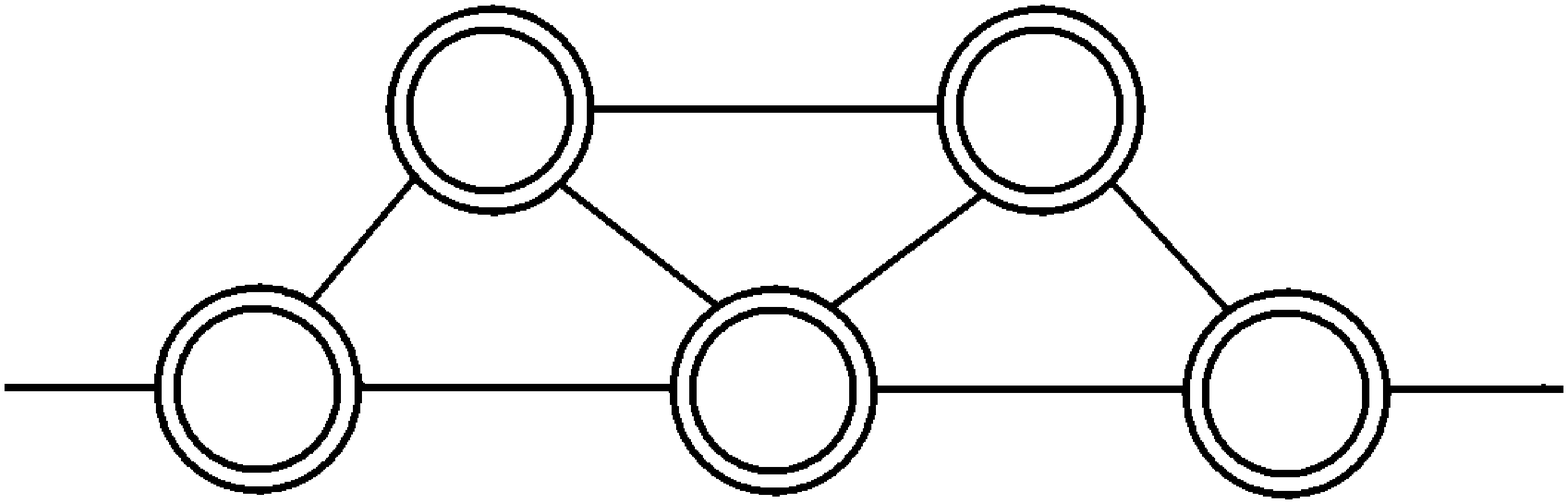}}
\caption{Of the following two diagrams with double trace loops the
first one will be leading.\label{multi}}
\end{figure}

Vertices, which generate unnecessary additional powers of
$\left(\frac1D\right)$, are suppressed in the large-$D$ limit;
multi-loop contributions with a minimal number of vertex lines
will dominate over diagrams with more vertex lines. That is, in
figure \ref{multi} the first diagram will dominate over the second
one.

\begin{figure}[h]
\parbox{3cm}{\includegraphics[height=0.7cm]{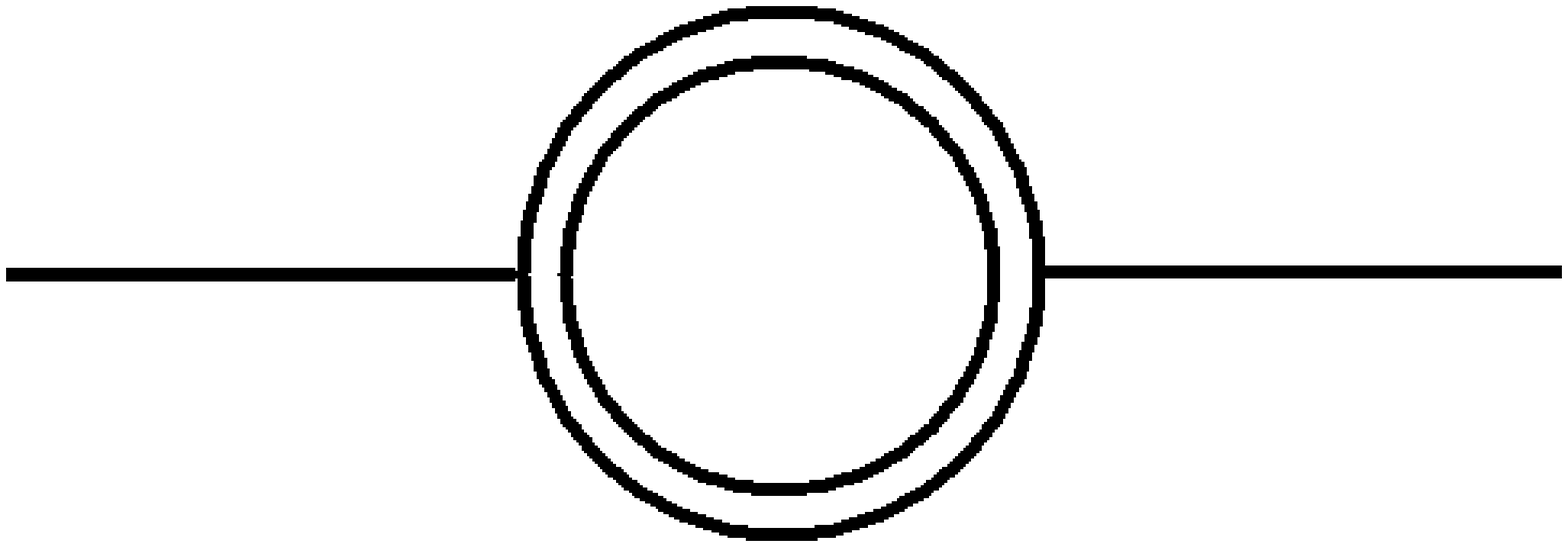}}
\parbox{4cm}{\includegraphics[height=0.7cm]{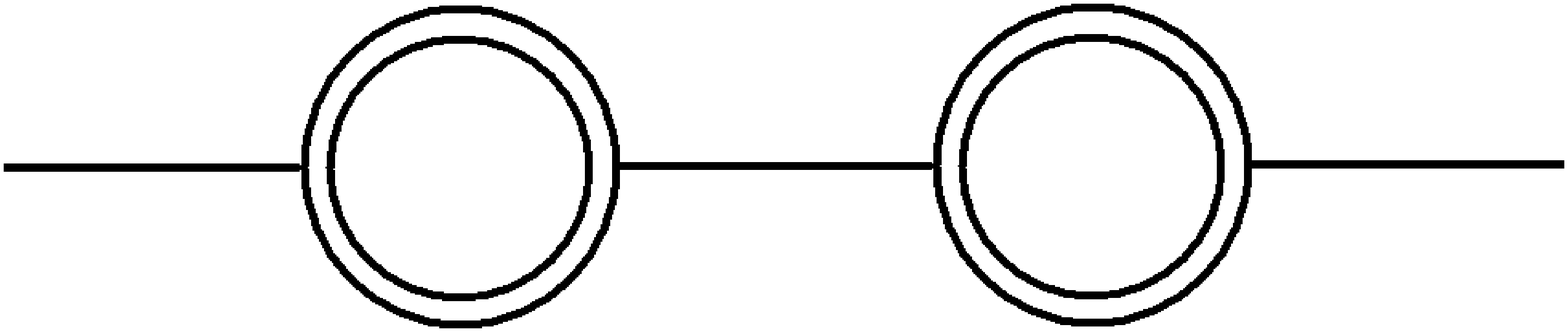}}
\parbox{4.5cm}{\includegraphics[height=0.7cm]{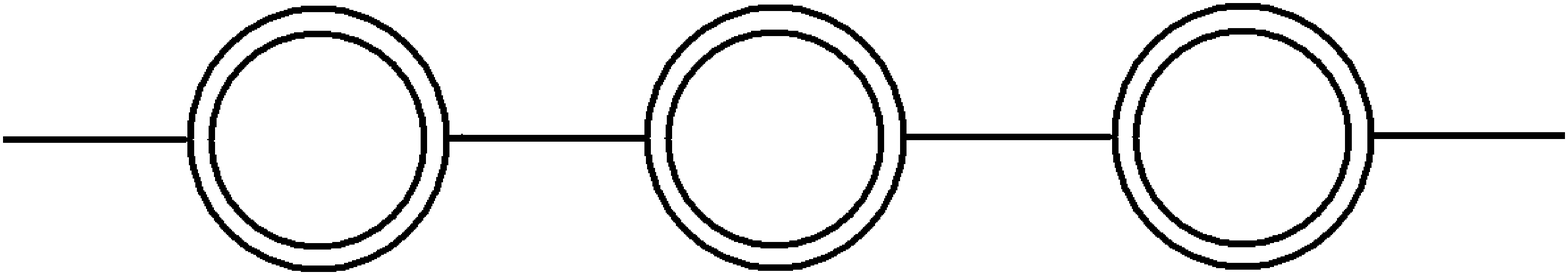}}
\parbox{5cm}{\includegraphics[height=0.7cm]{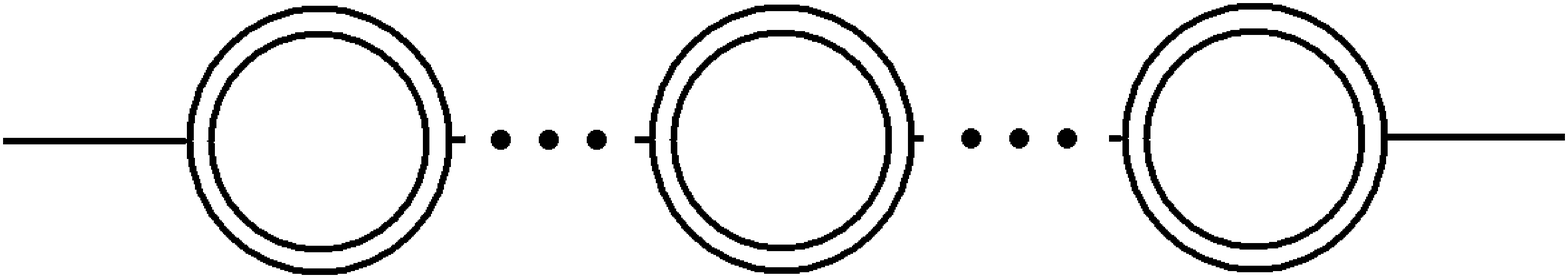}}
\caption{Leading bubble 2-point diagrams.\label{bubble2}}
\end{figure}

Thus, separated bubble diagrams will contribute to the large-$D$
limit in quantum gravity. The 2-point bubble diagram limit is
shown in figure \ref{bubble2}.
\begin{figure}[h]
\parbox{4cm}{\includegraphics[height=0.7cm]{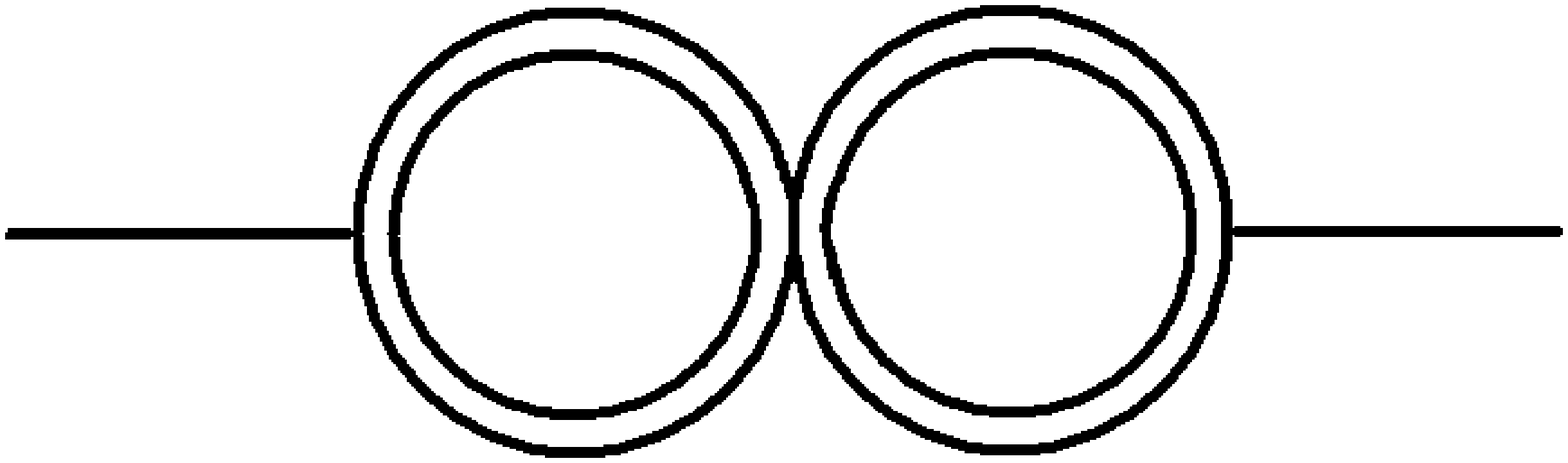}}
\parbox{4cm}{\includegraphics[height=0.7cm]{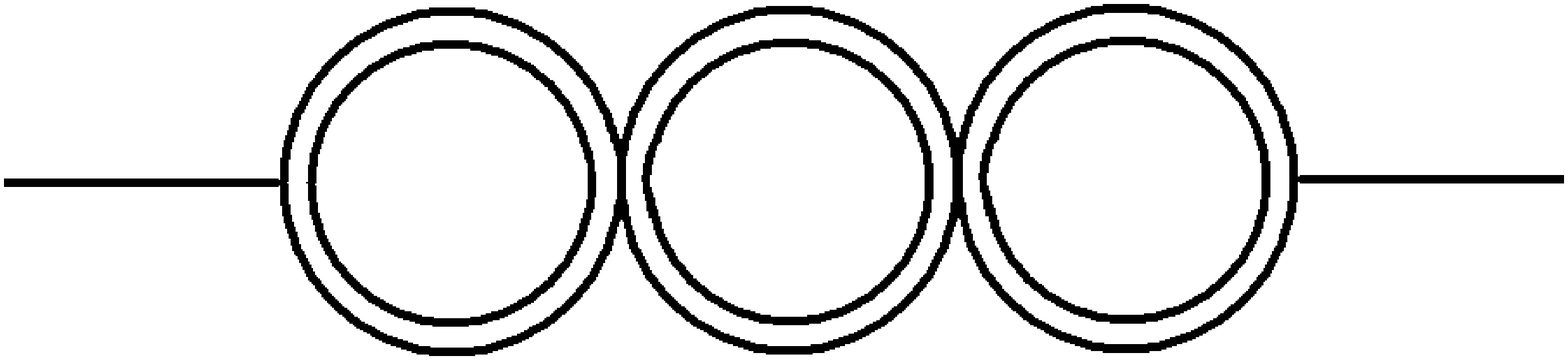}}
\parbox{4cm}{\includegraphics[height=0.7cm]{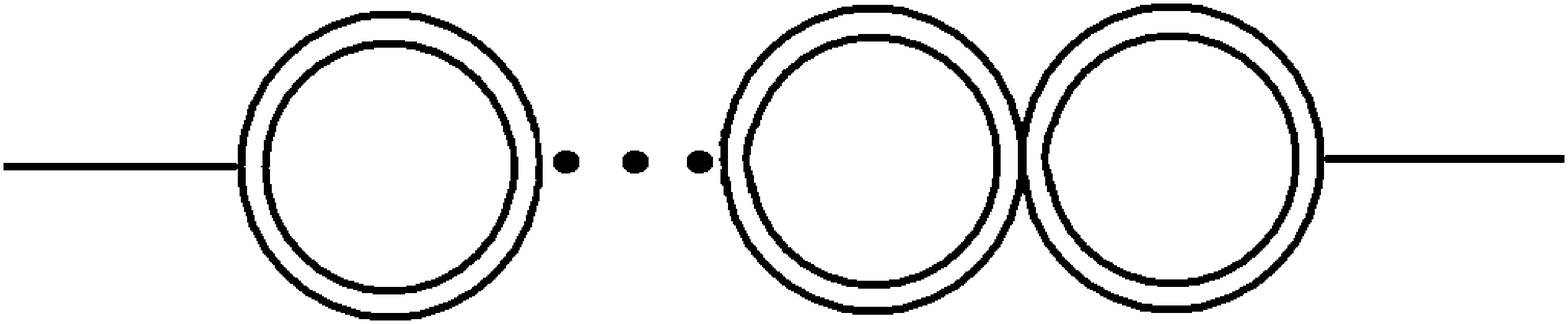}}
\caption{Another type of 2-point diagram build from the index
structure (4J), which will be leading in the large-$D$
limit.\label{4-vertex}}
\end{figure}

Another type of diagrams which can make leading contributions, is
the diagrams depicted in figure \ref{4-vertex}. These
contributions are of a type where a propagator directly combines
different legs in a vertex factor with a double trace loop.
Because of the index structure (4J) in the 4-point vertex such
contributions are possible. These types of diagrams will have the
same dimensional dependence: $\left(\frac{\kappa^4}{D^4}\times
D^4\sim 1\right)$, as the 2-loop separated bubble diagrams. It is
claimed in~\cite{Strominger:1981jg}, that such diagrams will be
non-leading. We have however found no evidence of this in our
investigations of the large-$D$ limit.

\begin{figure}[h]
\parbox{3cm}{\includegraphics[height=1.3cm]{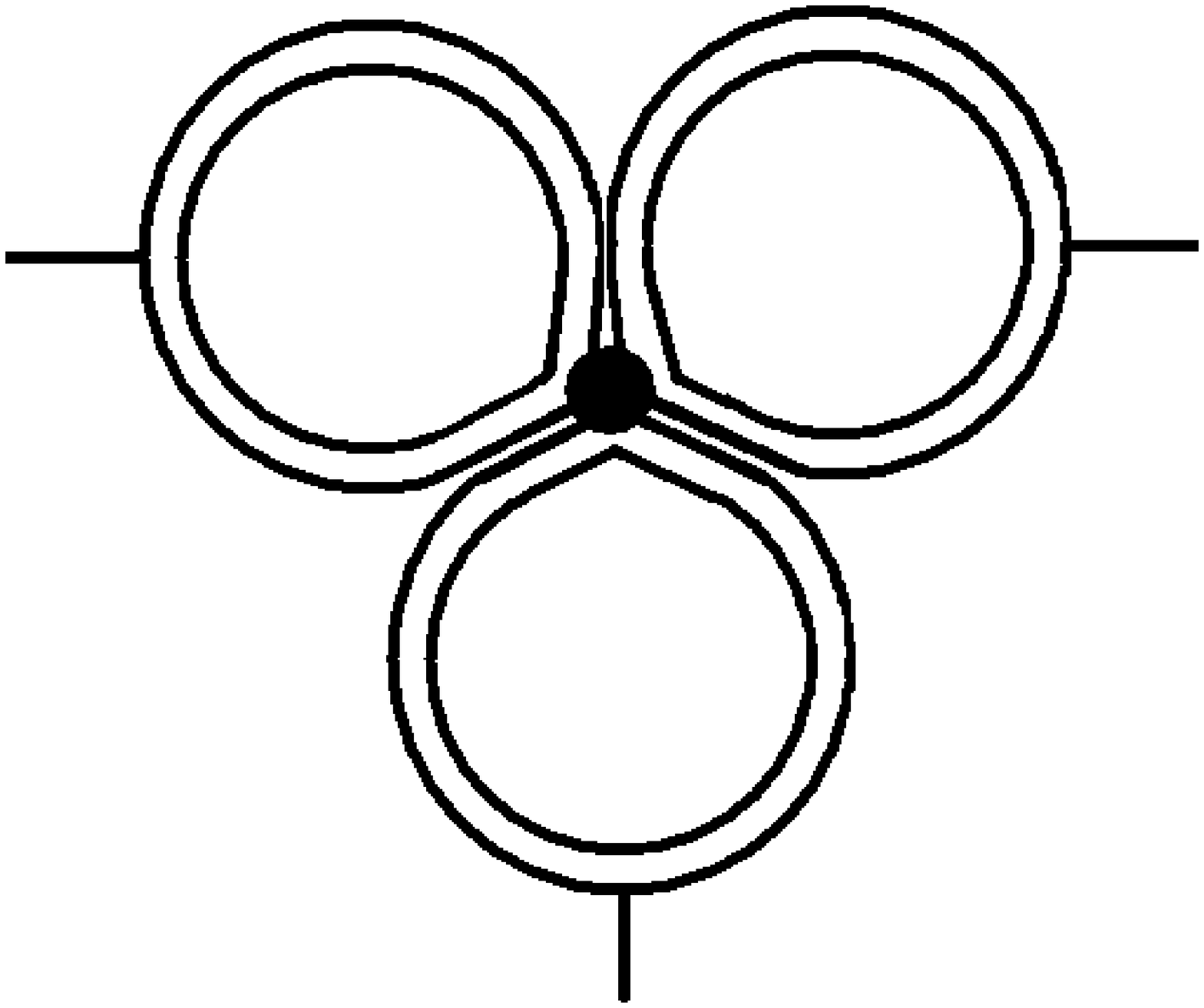}}
\parbox{3cm}{\includegraphics[height=1.3cm]{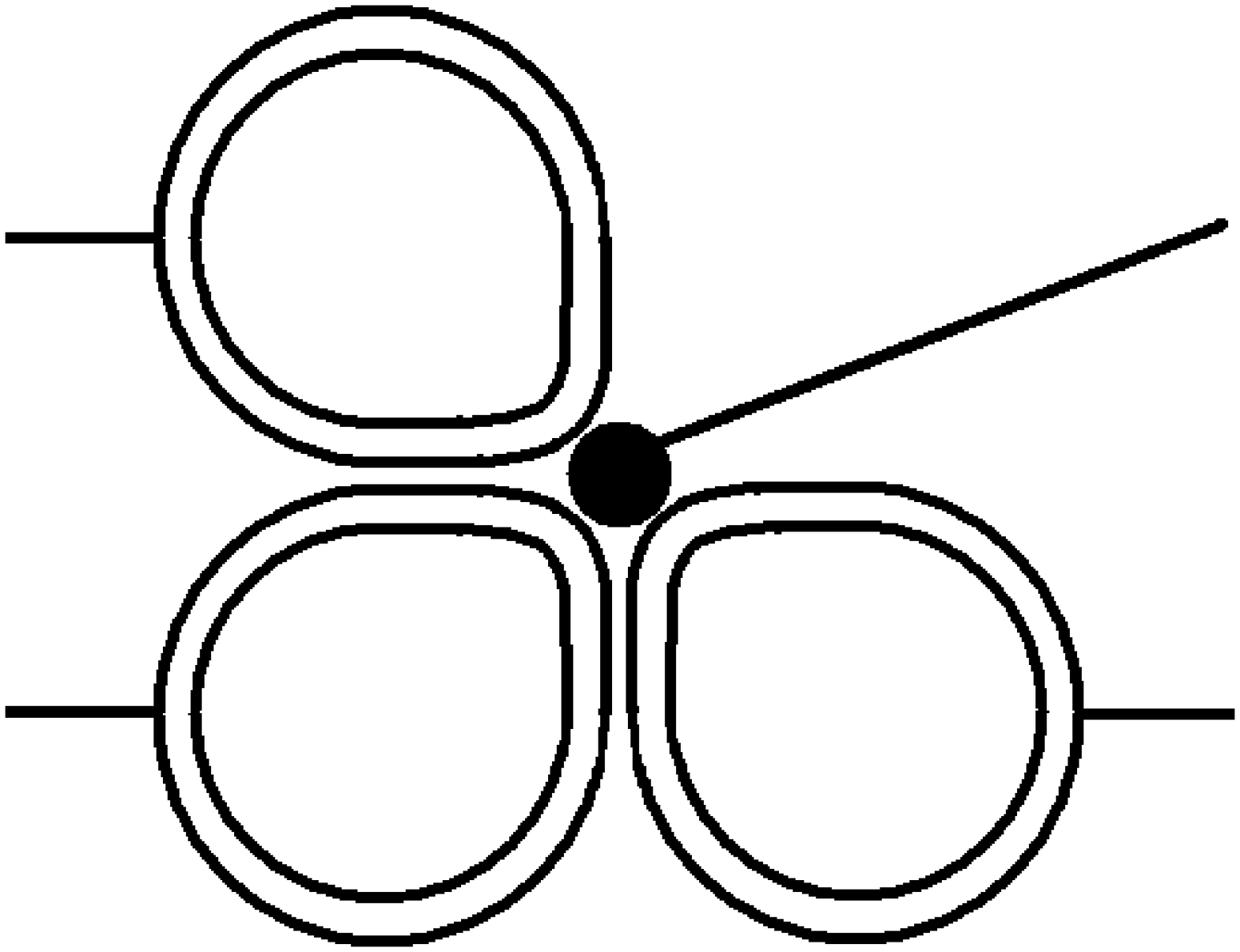}}
\parbox{3cm}{\includegraphics[height=1.3cm]{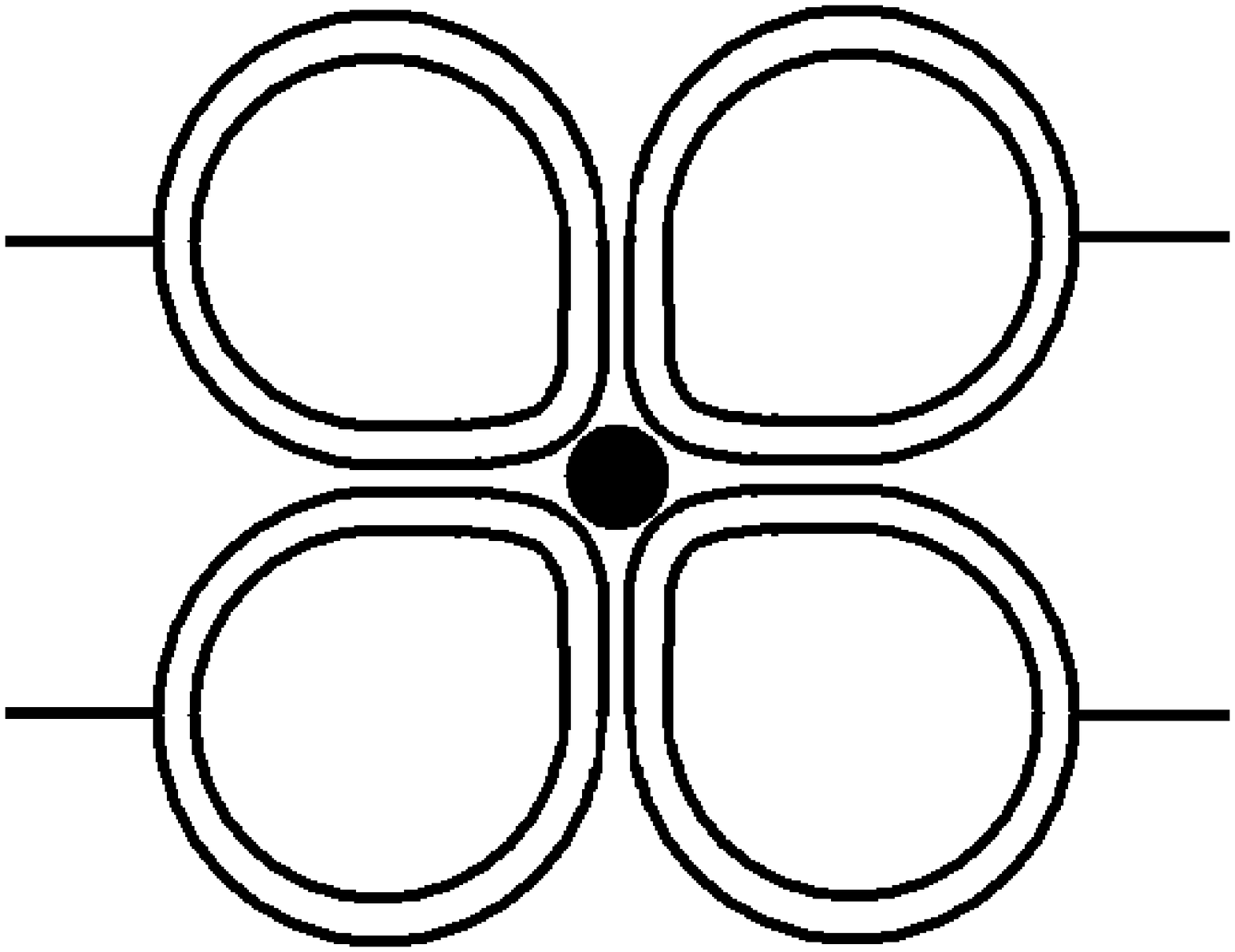}}
\caption{Particular examples of leading vertex-loop diagrams build
from 6-point and 8-point vertex factors
respectively.\label{6-point}}
\end{figure}

Higher order vertices with more external legs can generate
multi-loop contributions of the vertex-loop type, {\it e.g.}, a
6-point vertex can in principle generate a 3-loop contribution to
the 3-point function, a 8-point vertex a 4-loop contribution the
the 4-point function etc (see figure \ref{6-point}).

The 2-point large-$D$ limit will hence consist of the sum of the
propagator and all types of leading 2-point diagrams which one
considered above, {\it i.e.}, bubble diagrams and vertex-loops
diagrams and combinations of the two types of diagrams. Ghost
insertions will be suppressed in the large-$D$ limit, because such
diagrams will not carry two traces, however they will go into the
theory at non-leading order. The general $n$-point function will
consist of diagrams built from separated bubbles or vertex-loop
diagrams. We have devoted appendix \ref{app2} to a study of the
generic $n$-point functions.

\section{The large-$D$ limit and the effective field theory extension of gravity}
So far we have avoided the renormalization
difficulties of Einstein gravity. In order to obtain a effective
renormalizable theory up to the Planck scale we will have to introduce
the effective field theory description.

The most general effective Lagrangian of a $D$-dimensional
gravitational theory has the form:
\begin{equation}
{\cal L} = \int d^Dx \sqrt{-g}\bigg(\Big(\frac{2}{\kappa^2}R+
c_1R^2+c_2R_{\mu\nu}R^{\mu\nu}+\ldots\Big)+{\cal L}_{\rm eff.\
matter}\bigg),
\end{equation}
where ($R^\alpha_{\ \mu\nu\beta}$) is the curvature tensor ($g =
-{\rm det}(g_{\mu\nu})$) and the gravitational coupling is defined
as before, {\it i.e.}, ($\kappa^2 = 32\pi G_D$). The matter
Lagrangian includes in principle everything which couples to a
gravitational field~\cite{matter}, {\it i.e.}, any effective or
higher derivative couplings of gravity to bosonic and or fermionic
matter. In this paper we will however solely look at pure
gravitational interactions. The effective Lagrangian for the
theory is then reduced to invariants built from the Riemann
tensors:
\begin{equation}
{\cal L} = \int d^Dx \sqrt{-g}\Big(\frac{2R}{\kappa^2}+c_1R^2+
c_2R_{\mu\nu}R^{\mu\nu}+\ldots\Big).
\end{equation}
In an effective field theory higher derivative couplings of the
fields are allowed for, while the underlying physical symmetries
of the theory are kept manifestly intact. In gravity the general
action of the theory has to be covariant with respect to the
external gravitational fields.

The renormalization problems of traditional Einstein gravity are
trivially solved by the effective field theory approach. The
effective treatment of gravity includes all possible invariants
and hence all divergences occurring in the loop diagrams can be
absorbed in a renormalized effective action.

The effective expansion of the theory will be an expansion of the
Lagrangian in powers of momentum, and the minimal powers of
momentum will dominate the effective field theory at low energy
scales. In gravity this means that the ($R \sim ({\rm
momentum})^2$) term will be dominant at normal energies, {\it
i.e.}, gravity as an effective field theory at normal energies
will essentially still be general relativity. At higher energy
scales the higher derivative terms ($R^2 \sim ({\rm momentum})^4)
,\ \ldots, (\ R^3\sim ({\rm momentum})^6) \ldots$, corresponding
to higher powers of momentum will mix in and become increasingly
important.

The large-$D$ limit in Einstein gravity is arrived at by expanding
the theory in ($D$) and subsequently taking ($D\rightarrow
\infty$). This can be done in its effective extension too.
Treating gravity as an effective theory and deriving the large-$D$
limit results in a double expansion of the theory, {\it i.e.}, we
expand the theory both in powers of momentum and in powers of
$\left(\frac1D\right)$. In order to understand the large-$D$ limit
in the case of gravity as an effective field theory, we have to
examine the vertex structure for the additional effective field
theory terms. Clearly any vertex contribution from, {\it e.g.},
the effective field theory terms, ($R^2$), and, ($R_{\mu\nu}^2$),
will have four momentum factors and hence new index structures are
possible for the effective vertices.

\begin{figure}[h]
\begin{tabular}{cccccc}\vspace{0.1cm}
$V_3^{\rm eff}=$& $\kappa^3\left(\parbox{1cm}
{\includegraphics[height=0.8cm]{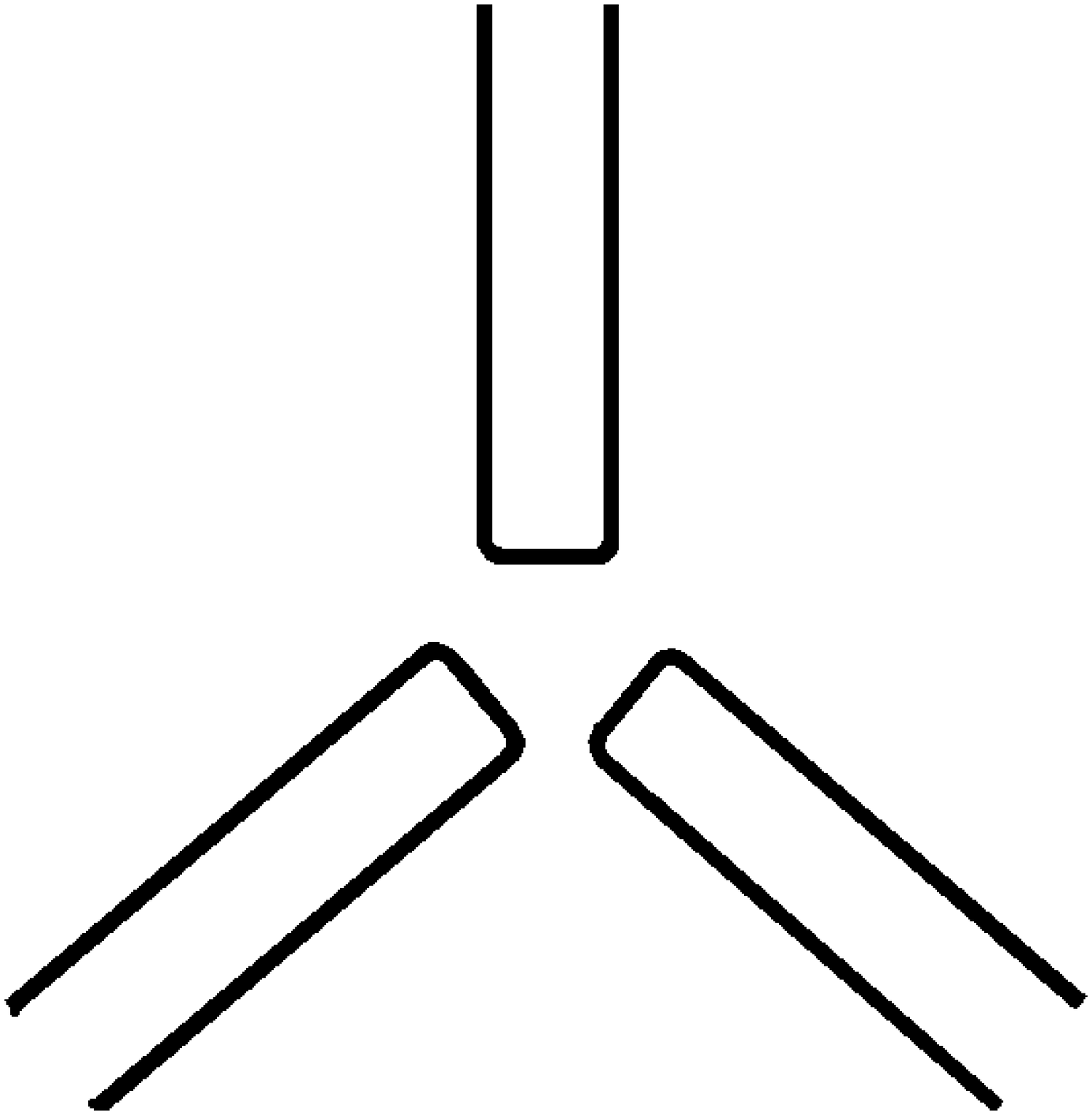}}\right)^{\rm eff}_{\rm
3A}$+&
$\kappa^3\left(\parbox{1cm}{\includegraphics[height=0.8cm]{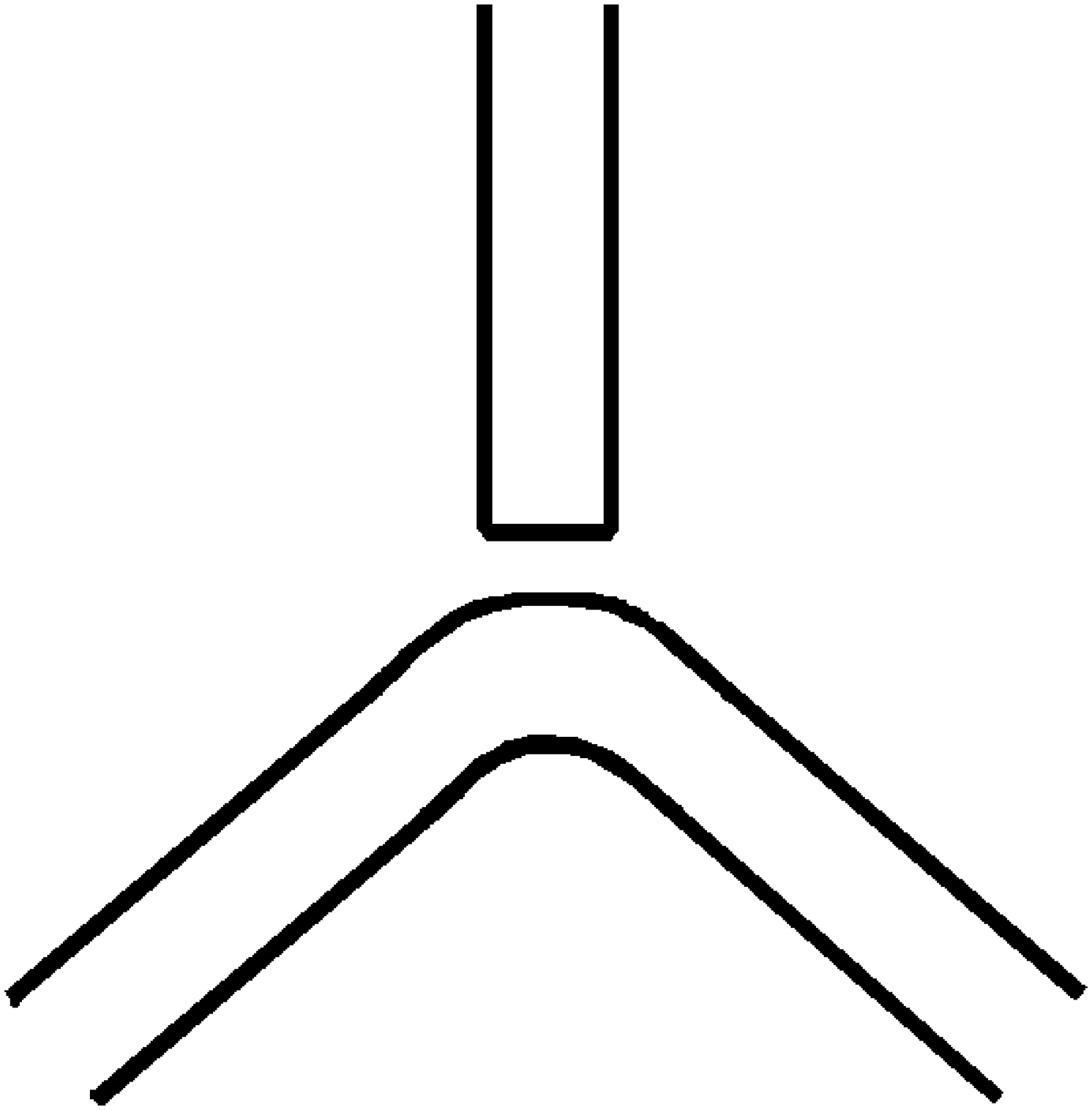}}\right)^{\rm
eff}_{\rm 3B}$+&
$\kappa^3\left(\parbox{1cm}{\includegraphics[height=0.8cm]{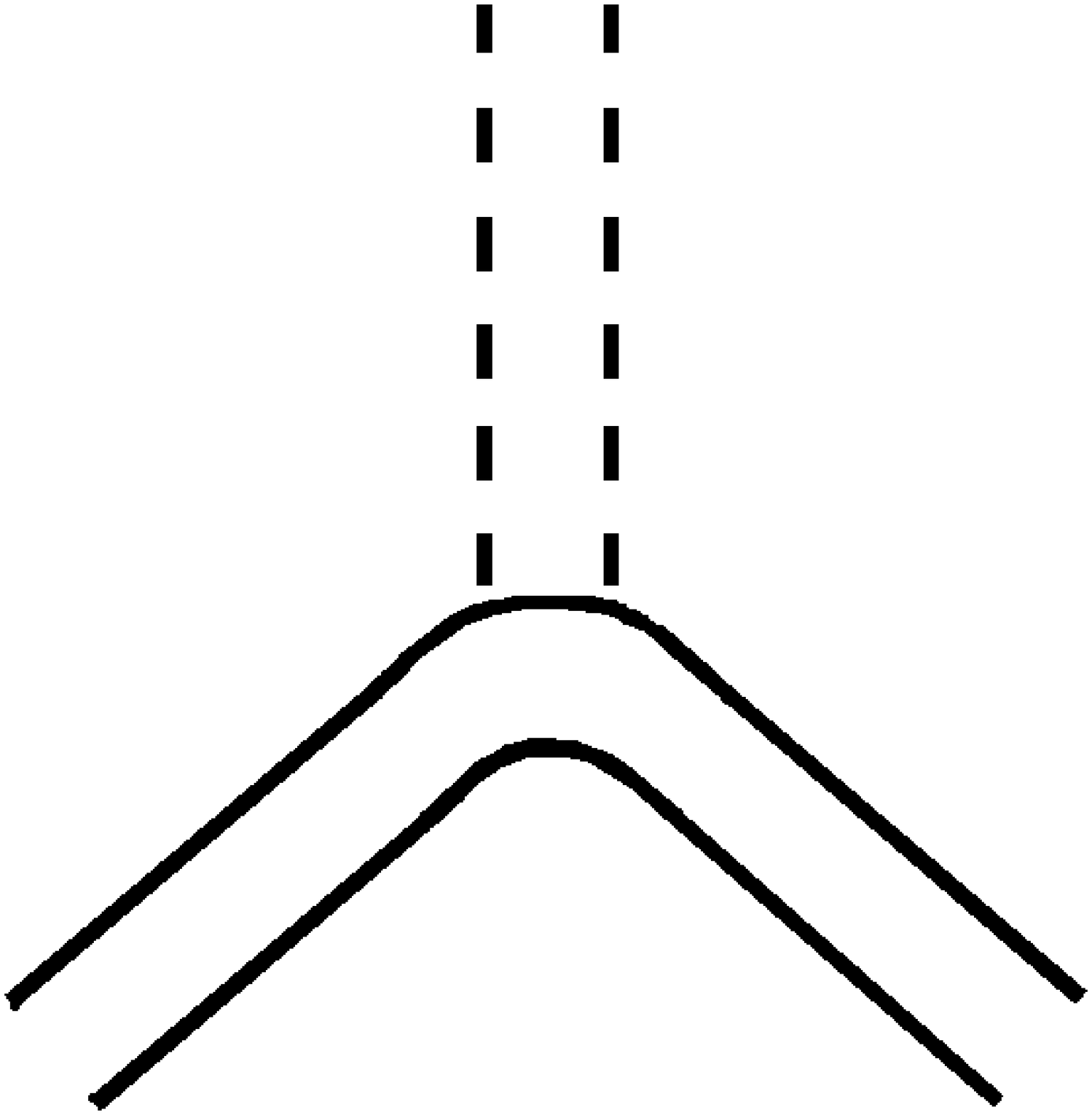}}\right)^{\rm
eff}_{\rm 3C}$+&
$\kappa^3\left(\parbox{1cm}{\includegraphics[height=0.8cm]{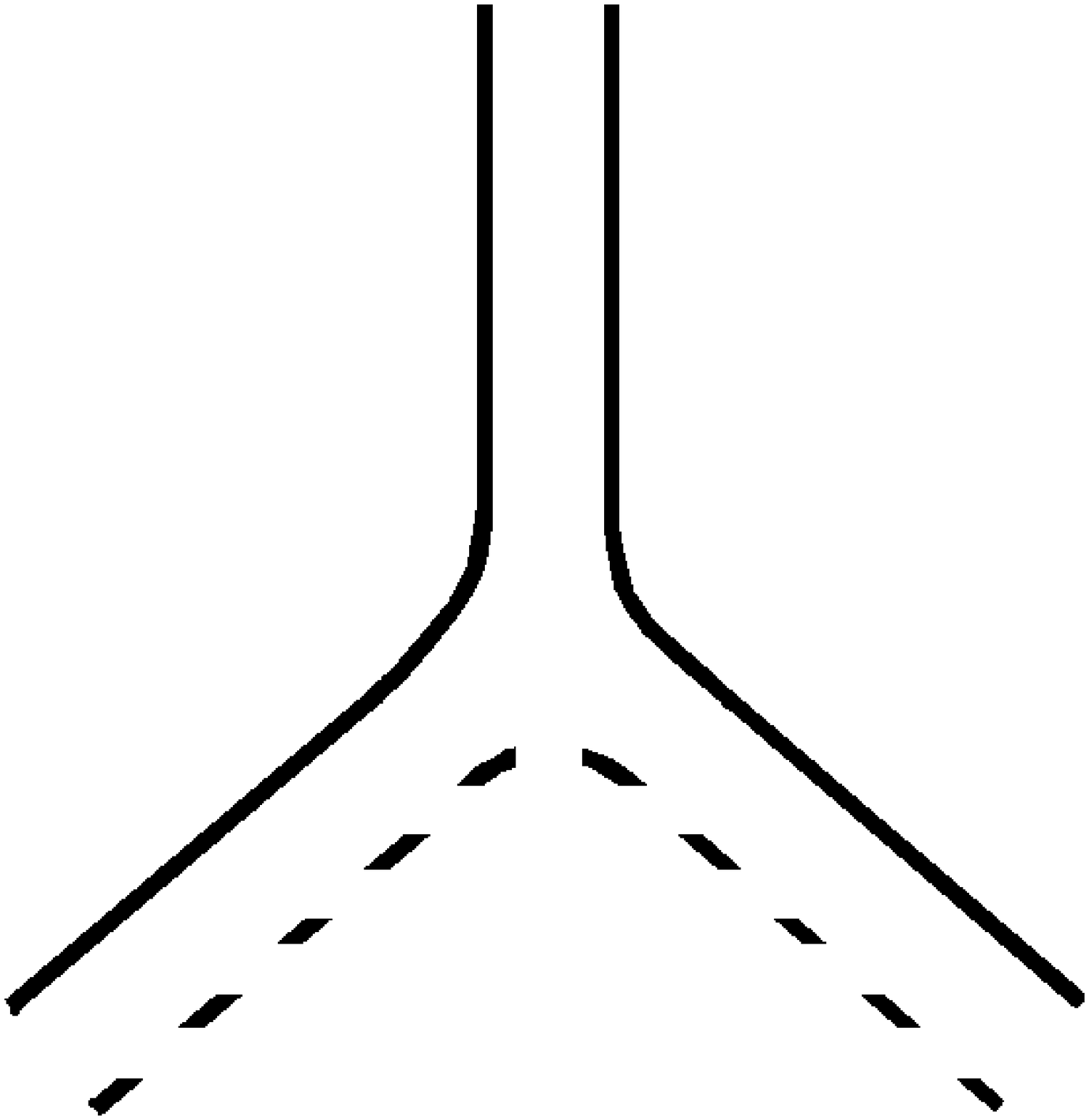}}\right)^{\rm
eff}_{\rm 3D}$+&
$\kappa^3\left(\parbox{1cm}{\includegraphics[height=0.8cm]{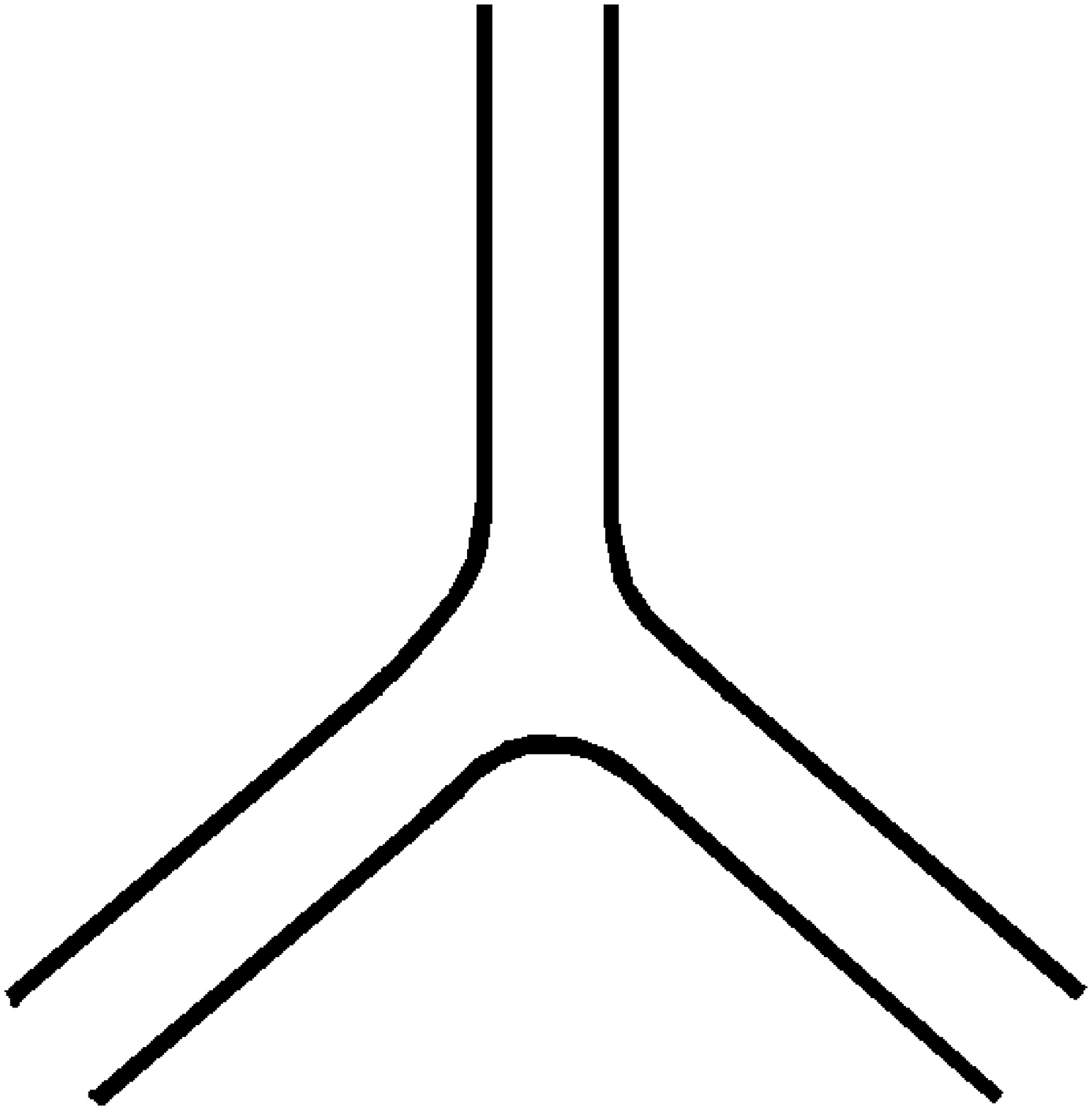}}\right)^{\rm eff}_{\rm 3E}$+\\
&$\kappa^3\left(\parbox{1cm}{\includegraphics[height=0.8cm]{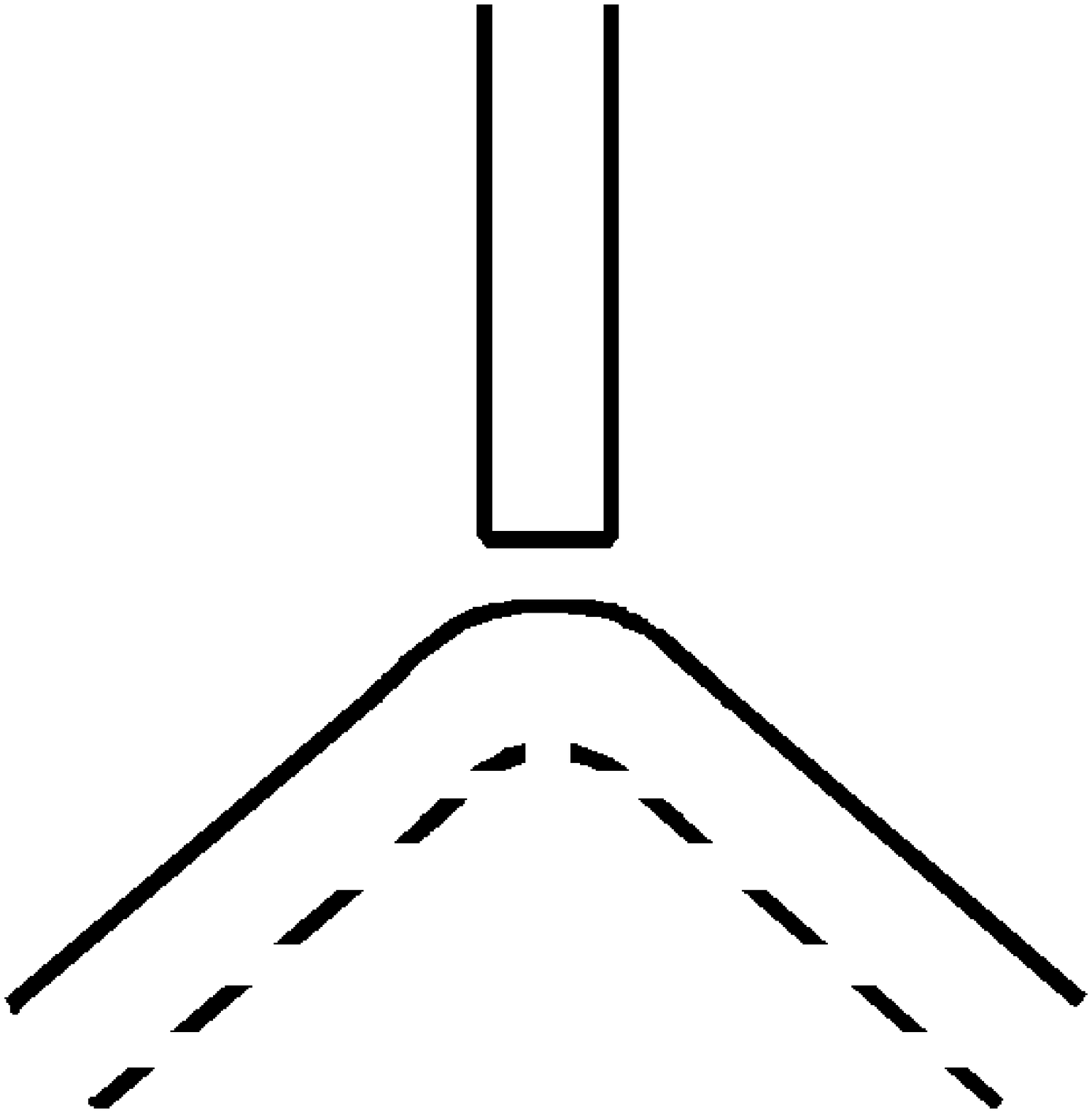}}\right)^{\rm
eff}_{\rm 3F}$+&
$\kappa^3\left(\parbox{1cm}{\includegraphics[height=0.8cm]{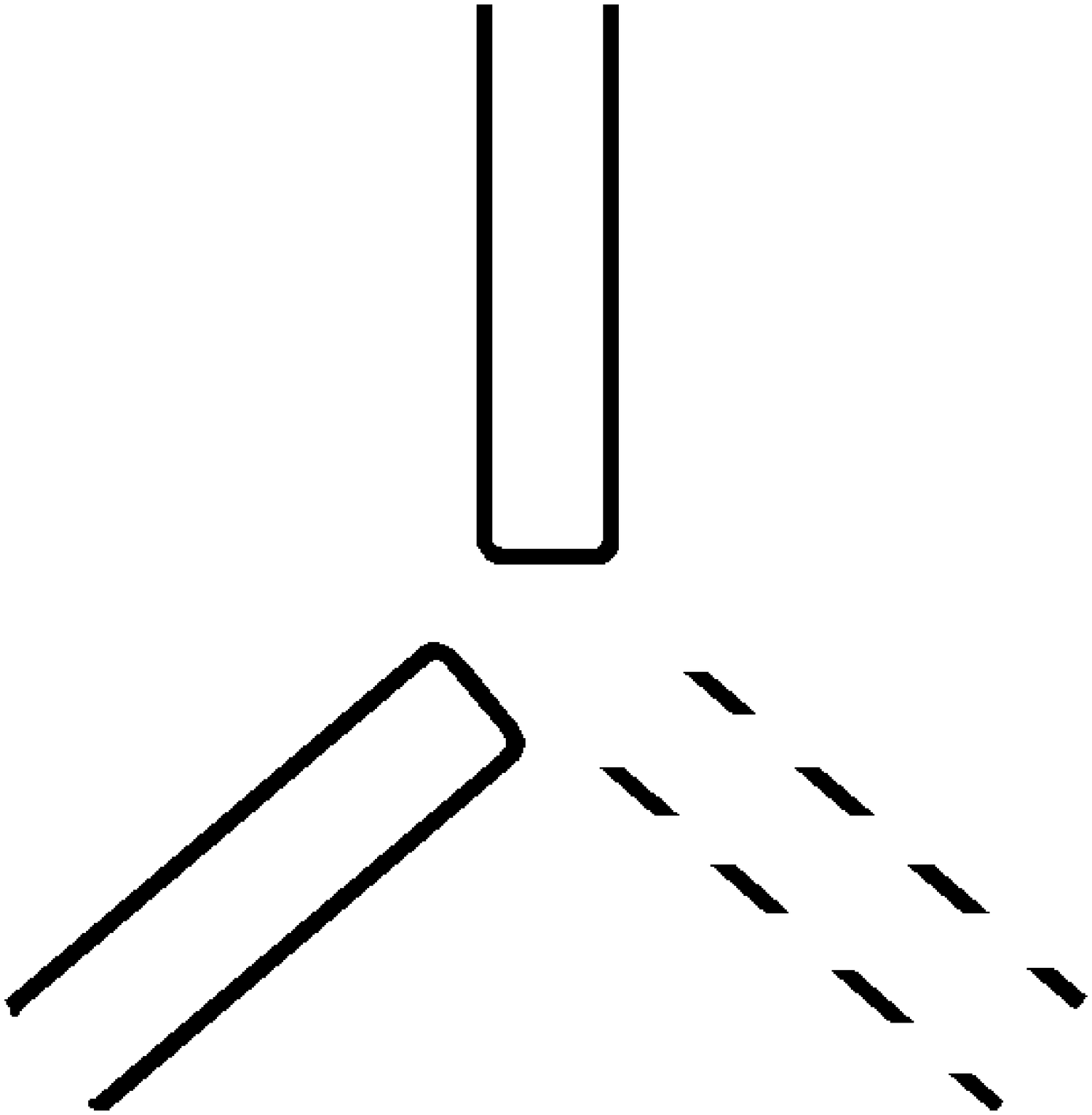}}\right)^{\rm
eff}_{\rm 3G}$+&
$\kappa^3\left(\parbox{1cm}{\includegraphics[height=0.8cm]{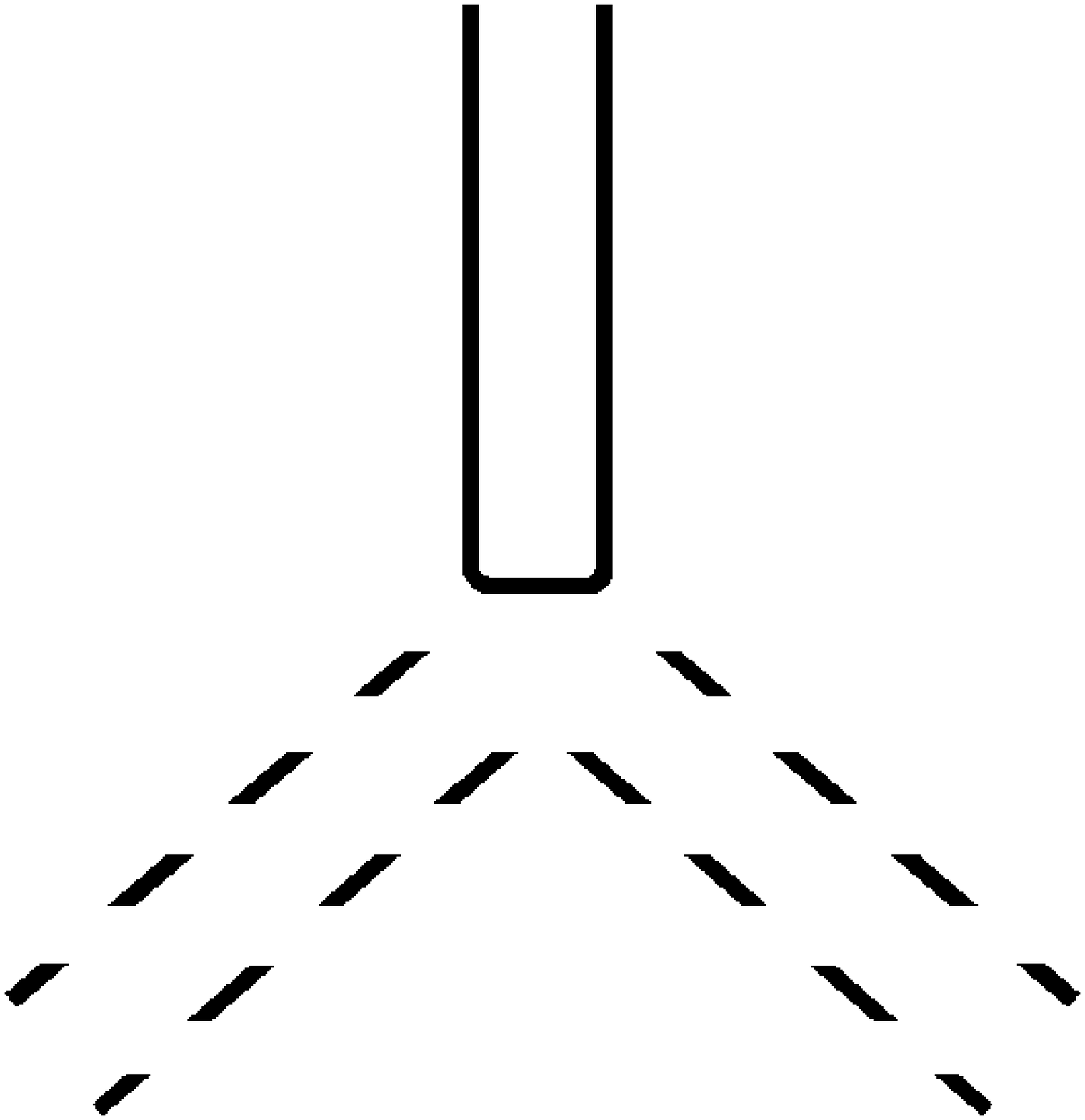}}\right)^{\rm
eff}_{\rm 3H}$+&
$\kappa^3\left(\parbox{1cm}{\includegraphics[height=0.8cm]{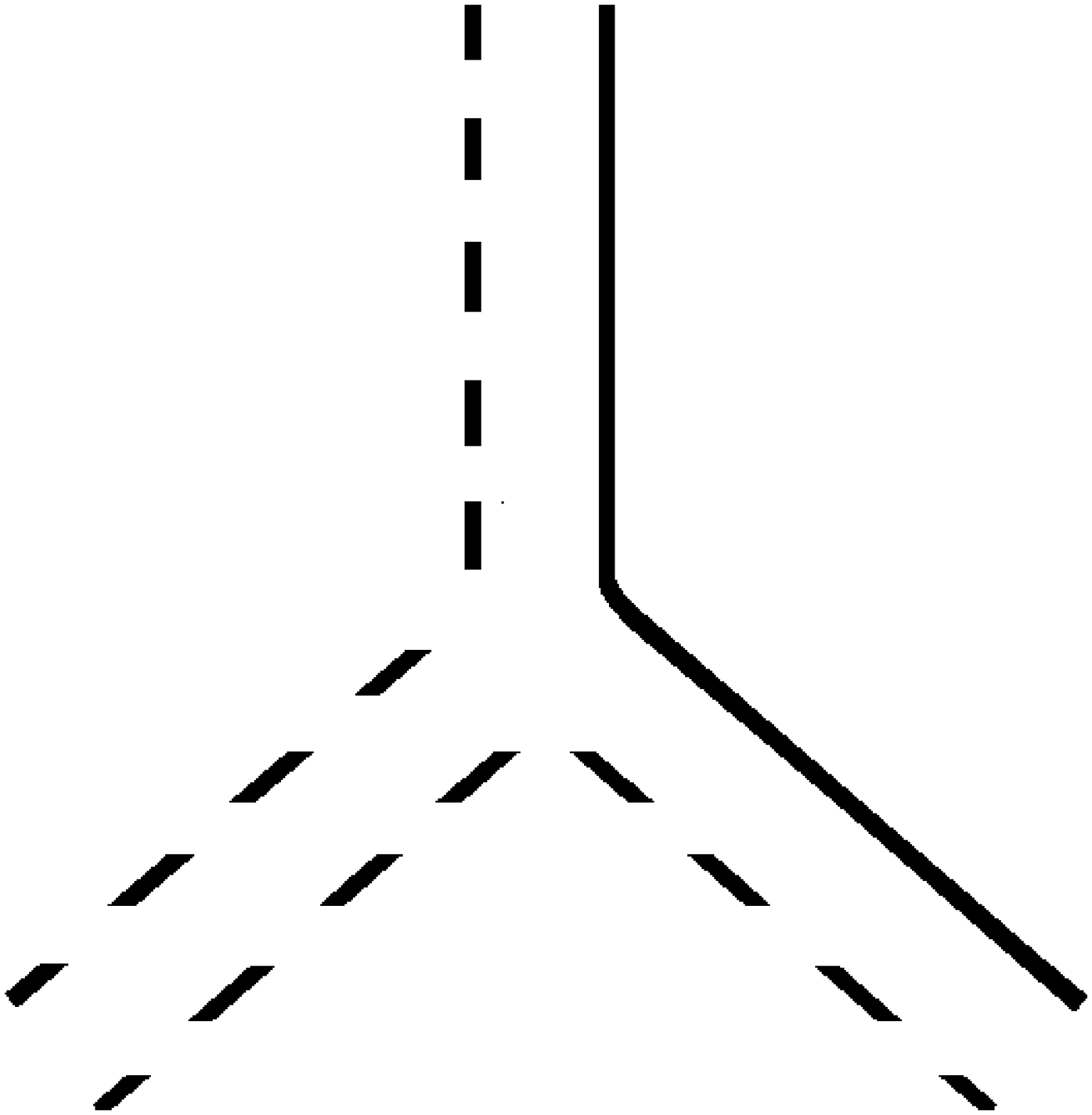}}\right)^{\rm
eff}_{\rm 3I}$.&
\end{tabular}
\caption{A graphical representation of the ($R^2$) and the
($R^2_{\mu\nu}$) index structures in the effective 3-point vertex
factor. A dashed line represents a contraction of a index with a
momentum line. A full line symbolizes a contraction of two index
lines.\label{efft}}
\end{figure}

We find the following vertex structure for the ($R^2$) and the
($R_{\mu\nu}^2$) terms of the effective action (see figure
\ref{efft} and appendix \ref{app1}).

\begin{figure}[h]
\begin{tabular}{ll}\vspace{0.1cm}
$\left(\parbox{1cm}{\includegraphics[height=0.8cm]{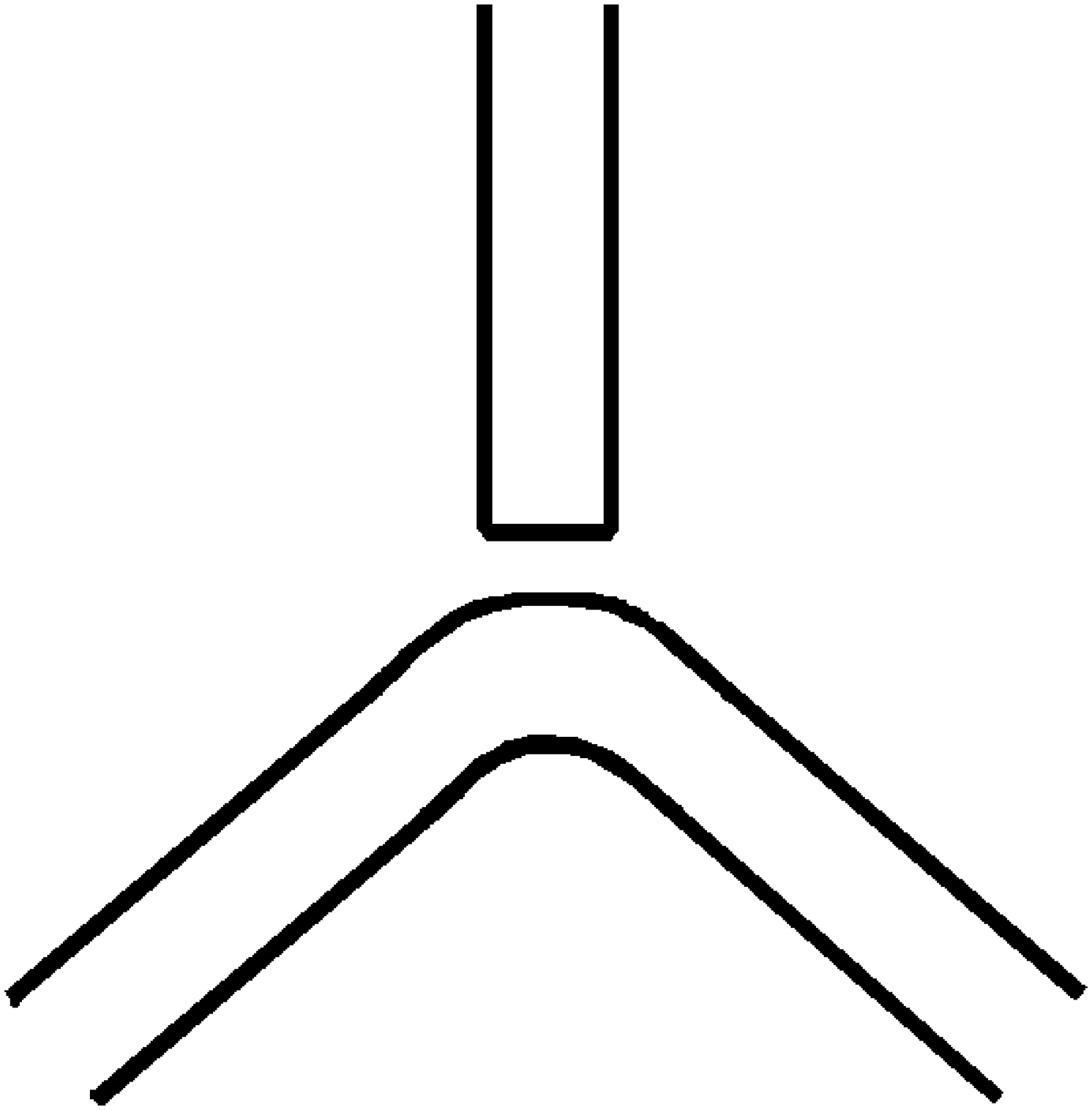}}\right)^{\rm
eff}_{\rm 3B}$ &$\sim {\rm
sym}[-P_3\left(\eta_{\mu\alpha}\eta_{\nu\sigma}\eta_{\beta\gamma}[3c_1k_1^2(k_2\cdot
k_3)+c_2(\frac12(k_1\cdot k_2)(k_1\cdot
k_3)-\frac14k_2^2k_3^2)]\right)$\\ &
$-P_6\left(\eta_{\mu\alpha}\eta_{\nu\sigma}\eta_{\beta\gamma}[2c_1k_1^2k_3^2+c_2(\frac12(k_1\cdot
k_3)^2-\frac14k_2^2(k_1\cdot k_3))]\right)],$
\\ \vspace{0.05cm}
$\left(\parbox{1cm}{\includegraphics[height=0.8cm]{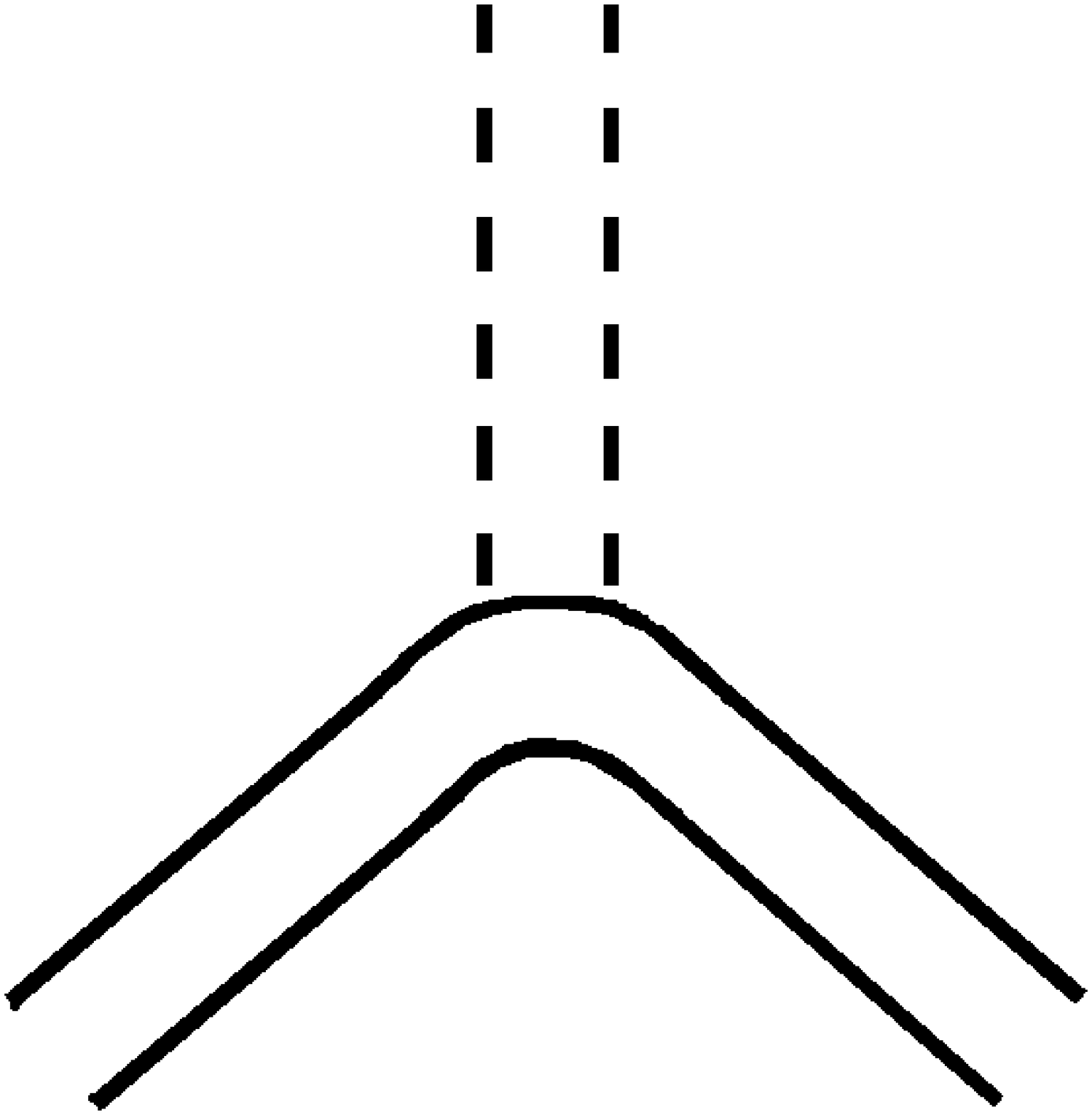}}\right)^{\rm
eff}_{\rm 3C}$ &$\sim {\rm
sym}[-P_3\left(k_{1\mu}k_{1\alpha}\eta_{\nu\sigma}\eta_{\beta\gamma}[3c_1k_2\cdot
k_3]\right)
-2P_6\left(k_{1\mu}k_{1\alpha}\eta_{\nu\sigma}\eta_{\beta\gamma}[c_1
k_3^2]\right)
+\frac12P_6\left(k_{1\mu}k_{2\alpha}\eta_{\nu\sigma}\eta_{\beta\gamma}[c_2k_1\cdot
k_3]\right)
$\\&+$P_6\left(k_{1\mu}k_{3\alpha}\eta_{\nu\sigma}\eta_{\beta\gamma}[c_2k_1\cdot
k_3]\right)
-\frac12P_3\left(k_{2\mu}k_{3\alpha}\eta_{\nu\sigma}\eta_{\beta\gamma}[c_2k_1^2]\right)
-\frac12P_6\left(k_{3\mu}k_{3\alpha}\eta_{\nu\sigma}\eta_{\beta\gamma}[c_2k_1^2]\right)$\\&
$
-\frac12P_6\left(k_{1\beta}k_{3\nu}\eta_{\mu\sigma}\eta_{\alpha\gamma}[c_2k_2^2]\right)
-\frac12P_6\left(k_{3\beta}k_{3\nu}\eta_{\mu\sigma}\eta_{\alpha\gamma}[c_2k_2^2]\right)].$
\end{tabular}
\caption{Explicit index structure terms in the effective vertex
3-point factor, which can contribute with leading contributions in
the large-$D$ limit.\label{3B3C}}
\end{figure}

We are particularly interested in the terms which will give
leading trace contributions. For the (3B)$^{\rm eff}$ and
(3C)$^{\rm eff}$ index structures of we have shown some explicit
results (see figure \ref{3B3C}).

It is seen that also in the effective theory there are terms which
will generate double trace structures. Hence an effective 1-loop
bubble diagram with two traces is possible. Such diagrams will go
as $\left(\frac{\kappa^4}{D^4}c_1\times D^2 \sim
\frac{c_1}{D^2}\right)$. In order to arrive at the same limit for
the effective tree diagrams as the tree diagrams in Einstein
gravity we need to rescale ($c_1$) and ($c_2$) by ($D^2$) i.e.,
($c_1 \rightarrow c_1D^2$) and ($c_2 \rightarrow c_2D^2$). Thereby
every $n$-point function involving the effective vertices will go
into the $n$-point functions of Einstein gravity. In fact this
will hold for any ($c_i$) provided that the index structures
giving double traces also exists for the higher order effective
terms. Other rescalings of the ($c_i$)'s are possible. But such
other rescalings will give a different tree limit for the
effective theory at large-$D$. The effective tree-level graphs
will be thus be scaled away from the Einstein tree graph limit.

The above rescaling with ($c_i \rightarrow c_iD^2$) corresponds to
the maximal rescaling possible. Any rescaling with, {\it e.g.},
($c_i \rightarrow c_iD^n$, $n > 2$) will not be possible, if we
still want to obtain a finite limit for the loop graphs at ($D
\rightarrow \infty$). A rescaling with $(D^2)$ will give maximal
support to the effective terms at large-$D$.

The effective terms in the action are needed in order to
renormalize infinities from the loop diagrams away. The
renormalization of the effective field theory can be carried out
at any particular ($D$). Of course the exact cancellation of pole
terms from the loops will depend on the integrals, and the algebra
will change with the dimension, but one can take this into account
for any particular order of $(D=D_p)$ by an explicit calculation
of the counter terms at dimension ($D_p$) followed by an
adjustment of the coefficients in the effective action. The exact
renormalization will in this way take place order by order in
$\left(\frac1D\right)$. For any rescaling of the $(c_i)$'s it is
thus always possible carry out a renormalization of the theory.
Depending on how the rescaling of the coefficients $(c_i)$ are
done, there will be different large-$D$ limits for the effective
field theory extension of the theory.

The effective field theory treatment does in fact not change the
large-$D$ of gravity as much as one could expect. The effective
enlargement of gravity does not introduce new exciting leading
diagrams. Most parts of our analysis of the large-$D$ limit
depends only on the index structure for the various vertices, and
the new index structures provided by the effective field theory
terms only add new momentum lines, which do not give additional
traces over index loops. Therefore the effective extension of
Einstein gravity is not a very radical change of its large-$D$
limit. In the effective extension of gravity the action is
trivially renormalizable up to a cut-off scale at ($M_{\rm
Planck}$) and the large-$D$ expansion just as well defined as an
large-$N$ expansion of a renormalizable planar expansion of a
Yang-Mills action.

As a subject for further investigations, it might be useful to
employ more general considerations about the index structures.
Seemingly all possible types of index-structures are present in
the effective field theory extension of gravity in the
conventional approach. As only certain index structures go into
the leading large-$D$ limit, the 'interesting' index structures
can be classified at any given loop order. In fact one could
consider loop amplitudes with only 'interesting' index structures
present. Such applications might useful in very complicated
quantum gravity calculations, where only the leading large-$D$
behavior is interesting.

\begin{figure}[h]
\parbox{2cm}{\includegraphics[height=0.5cm]{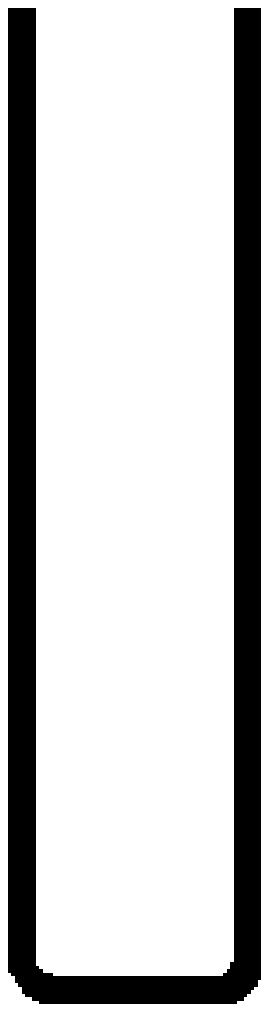}}
\parbox{2cm}{\includegraphics[height=0.5cm]{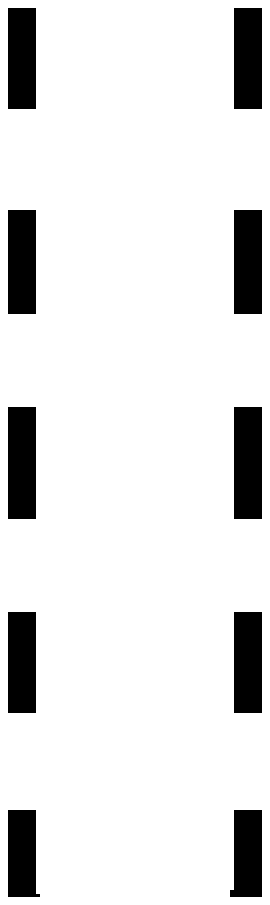}}
\parbox{2cm}{\includegraphics[height=0.5cm]{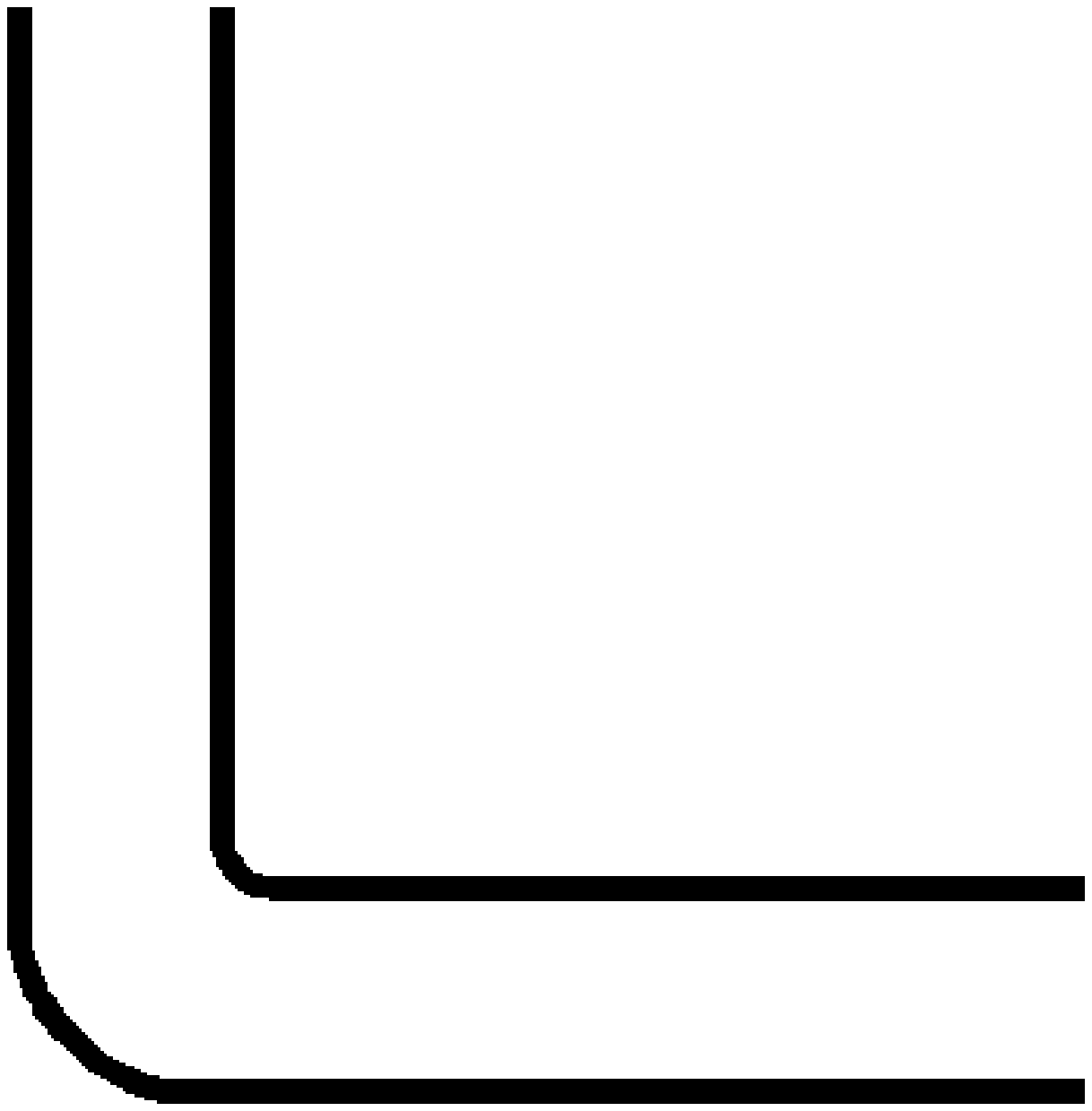}}
\parbox{2cm}{\includegraphics[height=0.5cm]{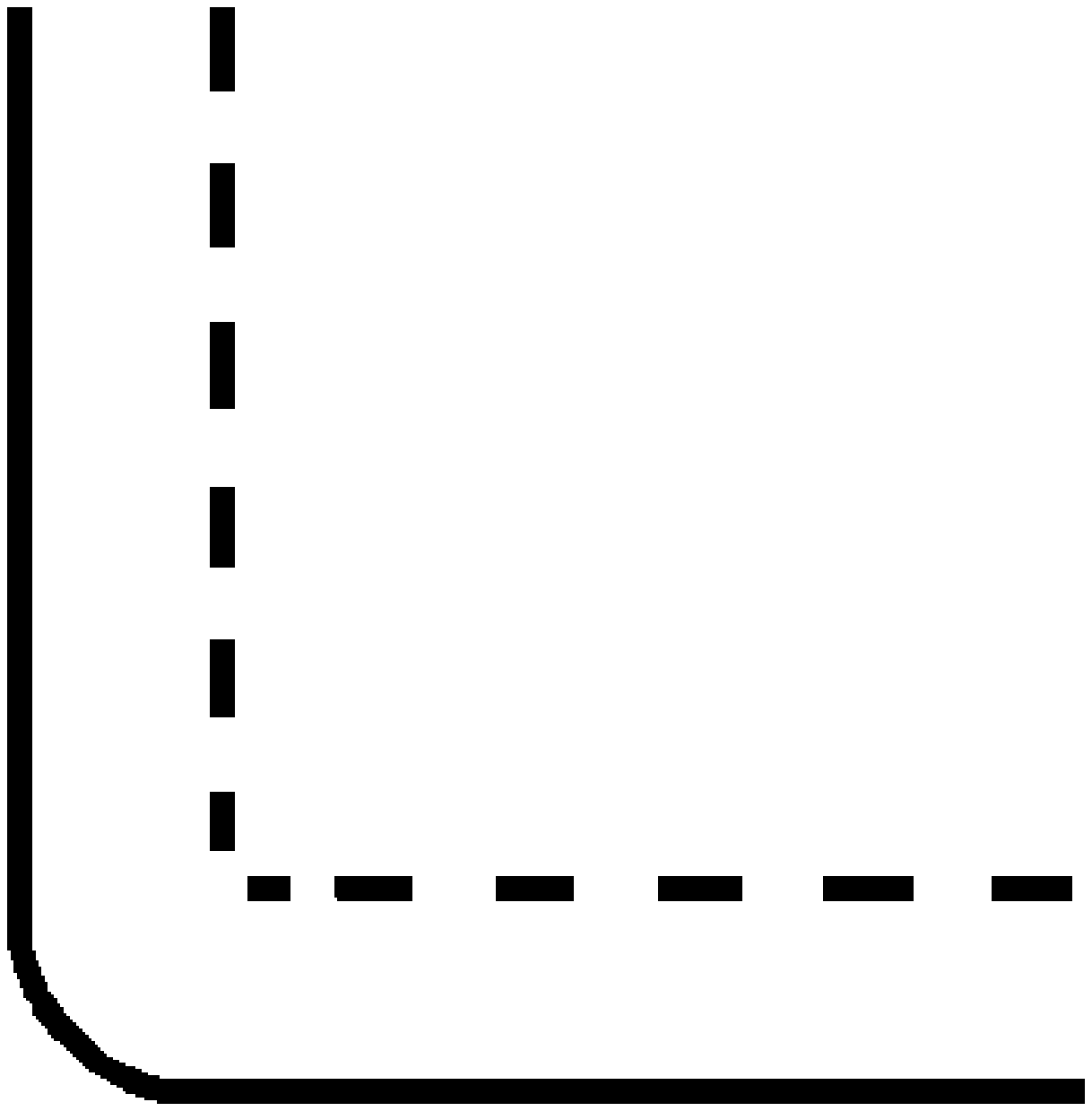}}
\parbox{2cm}{\includegraphics[height=1cm]{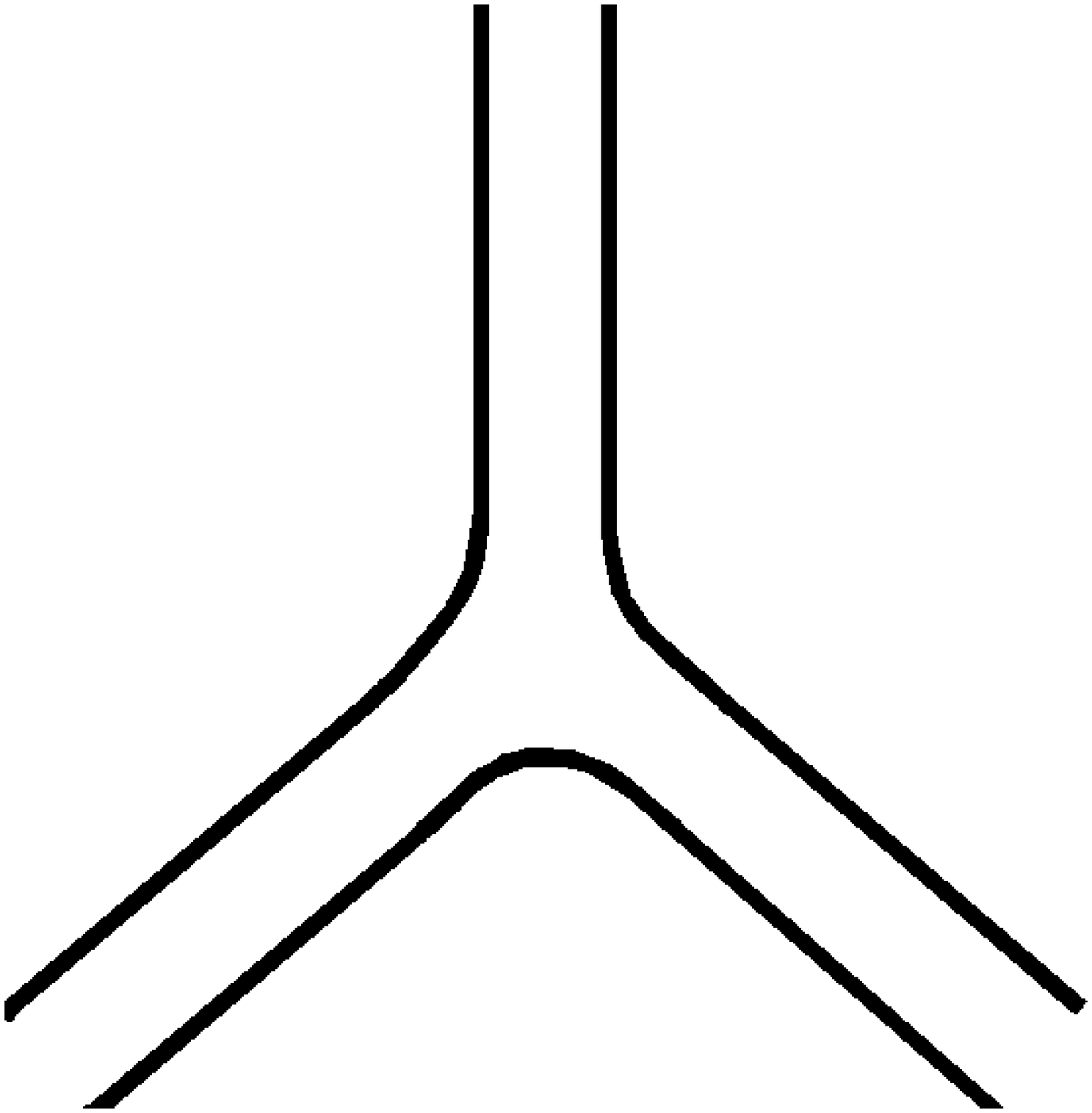}}
\parbox{2cm}{\includegraphics[height=1cm]{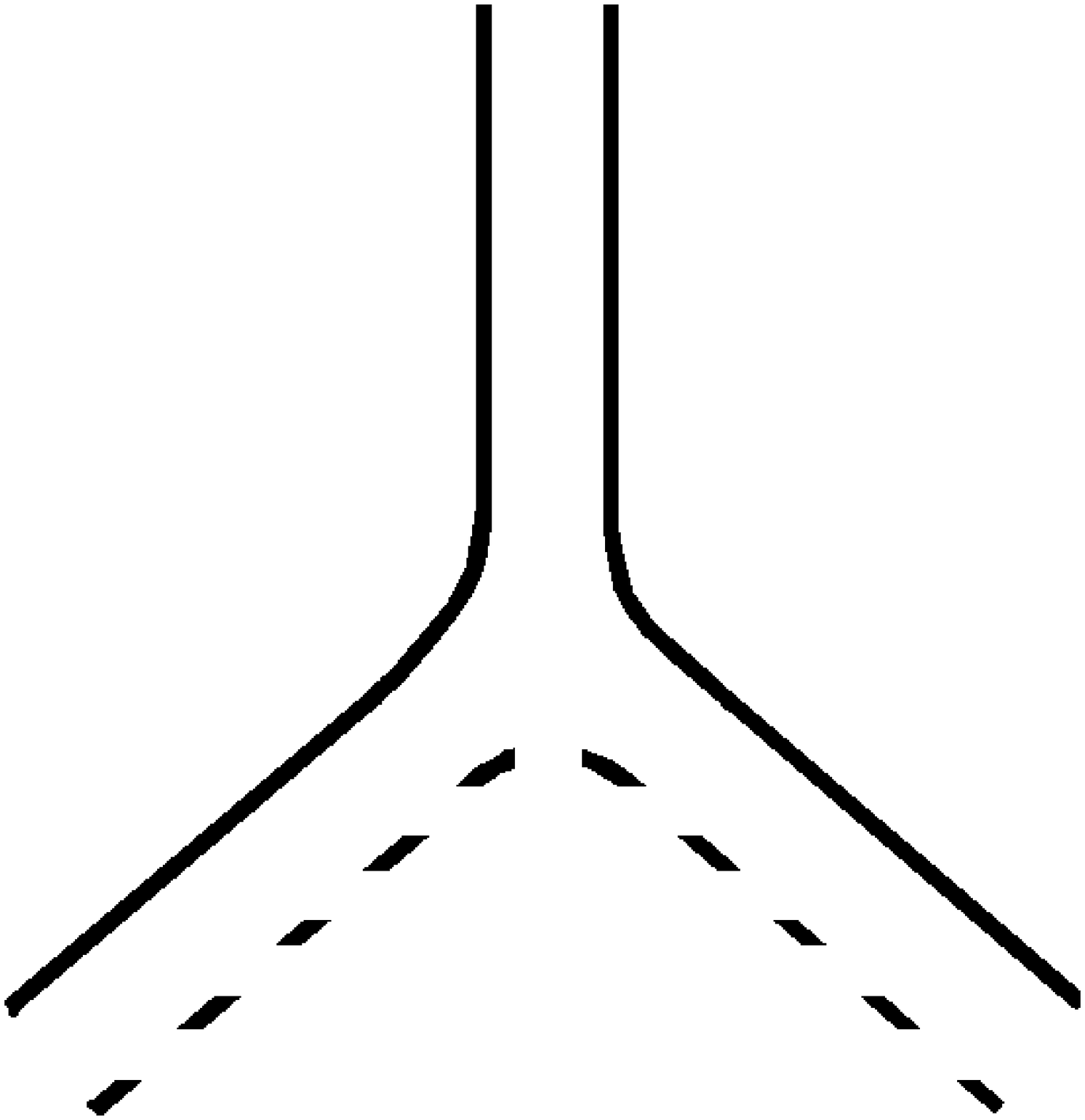}}
\parbox{2cm}{\includegraphics[height=1cm]{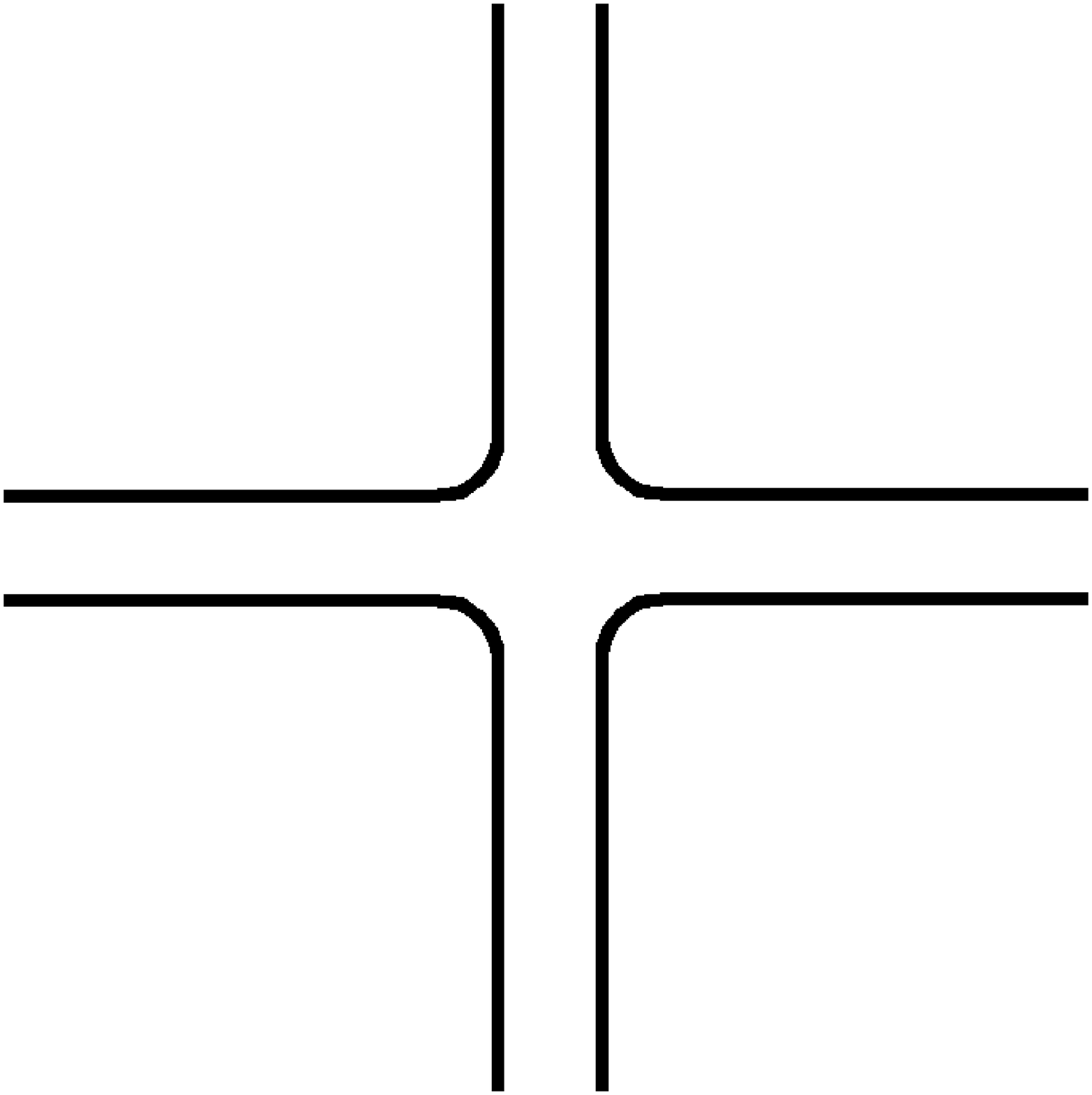}}
\parbox{2cm}{\includegraphics[height=1cm]{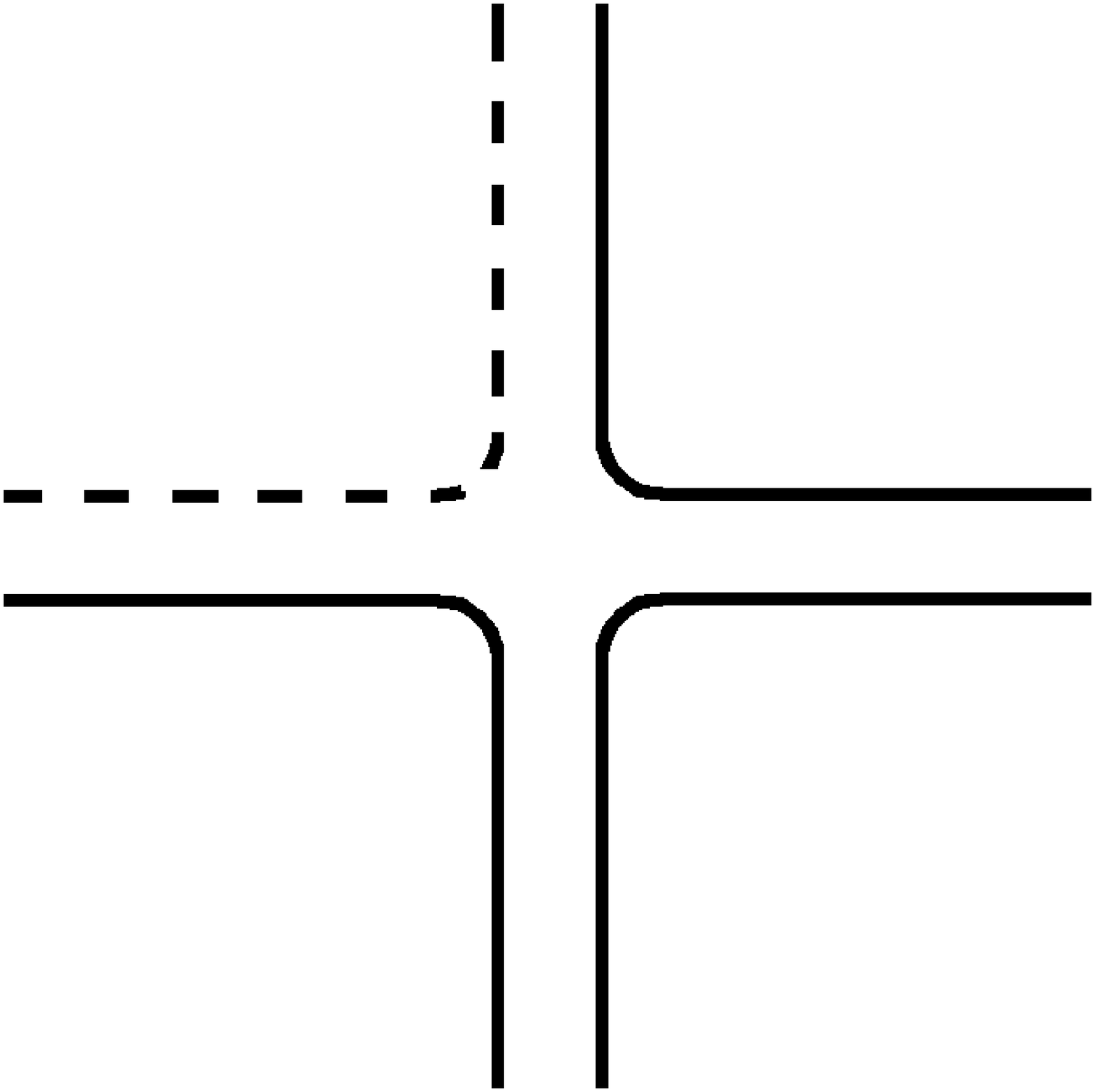}}
\caption{The first 8 basic index structures which builds up any
conventional vertex factor.\label{indexs}}
\end{figure}

A comparison of the index structures present in other field
expansions and the conventional expansion might also be a working
area for further investigations. Every index structure in the
conventional expansion of the field can be build up from the
following types of index structures, see figure \ref{indexs}.

Thus, it is not possible to consider any index structures in a 3-
or 4-point vertex, which cannot be decomposed into some product of
the above types of index contractions. It might be possible to
extend this to something useful in the analysis of the large-$D$
limit, {\it e.g.}, a simpler description of the large-$D$
diagrammatic truncation of the theory. This is another interesting
working area for further research.

\section{Considerations about the space-time integrals}
The space-time integrals pose certain fundamental restrictions to
our analysis of the large-$D$ limit in gravity. So far we have
considered only the algebraic trace structures of the $n$-point
diagrams. This analysis lead us to a consistent large-$D$ limit of
gravity, where we saw that only bubble and vertex-loop diagrams
will carry the leading contributions in $\left(\frac1D\right)$. In
a large-$N$ limit of a gauge theory this would have been adequate,
however the space-time dimension is much more that just a symmetry
index for the gauge group, and has a deeper significance for the
physical theory than just that of a symmetry index. Space-time
integrations in a $D$-dimensional world will contribute extra
dimensional dependencies to the graphs, and it thus has to be
investigated if the $D$-dimensional integrals could upset the
algebraic large-$D$ limit. The issue about the $D$-dimensional
integrals is not resolved in ref.~\cite{Strominger:1981jg}, only
explicit examples are discussed and some conjectures about the
contributions from certain integrals in the graphs at large-$D$
are stated. It is clear that in the case of the 1-loop separated
integrals the dimensional dependence will go as $\left(\sim
\Gamma(\frac{D-1}{2})\right)$. Such a dimensional dependence can
always be rescaled into an additional redefinition of $(\kappa)$,
{\it i.e.}, we can redefine $\left(\kappa\rightarrow
\frac{\kappa}{\sqrt{C}}\right)$ where $\left(C \sim
\Gamma(\frac{D-1}{2})\right)$. The dimensional dependence from
both bubble graphs and vertex-loop graphs will be possible to
scale away in this manner. Graphs which have a nested structure,
{\it i.e.}, which have shared propagator lines, will be however be
suppressed in $(D)$ compared to the separated loop diagrams. Thus
the rescaling of $(\kappa)$ will make the dimensional dependence
of such diagrams even worse.

In the lack of rigorously proven mathematical statements, it is
hard to be completely certain. But it is seen, that in the
large-$D$ limit the dimensional dependencies from the integrations
in $D$-dimensions will favor the same graphs as the algebraic
graphs limit.

The effective extension of the gravitational action does not pose
any problems in these considerations, concerning the dimensional
dependencies of the integrals and the extra rescalings of
$(\kappa)$. The conjectures made for the Einstein-Hilbert types of
leading loop diagrams hold equally well in the effective field
theory case. A rescaling of the effective coupling constants have
of course to be considered in light of the additional rescaling of
($\kappa$). We will have ($c_i\rightarrow Cc_i$) in this case.
Nested loop diagrams, which algebraically will be down in
$\left(\frac1D\right)$ compared to the leading graphs, will also
be down in $\left(\frac1D\right)$ in the case of an effective
field theory.

To directly resolve the complications posed by the space-time
integrals, in principle all classes of integrals going into the
$n$-point functions of gravity have to be investigated. Such an
investigation would indeed be a very ambitious task and it is
outside the scope of this investigation. However, as a point for
further investigations, this would be an interesting place to
begin a more rigorous mathematical justification of the large-$D$
limit.

To avoid any complications with extra factors of ($D$) arising
from the evaluation of the integrals, one possibility is to treat
the extra dimensions in the theory as compactified Kaluza-Klein
space-time dimensions. The extra dimensions are then only excited
above the Planck scale. The space-time integration will then only
have to be preformed over a finite number of space-time
dimensions, {\it e.g.}, the traditional first four. In such a
theory the large-$D$ behavior is solely determined by the
algebraic traces in the graphs and the integrals cannot not
contribute with any additional factors of ($D$) to the graphs.
Having a cutoff of the theory at the Planck scale is fully
consistent with effective treatment of the gravitational action.
The effective field theory description is anyway only valid up
until the Planck scale, so an effective action having a Planckian
Kaluza-Klein cutoff of the momentum integrations, is a
possibility.

\section{Comparison of the large-$N$ limit and the large-$D$ limit}
The large-$N$ limit of a gauge theory and the large-$D$ limit in
gravity has some similarities and some dissimilarities which we
will look upon here.

The large-$N$ limit in a gauge theory is the limit where the
symmetry indices of the gauge theory can go to infinity. The idea
behind the gravity large-$D$ limit is to do the same, where in
this case ($D$) will play the role of ($N$), we thus treat the
spatial dimension ($D$) as if it is only a symmetry index for the
theory, {\it i.e.}, a physical parameter, in which we are allowed
to make an expansion.

\begin{figure}[h]
\includegraphics[height=4cm]{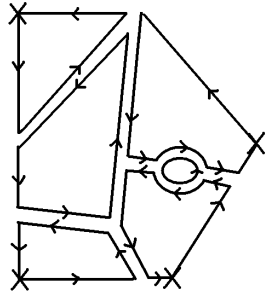}
\caption{A particular planar diagram in the large-$N$ limit. A
$(\times)$ symbolizes an external source, the full lines are index
lines, an full internal index loop gives a factor of
($N$).\label{planar}}
\end{figure}

As it was shown by 't Hooft in ref.~\cite{tHooft} the large-$N$
limit in a gauge theory will be a planar diagram limit, consisting
of all diagrams which can be pictured in a plane. The leading
diagrams will be of the type depicted in figure \ref{planar}.

The arguments for this diagrammatic limit follow similar arguments
to those we applied in the large-$D$ limit of gravity. However the
index loops in the gauge theory are index loops of the internal
symmetry group, and for each type of vertex there will only be one
index structure. Below we show the particular index structure for
a 3-point vertex in a gauge theory.
\begin{figure}[h]
\includegraphics[height=2.5cm]{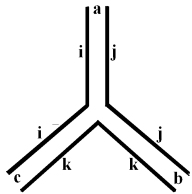}
\caption{The index structure for the 3-point in vertex in a gauge theory.}
\end{figure}

This index structure is also present in the gravity case, {\it
i.e.}, in the (3E) index structure, but various other index
structure are present too, {\it e.g.}, (3A), (3B), $\ldots$, and
each index structure will generically give dissimilar traces, {\it
i.e.}, dissimilar factors of ($D$) for the loops.

In gravity only certain diagrams in the large-$D$ limit carry
leading contributions, but only particular parts of the graph's
amplitudes will survive when ($D\rightarrow\infty$) because of the
different index structures in the vertex factor. The gravity
large-$D$ limit is hence more like a truncated graph limit, where
only some parts of the graph amplitude will be important, {\it
i.e.}, every leading graph will also carry contributions which are
non-leading.

In the Yang-Mills large-$N$ limit the planar graph's amplitudes
will be equally important as a whole. No truncation of the graph's
amplitudes occur. Thus, in a Yang-Mills theory the large-$N$
amplitude is found by summing the full set of planar graphs. This
is a major dissimilarity between the two expansions

The leading diagram limits in the two expansions are not identical
either. The leading graphs in gravity at large-$D$ are given by
the diagrams which have a possibility for having a closed double
trace structure, hence the bubble and vertex-loop diagrams are
favored in the large-$D$ limit. The diagrams contributing to the
gravity large-$D$ limit will thus only be a subset of the full
planar diagram limit.

\section{Other expansions of the fields}
In gravity it is a well-known fact that different definitions of
the fields will lead to different results for otherwise similar
diagrams. Only the full amplitudes will be gauge invariant and
identical for different expansions of the gravitational field.
Therefore comparing the large-$D$ diagrammatical expansion for
dissimilar definitions of the gravitational field, diagram for
diagram, will have no meaning. The analysis of the large-$D$ limit
carried out in this paper is based on the conventional definition
of the gravitational field, and the large-$D$ diagrammatic
expansion should hence be discussed in this light.

Other field choices may be considered too, and their diagrammatic
limits should be investigated as well. It may be worth some
efforts and additional investigations to see, if the diagrammatic
large-$D$ limit considered here is an unique limit for every
definition of the field, {\it i.e.}, to see if the same
diagrammatic limit always will occur for any definition of the
gravitational field. A scenario for further investigations of
this, could e.g., be to look upon if the field expansions can be
dressed in such a way that we reach a less complicated
diagrammatic expansion at large-$D$, at the cost of having a more
complicated expansion of the Lagrangian. For example, it could be
that vertex-loop corrections are suppressed for some definitions
of the gravitational field, but that more complicated trace
structures and cancellations between diagrams occurs in the
calculations. In the case of the Goldberg definition of the
gravitational field the vertex index structure is less
complicated~\cite{Strominger:1981jg}, however the trace
contractions of diagrams appears to much more complicated than in
the conventional method. It should be investigated more carefully
if the Goldberg definition of the field may be easier to employ in
calculations. Indeed the diagrammatical limit considered
by~\cite{Strominger:1981jg} is easier, but it is not clear if
results from two different expansions of the gravitational field
are used to derive this. Further investigations should sheet more
light on the large-$D$ limit, and on how a field redefinition may
or may not change its large-$D$ diagrammatic expansion.

Furthermore one can consider the background field method in the
context of the large-$D$ limit as well. This is another working
area for further investigations. The background field method is
very efficient in complicated calculations and it may be useful to
employ it in the large-$D$ analysis of gravity as well. No
problems in using background field methods together with a
large-$D$ quantum gravity expansion seems to be present.

\section{Discussion}
In this paper we have discussed the large-$D$ limit of effective
quantum gravity. In the large-$D$ limit, a particular subset of
planar diagrams will carry all leading $\left(\frac1D\right)$
contributions to the $n$-point functions. The large-$D$ limit for
any given $n$-point function will consist of the full tree
$n$-point amplitude, together with a set of loop corrections which
will consist of the leading bubble and vertex-loop graphs we have
considered. The effective treatment of the gravitational action
insures that a renormalization of the action is possible and that
none of the $n$-point renormalized amplitudes will carry
uncancelled divergent pole terms. An effective renormalization of
the theory can be preformed at any particular dimensionality.

The leading $\left(\frac1D\right)$ contributions to the theory
will be algebraically less complex than the graphs in the full
amplitude. Calculating only the $\left(\frac1D\right)$ leading
contributions simplifies explicit calculations of graphs in the
large-$D$ limit. The large-$D$ limit we have found is completely
well defined as long as we do not extend the space-time
integrations to the full $D$-dimensional space-time. That is, we
will always have a consistent large-$D$ theory as long as the
integrations only gain support in a finite dimensional space-time
and the remaining extra dimensions in the space-time is left as
compactified, {\it e.g.}, below the Planck scale.

We hence have a renormalizable, definite and consistent large-$D$
limit for effective quantum gravity -- good below the Planck
scale!

No solution to the problem of extending the space-time integrals
to a full $D$-dimensional space-time has been presented in this
paper, however it is clear that the effective treatment of gravity
do not impose any additional problems in such investigations. The
considerations about the $D$-dimensional integrals discussed in
ref.~\cite{Strominger:1981jg} hold equally well in a effective
field theory, however as well as we will have to make an
additional rescaling of ($\kappa$) in order to account for the
extra dimensional dependence coming from the $D$-dimensional
integrals, we will have make an extra rescaling of the
coefficients ($c_i$) in the effective Lagrangian too.

Our large-$D$ graph limit is not in agreement with the large-$D$
limit of ref.~\cite{Strominger:1981jg}, the bubble graph are
present in both limits, but vertex-loops are not allowed and
claimed to be down in $\left(\frac1D\right)$ the latter. We do not
agree on this point, and believe that this might be a consequence
of comparing similar diagrams with different field definitions. It
is a well-known fact in gravity that dissimilar field definitions
lead to different results for the individual diagrams. Full
amplitudes are of course unaffected by any particular definition
of the field, but results for similar diagrams with different
field choices cannot be immediately compared.

Possible extensions of the large-$D$ considerations discussed here
would be to allow for external matter in the Lagrangian. This
might present some new interesting aspects in the analysis, and
would relate the theory more directly to external physical
observables, {\it e.g.}, scattering amplitudes, and corrections to
the Newton potential at large-$D$ or to geometrical objects such
as a space-time metric. The investigations carried out
in~\cite{B1,B2} could be discussed from a large-$D$ point of view.

Investigations in quantum gravitational cosmology may also present
an interesting work area for applications of the large-$D$ limit,
e.g. quantum gravity large-$D$ big bang models etc.

The analysis of the large-$D$ limit in quantum gravity have
prevailed that using the physical dimension as an expansion
parameter for the theory, is not as uncomplicated as expanding a
Yang-Mills theory at large-$N$. The physical dimension goes into
so many aspects of the theory, {\it e.g.}, the integrals, the
physical interpretations etc, that it is far more complicated to
understand a large-$D$ expansion in gravity than a large-$N$
expansion in a Yang-Mills theory. In some sense it is hard to tell
what the large-$D$ limit of gravity really physically relates to,
{\it i.e.}, what describes a gravitational theory with infinitely
many spatial dimensions? One way to look at the large-$D$ limit in
gravity, might be to see the large-$D$ as a physical phase
transition of the theory at the Planck scale. This suggests that
the large-$D$ expansion of gravity should be seen as a very high
energy scale limit for a gravitational theory.

Large-$N$ expansions of gauge theories~\cite{tHooft} have many
interesting calculational and fundamental applications in, {\it
e.g.}, string theory, high energy, nuclear and condensed matter
physics. A planar diagram limit for a gauge theory will be a
string theory at large distance, this can be implemented in
relating different physical limits of gauge theory and fundamental
strings. In explicit calculations, large-$N$ considerations have
many important applications, {\it e.g.}, in the approximation of
QCD amplitudes from planar diagram considerations or in the theory
of phase-transitions in condensed matter physics, {\it e.g.}, in
theories of superconductivity, where the number of particles will
play the role of ($N$).

High energy physics and string theory are not the only places
where large symmetry investigations can be employed to obtain
useful information. As a calculational tool large-$D$
considerations can be imposed in quantum mechanics to obtain
useful information about eigenvalues and wave-functions of
arbitrary quantum mechanical potentials with a extremely high
precision~\cite{Bjerrum-Bohr:qr}.

Practical applications of the large-$D$ limit in explicit
calculations of $n$-point functions could be the scope for further
investigations as well as certain limits of string theory might be
related to the large-$D$ limit in quantum gravity. The planar
large-$D$ limit found in our investigation might resemble that of
a truncated string limit at large distances? Investigations of the
(Kawai-Lewellen-Tye~\cite{KLT}) open/closed string, gauge
theory/gravity relations could also be interesting to preform in
the context of a large-$D$ expansion. (See
e.g.,~\cite{Bern3,Bern:2002kj,BjerrumBohr:2003af,Bjerrum-Bohr:2003vy}
for resent work on this subject. See also~
\cite{Deser:1998jz,Deser:2000xz}) Such questions are outside the
scope of this investigation, however knowing the huge importance
of large-$N$ considerations in modern physics, large-$D$
considerations of gravity might pose very interesting tasks for
further research.

\begin{acknowledgments}
I would like to thank Poul Henrik Damgaard for suggesting this
investigation and for interesting discussions and advise.
\end{acknowledgments}

\begin{appendix}
\section{Effective 3- and 4-point vertices}\label{app1}
The effective Lagrangian takes the form:
\begin{equation}
{\cal L} = \int d^Dx
\sqrt{-g}\Big(\frac{2R}{\kappa^2}+c_1R^2+c_2R_{\mu\nu}R^{\mu\nu}+\ldots\Big).
\end{equation}
In order to find the effective vertex factors we need to expand
$(\sqrt{-g}R^2)$ and $(\sqrt{-g}R_{\mu\nu}^2)$.

We are working in the conventional expansion of the field, so we
define:
\begin{equation}
g_{\mu\nu} \equiv \eta_{\mu\nu}+\kappa h_{\mu\nu}.
\end{equation}
To second order we find the following expansion for ($\sqrt{-g}$):
\begin{equation}
\sqrt{-g} = \exp\left(\frac12\log (\eta_{\mu\nu}+\kappa
h_{\mu\nu})\right) =
\left(1+\frac\kappa2h_\alpha^\alpha-\frac{\kappa^2}4h_\alpha^\beta
h^\alpha_\beta+\frac{\kappa^2}8(h_\alpha^\alpha)^2\ldots\right).
\end{equation}
To first order in ($\kappa$) we have the following expansion for
($R$):
\begin{equation}\begin{split}
R^{(1)}=\kappa\left[\partial_{\alpha}\partial^{\alpha}h_\beta^\beta-\partial^\alpha\partial^\beta
h_{\alpha\beta}\right],
\end{split}\end{equation}
and to second order in ($\kappa^2$) we find:
\begin{equation}\begin{split}
R^{(2)}&=\kappa^2\Big[-\frac12\partial_\alpha\left[h_\mu^\beta\partial^\alpha
h_\beta^\mu\right]+\frac12\partial_\beta\left[h_\nu^\beta(2\partial_\alpha
h^{\nu\alpha} - \partial_\nu
h_\alpha^\alpha)\right]+\frac14\left[\partial_\alpha h_\beta^\nu +
\partial_\beta h_\alpha^\nu - \partial^\nu h_{\beta\alpha}\right]\left[\partial^\alpha h_\nu^\beta + \partial_\nu h^{\beta\alpha}-\partial^\beta
h_\nu^\alpha\right]\\
&-\frac14\left[2\partial_\alpha h^{\nu\alpha}-\partial^\nu
h_\alpha^\alpha\right]\partial_\nu h_\beta^\beta
-\frac12h^{\nu\alpha}\partial_\nu\partial_\alpha h^\beta_\beta +
\frac12h_\alpha^\nu\partial_\beta\left[\partial^\alpha
h_\nu^\beta+\partial_\nu h^{\beta\alpha} - \partial^\beta
h_\nu^\alpha\right]\Big]
\end{split}\end{equation}
In the same way we find for ($R_{\mu\nu}$):
\begin{equation}\begin{split}
R^{(1)}_{\nu\alpha} = \frac{\kappa}2
\left[\partial_\nu\partial_\alpha h_\beta^\beta-
\partial_\beta\partial_\alpha h_\nu^\beta-
\partial_\beta\partial_\nu h_\alpha^\beta + \partial^2
h_{\nu\alpha}\right],
\end{split}\end{equation}
and
\begin{equation}\begin{split}
R^{(2)}_{\nu\alpha} &=\kappa^2\Big[
-\frac12\partial_\alpha\left[h^{\beta\lambda}\partial_\nu
h_{\lambda\beta}\right] +
\frac12\partial_\beta\left[h^{\beta\lambda}(\partial_\nu
h_{\lambda\alpha}+\partial_\alpha h_{\nu\lambda} -
\partial_\lambda h_{\nu\alpha})\right]\\
&+\frac14\left[\partial_\beta h_\nu^\lambda +
\partial_\nu h_\beta^\lambda - \partial^\lambda h_{\nu\beta}\right]\left[\partial^\lambda h^\beta_\alpha + \partial_\alpha h^{\beta}_\lambda-\partial^\beta h_{\nu\alpha}\right]
-\frac14\left[\partial_\alpha h_\nu^\lambda + \partial_\nu
h_\alpha^\lambda -
\partial^\lambda_{\alpha\nu}\right]\partial_\lambda
h^\beta_\beta\Big].
\end{split}\end{equation}
From these equations we can expand to find  ($R^2$) and
($R^2_{\mu\nu}$).

Formally we can write:
\begin{equation}\begin{split}
\sqrt{-g}R^2 &= \frac12h_\alpha^\alpha R^{(1)}R^{(1)} + 2
R^{(1)}R^{(2)}\\ &=\frac{\kappa^3}2 h_\gamma^\gamma
\Big[\partial_{\alpha}\partial^{\alpha}h_\beta^\beta-\partial^\alpha\partial^\beta
h_{\alpha\beta}\Big]
\Big[\partial_{\sigma}\partial^{\sigma}h_\rho^\rho-\partial^\sigma\partial^\rho
h_{\sigma\rho}\Big] +
2\kappa^3\Big[\partial_{\sigma}\partial^{\sigma}h_\rho^\rho-\partial^\sigma\partial^\rho
h_{\sigma\rho}\Big]\bigg[-\frac12\partial_\alpha\left[h_\mu^\beta\partial^\alpha
h_\beta^\mu\right]\\&+\frac12\partial_\beta\left[h_\nu^\beta(2\partial_\alpha
h^{\nu\alpha} - \partial_\nu
h_\alpha^\alpha)\right]+\frac14\left[\partial_\alpha h_\beta^\nu +
\partial_\beta h_\alpha^\nu - \partial^\nu h_{\beta\alpha}\right]\left[\partial^\alpha h_\nu^\beta + \partial_\nu h^{\beta\alpha}-\partial^\beta
h_\nu^\alpha\right]\\
&-\frac14\left[2\partial_\alpha h^{\nu\alpha}-\partial^\nu
h_\alpha^\alpha\right]\partial_\nu h_\beta^\beta
-\frac12h^{\nu\alpha}\partial_\nu\partial_\alpha h^\beta_\beta +
\frac12h_\alpha^\nu\partial_\beta\left[\partial^\alpha
h_\nu^\beta+\partial_\nu h^{\beta\alpha} - \partial^\beta
h_\nu^\alpha\right] \bigg],
\end{split}\end{equation} and
\begin{equation}\begin{split}
\sqrt{-g}R^2_{\mu\nu} &= \frac12h_\gamma^\gamma
R_{\mu\nu}^{(1)}R^{(1) \mu\nu} -
2h^{\alpha\beta}R_{\mu\alpha}^{(1)}R^{(1) \mu}_\beta+ 2
R_{\mu\nu}^{(1)}R^{(2) \mu\nu} \\ &=\frac{\kappa^3}8
h_\gamma^\gamma \Big[\partial_\nu\partial_\alpha h_\beta^\beta-
\partial_\beta\partial_\alpha h_\nu^\beta-
\partial_\beta\partial_\nu h_\alpha^\beta + \partial^2 h_{\nu\alpha}\Big]
\Big[\partial^\nu\partial^\alpha h_\rho^\rho-
\partial_\rho\partial^\alpha h^{\nu\rho}-
\partial_\rho\partial^\nu h^{\alpha\rho} + \partial^2
h^{\nu\alpha}\Big]\\ & -\frac{\kappa^3}2h^{\rho\sigma}
\Big[\partial_\mu\partial_\sigma h_\gamma^\gamma-
\partial_\gamma\partial_\sigma h_\mu^\gamma-
\partial_\gamma\partial_\mu h_\sigma^\gamma + \partial^2 h_{\mu\sigma}\Big]
\Big[\partial^\mu\partial_\rho h_\beta^\beta-
\partial_\beta\partial_\rho h^{\mu\beta}-
\partial_\beta\partial^\mu h_\rho^\beta + \partial^2 h^{\mu}_{\rho}\Big]
\\&+\kappa^3 \Big[\partial_\nu\partial_\alpha h_\gamma^\gamma-
\partial_\gamma\partial_\alpha h_\nu^\gamma-
\partial_\gamma\partial_\nu h_\alpha^\gamma + \partial^2 h_{\nu\alpha}\Big]
\Big[ -\frac12\partial^\alpha\left[h^{\beta\lambda}\partial^\nu
h_{\lambda\beta}\right] +
\frac12\partial_\beta\left[h^{\beta\lambda}(\partial^\nu
h_{\lambda}^{\alpha}+\partial^\alpha h^{\nu}_{\lambda} -
\partial_\lambda h^{\nu\alpha})\right]\\
&+\frac14\Big[\partial_\beta h^{\nu\lambda} +
\partial^\nu h_\beta^\lambda - \partial^\lambda h^{\nu}_{\beta}\Big]\left[\partial_\lambda h^{\beta\alpha} + \partial^\alpha h^{\beta}_\lambda-\partial^\beta h^{\alpha}_\lambda\right]
-\frac14\left[\partial^\alpha h^{\nu\lambda} + \partial^{\nu}
h^{\alpha\lambda} -
\partial^\lambda h^{\alpha\nu}\right]\partial_\lambda
h^\beta_\beta\Big].
\end{split}\end{equation}

Contracting only on terms which go into the $(3B)^{\rm eff}$ and
$(3C)^{\rm eff}$ index structures, and putting ($\kappa = 1$) for
simplicity, we have found for the ($R^2$) contribution:
\begin{equation}
(R^2)_{(3B)^{\rm eff}} = \partial^2 h^\rho_\rho\Big (-\frac32
\partial_\alpha h^\beta_\mu\partial^\alpha h_\beta^\mu-2h_\beta^\mu
\partial^2 h^\beta_\mu\Big),
\end{equation}
and
\begin{equation}
(R^2)_{(3C)^{\rm eff}} = -\partial^\sigma\partial^\rho
h_{\sigma\rho}\Big (-\frac32
\partial_\alpha h^\beta_\mu\partial^\alpha h_\beta^\mu-2h_\beta^\mu
\partial^2 h^\beta_\mu\Big),
\end{equation}
For the ($R^2_{\mu\nu}$) contribution we have found:
\begin{equation}
(R^2_{\mu\nu})_{(3B)^{\rm eff}} = \frac18h_\gamma^\gamma
\partial^2 h_{\nu\alpha}\partial^2
h^{\nu\alpha}-\frac14\partial_\lambda\partial_\beta
h_\gamma^\gamma
\partial^\beta h^{\nu\alpha}\partial^\lambda h_{\nu\alpha}-\frac12\partial_\lambda\partial_\beta h_\gamma^\gamma
 h^{\nu\alpha}\partial^\beta\partial^\lambda h_{\nu\alpha} +\frac14 \partial_\lambda h_\gamma^\gamma \partial^2 h_{\nu\alpha} \partial^\lambda
 h^{\nu\alpha},
\end{equation}
and
\begin{equation}\begin{split}
(R^2_{\mu\nu})_{(3C)^{\rm eff}} &=
\frac12\partial_\lambda\partial_\beta h^\lambda_\sigma
\partial^\sigma h^{\nu\alpha} \partial^\beta h_{\nu\alpha}
+\partial_\lambda\partial_\beta h_\sigma^\lambda
h^{\nu\alpha}\partial^\sigma\partial^\beta h_{\nu\alpha}
-\frac14\partial^2h_{\lambda\beta}\partial^\lambda h^{\nu\alpha}
\partial^\beta h_{\nu\alpha} -\frac12 \partial^2 h_{\lambda\beta}
h^{\nu\alpha} \partial^\lambda\partial^\beta h_{\nu\alpha}\\ &
-\frac12\partial_\beta h^{\beta\lambda}\partial^2 h_{\nu\alpha}
\partial_\lambda h^{\nu\alpha}
-\frac12h^{\beta\lambda}\partial^2 h_{\nu\alpha} \partial_\lambda
\partial_\beta h^{\nu\alpha}.
\end{split}\end{equation}
This leads to the following results for the (3B)
and (3C) terms:
\begin{figure*}[h]
\begin{tabular}{ll}\vspace{0.1cm}
$\left(\parbox{1cm}{\includegraphics[height=0.8cm]{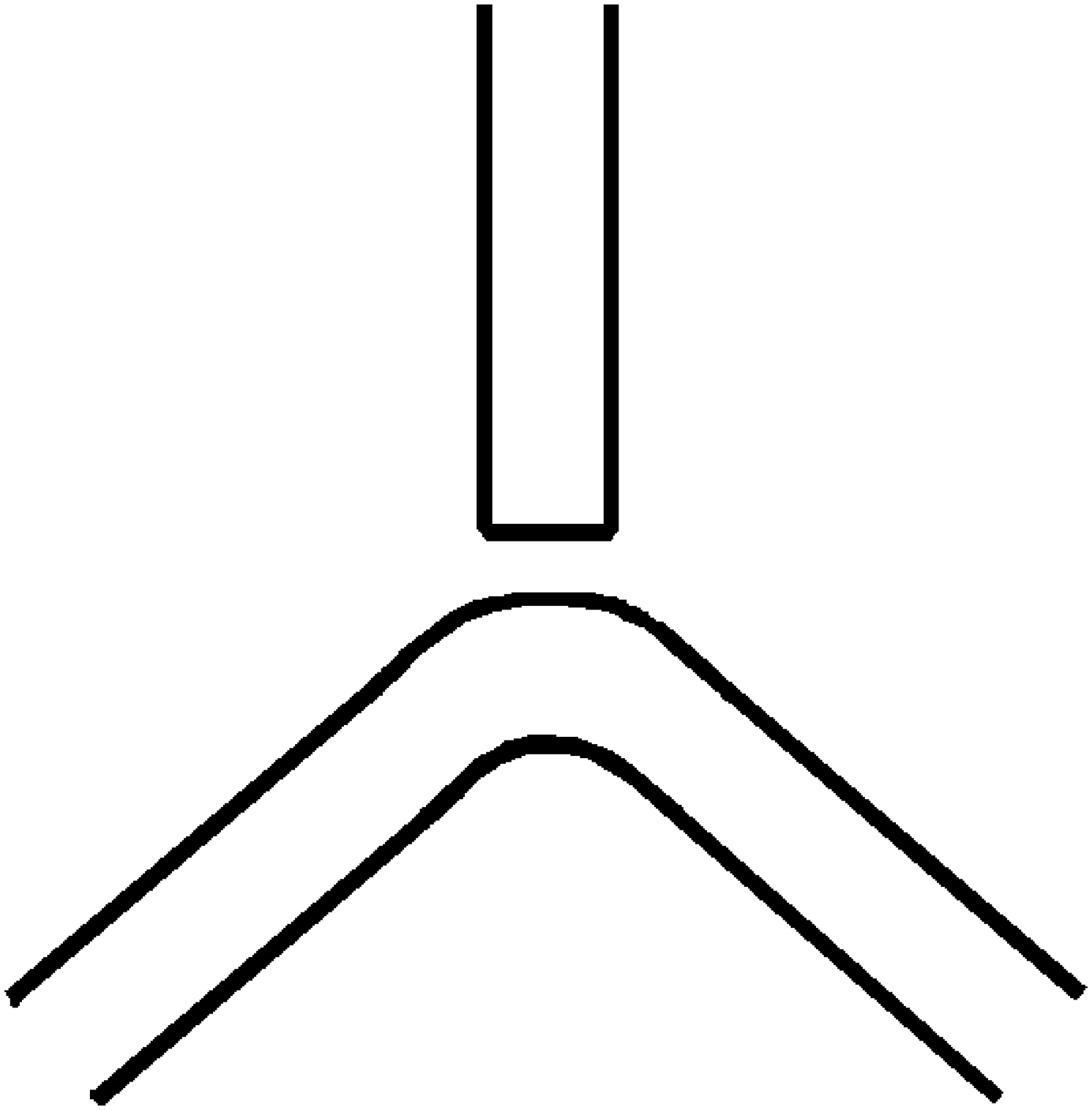}}\right)^{\rm
eff}_{\rm 3B}$ &$\sim {\rm
sym}[-P_3\left(\eta_{\mu\alpha}\eta_{\nu\sigma}\eta_{\beta\gamma}[3c_1k_1^2(k_2\cdot
k_3)+c_2(\frac12(k_1\cdot k_2)(k_1\cdot
k_3)-\frac14k_2^2k_3^2)]\right)$\\ &
$-P_6\left(\eta_{\mu\alpha}\eta_{\nu\sigma}\eta_{\beta\gamma}[2c_1k_1^2k_3^2+c_2(\frac12(k_1\cdot
k_3)^2-\frac14k_2^2(k_1\cdot k_3))]\right)],$
\\ \vspace{0.05cm}
$\left(\parbox{1cm}{\includegraphics[height=0.8cm]{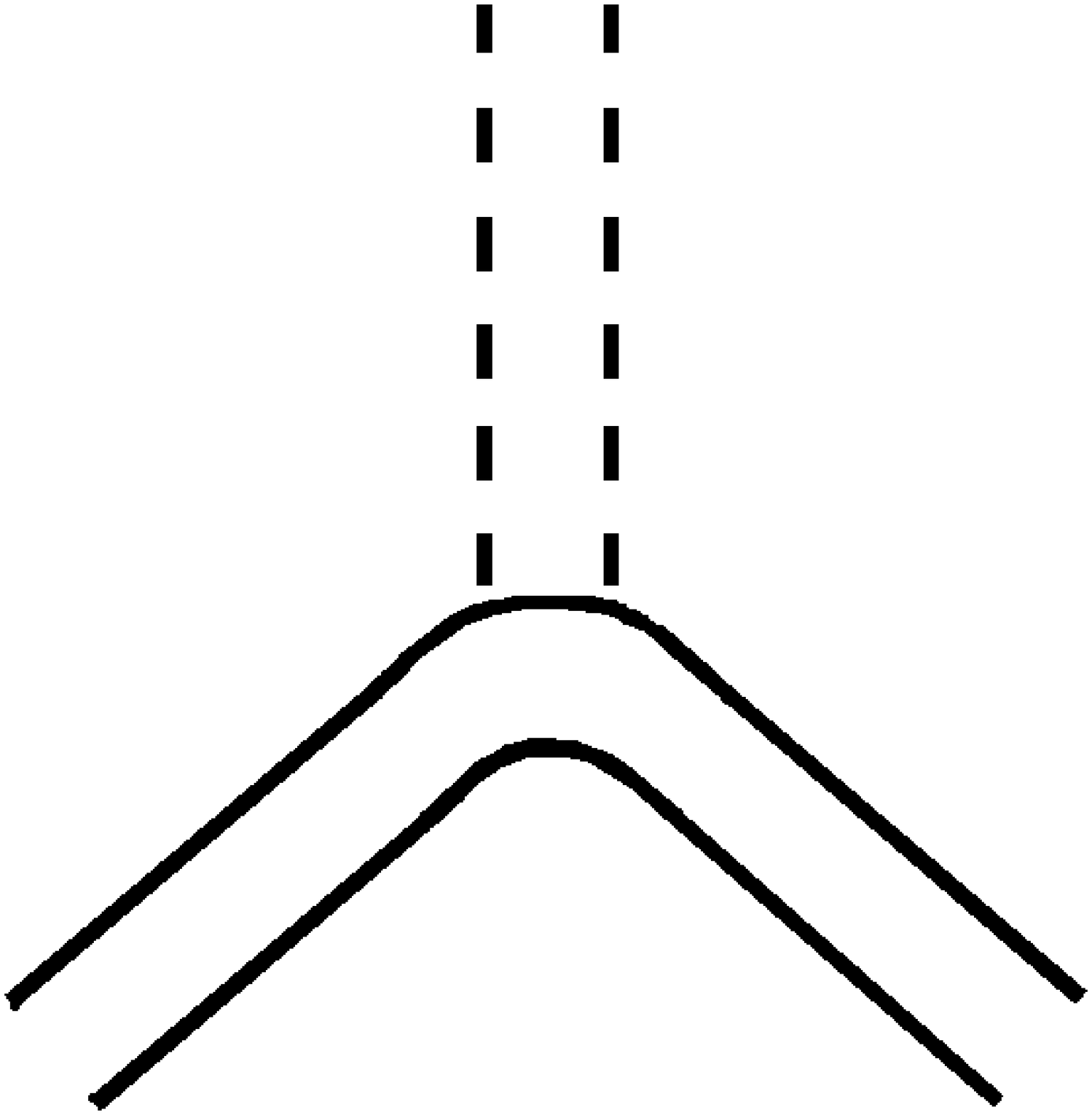}}\right)^{\rm
eff}_{\rm 3C}$ &$\sim {\rm
sym}[-P_3\left(k_{1\mu}k_{1\alpha}\eta_{\nu\sigma}\eta_{\beta\gamma}[3c_1k_2\cdot
k_3]\right)
-2P_6\left(k_{1\mu}k_{1\alpha}\eta_{\nu\sigma}\eta_{\beta\gamma}[c_1
k_3^2]\right)
+\frac12P_6\left(k_{1\mu}k_{2\alpha}\eta_{\nu\sigma}\eta_{\beta\gamma}[c_2k_1\cdot
k_3]\right)
$\\&+$P_6\left(k_{1\mu}k_{3\alpha}\eta_{\nu\sigma}\eta_{\beta\gamma}[c_2k_1\cdot
k_3]\right)
-\frac12P_3\left(k_{2\mu}k_{3\alpha}\eta_{\nu\sigma}\eta_{\beta\gamma}[c_2k_1^2]\right)
-\frac12P_6\left(k_{3\mu}k_{3\alpha}\eta_{\nu\sigma}\eta_{\beta\gamma}[c_2k_1^2]\right)$\\&
$
-\frac12P_6\left(k_{1\beta}k_{3\nu}\eta_{\mu\sigma}\eta_{\alpha\gamma}[c_2k_2^2]\right)
-\frac12P_6\left(k_{3\beta}k_{3\nu}\eta_{\mu\sigma}\eta_{\alpha\gamma}[c_2k_2^2]\right)].$
\end{tabular}
\end{figure*}

Using the same procedure, we can find expression for non-leading
effective contributions, however this will have no implications
for the loop contributions. We have only calculated 3-point
effective vertex index structures in this appendix, 4-point index
structures are tractable too by the same methods, but the algebra
gets much more complicated in this case.

\section{Comments on general $n$-point functions}\label{app2}
In the large-$D$ limit only certain graphs will give leading
contributions to the $n$-point functions. The diagrams favored in
the large-$D$ will have a simpler structure than the full
$n$-point graphs. Calculations which would be too complicated to
do in a full graph limit, might be tractable in the large-$D$
regime.

The $n$-point contributions in the large-$D$ limit are possible to
employ in approximations of $n$-point functions at any finite
dimension, exactly as the $n$-point planar diagrams in gauge
theories can be used, {\it e.g.}, at ($N=3$), to approximate
$n$-point functions. Knowing the full $n$-point limit is hence
very useful in the course of practical approximations of generic
$n$-point functions at finite ($D$). This appendix will be devoted
to the study of what is needed in order calculate $n$-point
functions in the large-$D$ limit.

\begin{figure}[h]
\includegraphics[height=1.2cm]{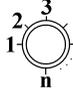}
\caption{The leading 1-loop $n$-point diagram, the index structure
for the external lines is not decided, they can have the momentum
structure, (i.e., as in (3C)), or they can be of the contracted
type, (i.e., as in (3B)). External lines can of course also
originate from 4-point vertex or a higher vertex index structure
such as, e.g., the index structures (4D) or (4E). This possibility
will however not affect the arguments in this appendix, so we will
leave this as a technical issue to be dealt with in explicit
calculations of $n$-point functions.\label{n-point}}
\end{figure}

In the large-$D$ limit the leading 1-loop $n$-point diagrams will
be of the type, as displayed in figure \ref{n-point}.

\begin{figure}[h]
\includegraphics[height=4cm]{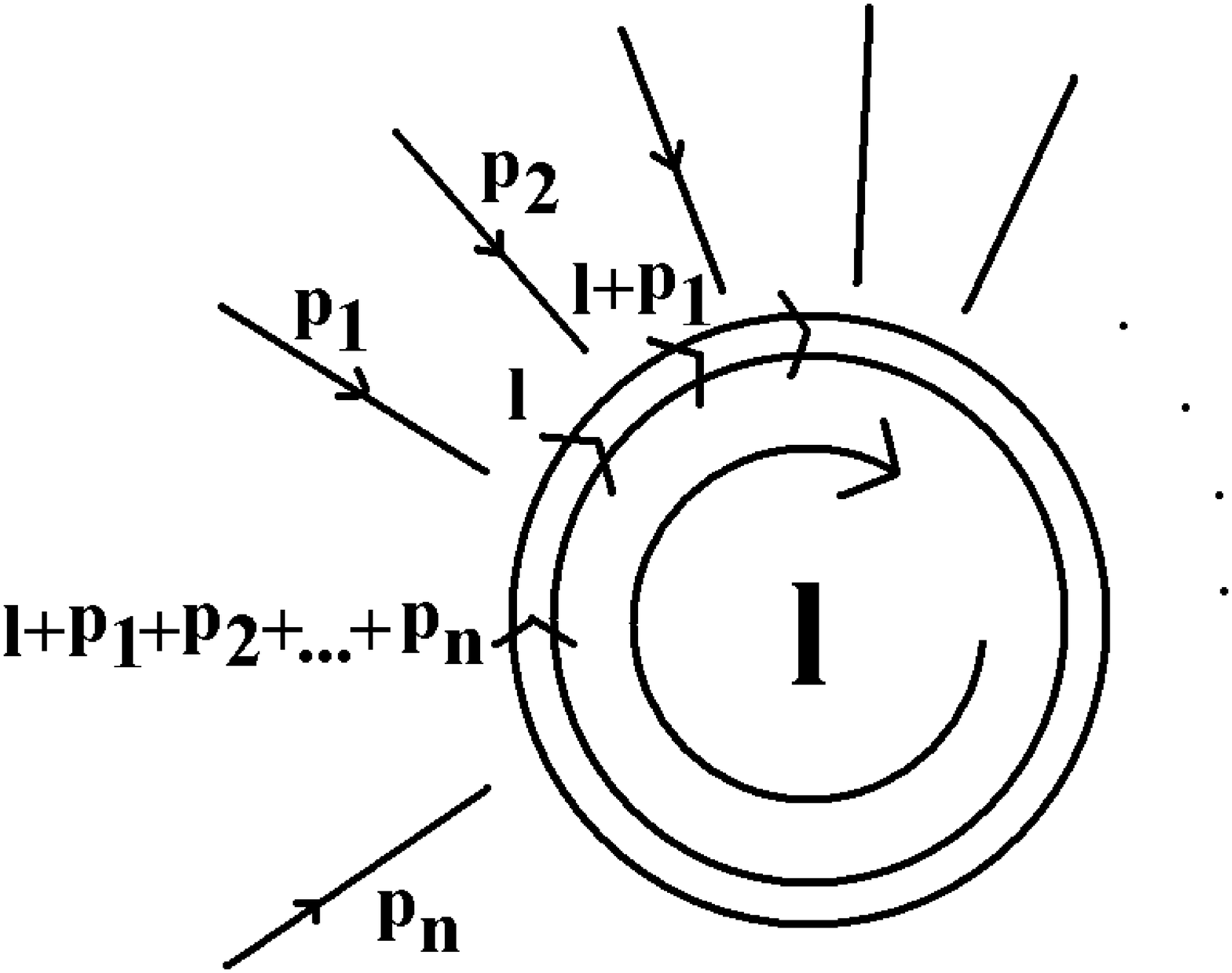}
\caption{A diagrammatic expression for the generic $n$-point
1-loop correction.\label{n-point3}}
\end{figure}

The completely generic $n$-point 1-loop correction will have the
diagrammatic expression, as shown in figure \ref{n-point3}.

\begin{figure}[h]
\parbox{4cm}{\includegraphics[height=1.2cm]{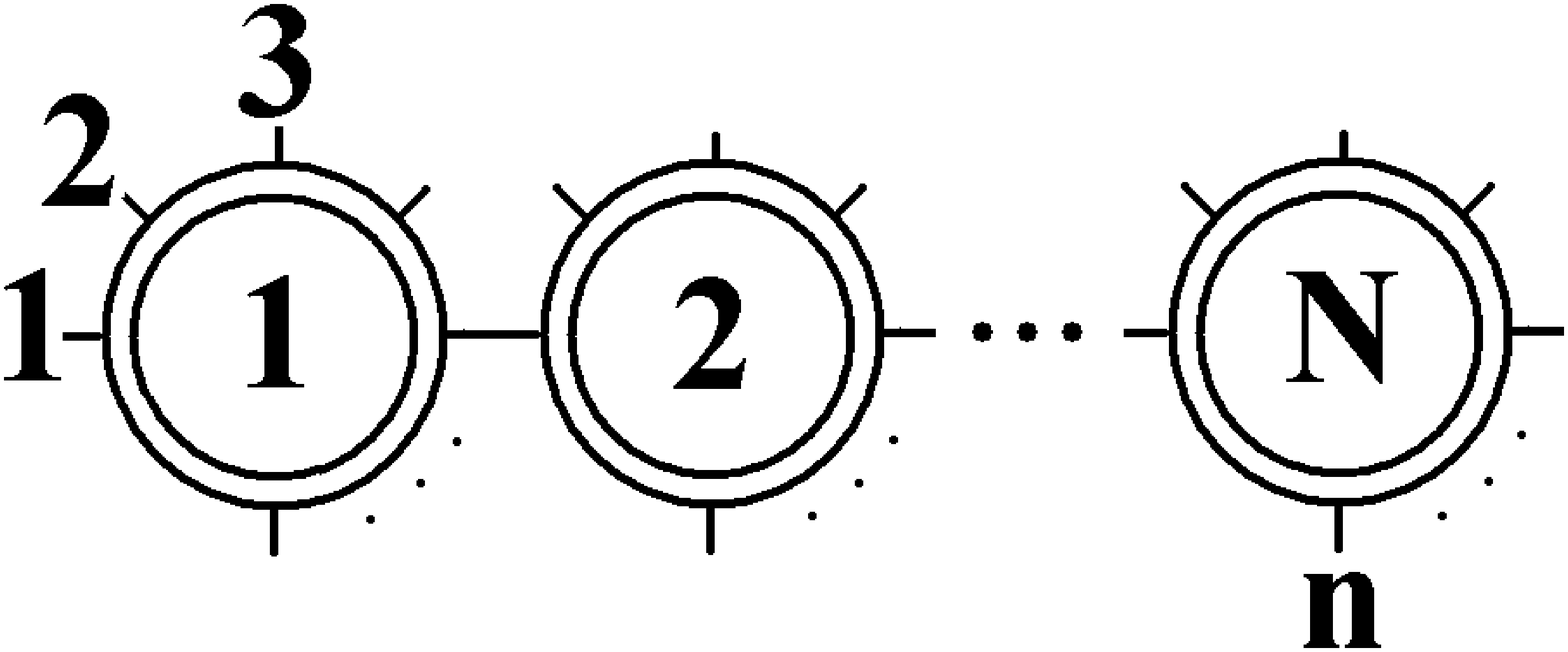}}
\parbox{4cm}{\includegraphics[height=1.2cm]{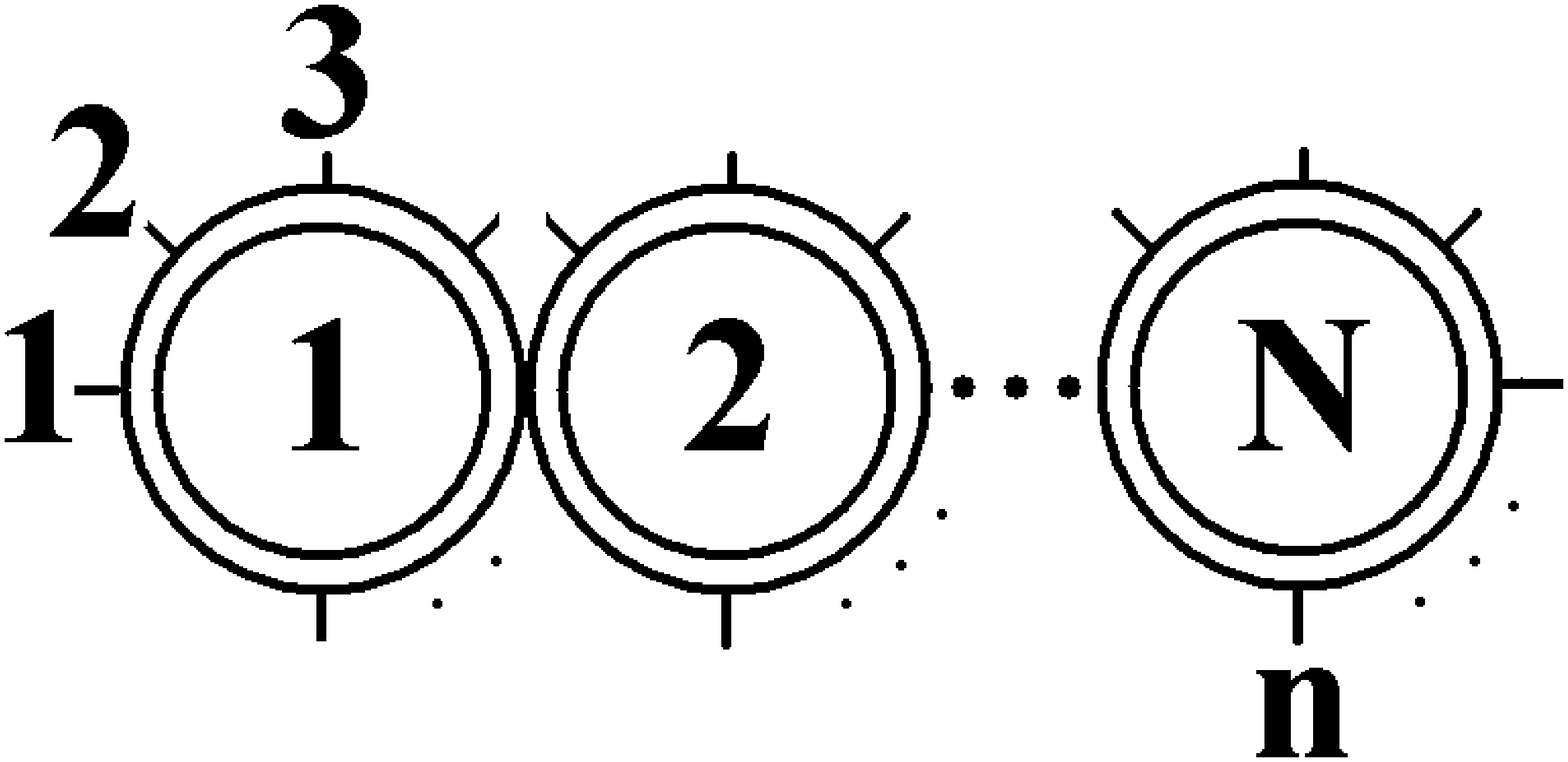}}
\parbox{4cm}{\includegraphics[height=1.2cm]{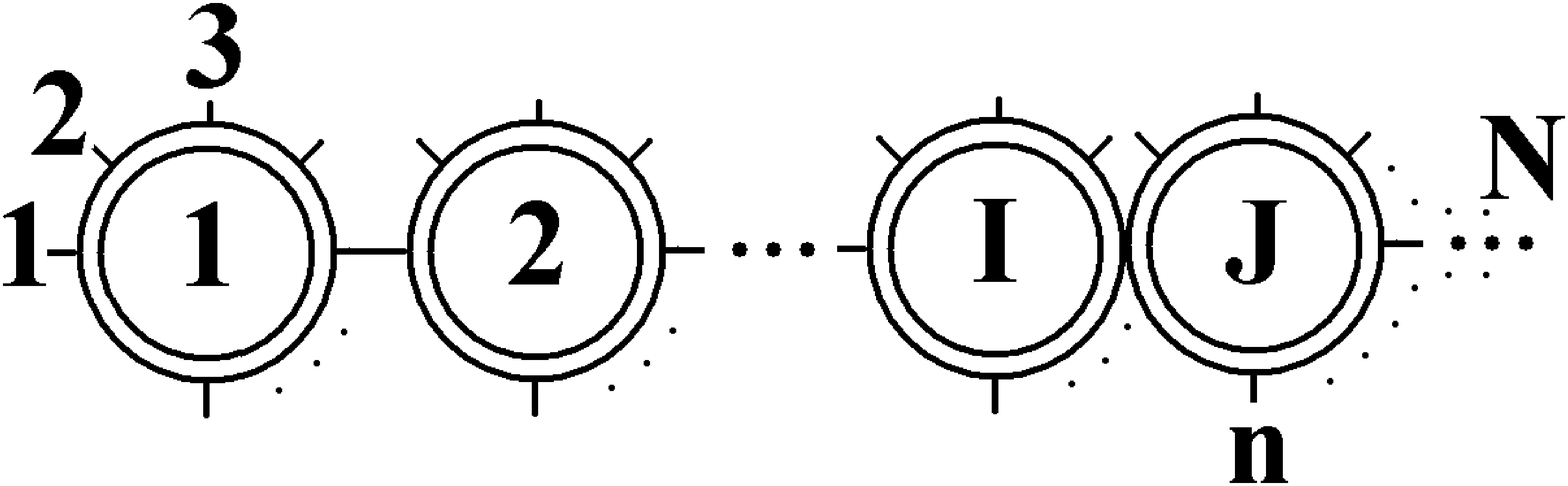}}
\parbox{4cm}{\includegraphics[height=1.8cm]{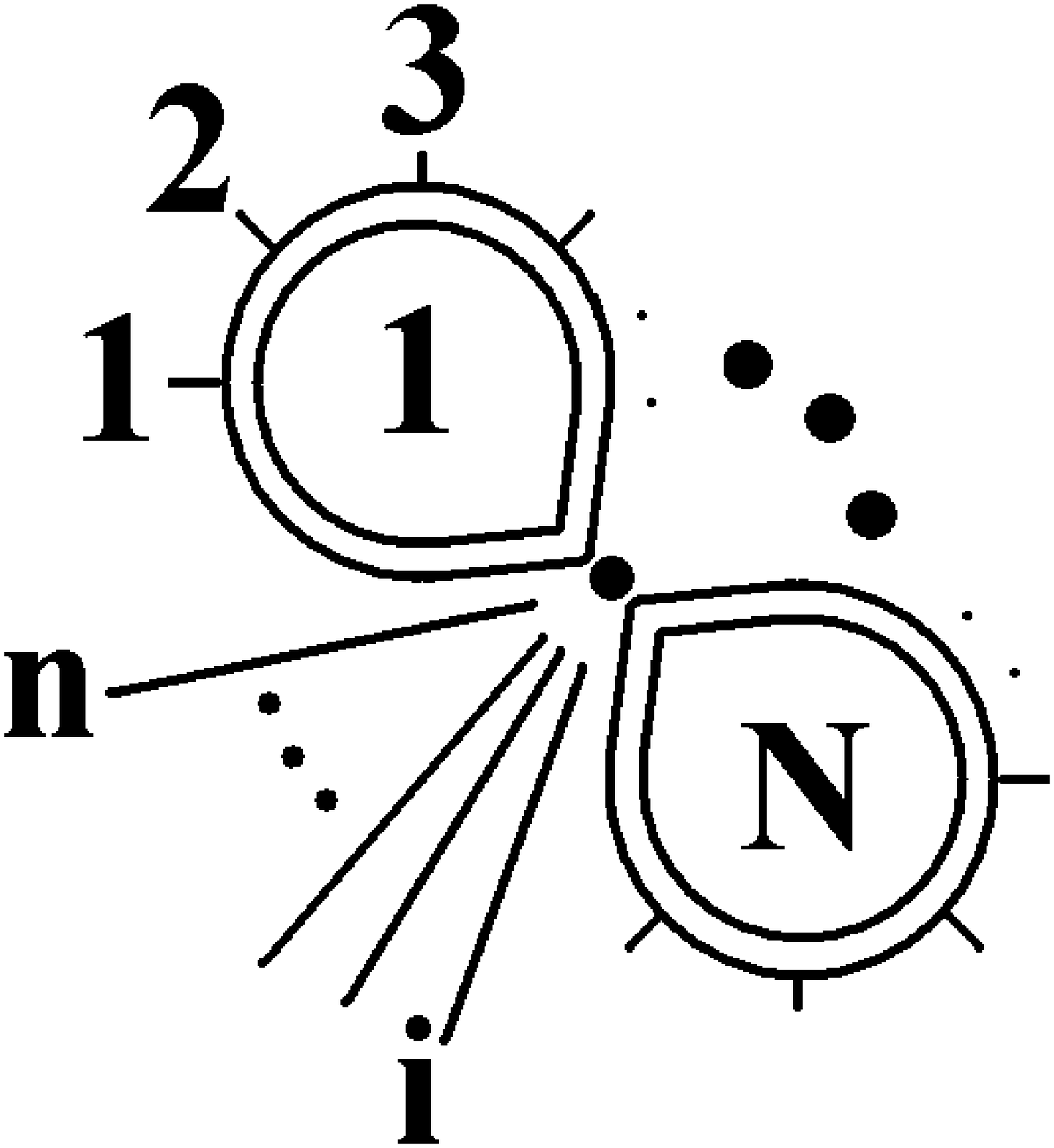}}
\caption{Some generic examples of graphs with leading
contributions in the gravity large-$D$ limit. The external lines
originating from a loop can only have two leading index
structures, (i.e., they can be of the same type as the external
lines originating from a (3B) or a (3C)) loop contribution. Tree
external lines are not restricted to any particular structure,
here the full vertex factor will contribute.\label{larged}}
\end{figure}

More generically the large-$D$, $N$-loop corrections to the
$n$-point function will consist of combinations of leading bubble
and vertex-loop contributions, such as shown in figure
\ref{larged}.

For the bubble contributions it is seen that all loops are
completely separated. Therefore everything is known once the
1-loop contributions are calculated. That is, in order to derive
the full $n$-point sole bubble contribution it is seen that if we
know all $j$-point tree graphs, as well as all $i$-point 1-loop
graphs (where, $i,j \leq n$) are known, the derivation the full
bubble contribution is a matter of contractions of diagrams and
combinatorics.

For the vertex-loop combinations the structure of the
contributions are a bit more complicated to analyze. The loops
will in this case be connected through the vertices, not via
propagators, and the integrals can therefore be joined together by
contractions of the their loop momenta. However, this does in fact
not matter; we can still do the diagrams if we know all integrals
which occurs in the generic 1-loop $n$-point function. The
integrals in the vertex-loop diagrams will namely be of completely
the same type as the bubble diagram integrals. There will be no
shared propagator lines which will combine the denominators of the
integrals. The algebraic contractions of indices will however be
more complicated to do for vertex-loop diagrams than for sole
bubble loop diagrams. Multiple index contractions from products of
various integrals will have to be carried out in order to do these
types of diagrams.

Computer algebraic manipulations can be employed to do the algebra
in the diagrams, so we will not focus on that here. The real
problem lays in the mathematical problem calculating generic
$n$-point integrals.

In the Einstein-Hilbert case, where each vertex can add only two
powers of momentum, we see that problem of this diagram is about
doing integrals such as:
\begin{equation}\begin{split}
I_n^{(\mu_1\nu_1\mu_2\nu_2\cdots\mu_n\nu_n)}= \int d^D l
\frac{l^{\mu_1}l^{\nu_1}l^{\mu_2}l^{\nu_2}\cdots
l^{\mu_n}l^{\nu_n}}{l^2
(l+p_1)^2(l+p_1+p_2)^2\cdots(l+p_1+p_2+\ldots p_n)^2},
\end{split}\end{equation}
\begin{equation}\begin{split}
I_n^{(\nu_1\mu_2\nu_2\cdots\mu_n\nu_n)}= \int d^D l
\frac{l^{\nu_1}l^{\mu_2}l^{\nu_2}\cdots l^{\mu_n}l^{\nu_n}}{l^2
(l+p_1)^2(l+p_1+p_2)^2\cdots(l+p_1+p_2+\ldots p_n)^2},
\end{split}\end{equation}
$$\vdots$$
\begin{equation}\begin{split}
I_n= \int d^D l \frac{1}{l^2
(l+p_1)^2(l+p_1+p_2)^2\cdots(l+p_1+p_2+\ldots p_n)^2},
\end{split}\end{equation}
Because the graviton is a massless particle such integrals will be
rather badly defined. The denominators in the integrals are close
to zero, and this makes the integrals divergent. One way to deal
with these integrals is to use the following parametrization of
the propagator:
\begin{equation}
1/q^2 = \int^\infty_0 dx \exp (-x q^2), {\rm for} \ \ \ q^2>0,
\end{equation}
and define the momentum integral over a complex $(2w)$-dimensional
Euclidian space-time. Thus, what is left to do is gaussian
integrals in a $(2w)$-dimensional complex space-time. Methods to
do such types of integrals are investigated
in~\cite{Capper:vb,Leibbrandt:1975dj,Capper:pv,Capper:dp}. The
results for: ($I_2$), ($I_2^{(\mu)}$), ($I_2^{(\mu\nu)}$) and
($I_3$) are in fact explicitly stated there. It is outside our
scope to actually delve into technicalities about explicit
mathematical manipulations of integrals, but it should be clear
that mathematical methods to deal with the 1-loop $n$-point
integral types exist and that this could an working area for
further research. Effective field theory will add more derivatives
to the loops, thus the nominator will carry additional momentum
contributions. Additional integrals will hence have to be carried
out in order to do explicit diagrams in an effective field theory
framework.
\end{appendix}

\end{document}